\documentclass[pdflatex,sn-basic]{sn-jnl}

\usepackage{subcaption}
\usepackage{makecell}
\usepackage{rotating}
\usepackage{color}
\usepackage[table]{xcolor}
\usepackage{multirow}
\usepackage{hhline}
\usepackage{pifont}
\newcommand{\cmark}{\ding{51}}
\newcommand{\xmark}{\ding{55}}
\usepackage{footmisc}

\usepackage{booktabs}

\usepackage[normalem]{ulem}

\makeatletter

\usepackage{color,array}

\makeatletter

\def\zapcolorreset{\let\reset@color\relax\ignorespaces}
\def\colorrows#1{\noalign{\aftergroup\zapcolorreset#1}\ignorespaces}

\makeatother

\jyear{2023}%

\theoremstyle{thmstyleone}%
\theoremstyle{thmstyletwo}%
\theoremstyle{thmstylethree}%

\begin{document}

\title[KG Embedding Methods for EA: An Experimental Review]{Knowledge Graph Embedding Methods for Entity Alignment: An Experimental Review
\vspace{-0.3cm}}

\author*[1,2]{\fnm{Nikolaos} \sur{Fanourakis}}\email{fanourakis@ics.forth.gr}

\author[1]{\fnm{Vasilis} \sur{Efthymiou}}\email{vefthym@ics.forth.gr}

\author[3]{\fnm{Dimitris} \sur{Kotzinos}}\email{Dimitrios.Kotzinos@cyu.fr}

\author[3]{\fnm{Vassilis} \sur{Christophides}}\email{Vassilis.Christophides@ensea.fr}

\affil[1]{\orgdiv{Institute of Computer Science}, \orgname{FORTH}, \orgaddress{\street{N. Plastira 100}, \city{Heraklion}, \country{Greece}}}

\affil[2]{\orgdiv{Computer Science Department}, \orgname{University of Crete}, \orgaddress{\street{Voutes Campus}, \city{Heraklion}, \country{Greece}}}

\affil[3]{\orgname{Lab. ETIS, CY Cergy Paris University, ENSEA, CNRS UMR 8051}, \orgaddress{\street{2 av. A. Chauvin}, \city{Cergy-Pontoise}, \country{France}}
\vspace{-0.7cm}}

\abstract{In recent years, we have witnessed the proliferation of knowledge graphs (KG) in various domains, aiming to support applications like question answering, recommendations, etc. A frequent task when integrating knowledge from different KGs is to find which subgraphs refer to the same real-world entity, a task largely known as the Entity Alignment. Recently, embedding methods have been used for entity alignment tasks, that learn a vector-space representation of entities which preserves their similarity in the original KGs. A wide variety of supervised, unsupervised, and semi-supervised methods have been proposed that exploit both factual (attribute based) and structural information (relation based) of entities in the KGs. Still, a quantitative assessment of their strengths and weaknesses in real-world KGs according to different performance metrics and KG characteristics is missing from the literature. 
In this work, we conduct the first meta-level analysis of popular embedding methods for entity alignment, based on a statistically sound methodology. Our analysis reveals statistically significant correlations of different embedding methods with various meta-features extracted by KGs and rank them in a statistically significant way according to their effectiveness across all real-world KGs of our testbed. Finally, we study interesting trade-offs in terms of methods' effectiveness and efficiency.
}

\keywords{
\vspace{-0.2cm} Knowledge Graph Embeddings, Entity Alignment}

\maketitle

\section{Introduction}\label{section:intro}

In recent years, we have witnessed the proliferation of \emph{knowledge graphs} (KGs) in various domains, aiming to support applications like entity search~\citep{DBLP:conf/kdd/0001GHHLMSSZ14}, question answering~\citep{DBLP:conf/aaai/AhmetajEFKLO021}, and recommendations~\citep{DBLP:journals/air/TarusNM18}. Typically, KGs store machine-readable \emph{descriptions} of real-world \emph{entities} (e.g., people, movies, books) that capture both \emph{relational} and \emph{factual} information. In this work, we refer to an \emph{entity description} as an identifiable set of property-value pairs that abstracts several data formats, such as relational, RDF, or property graphs. 

As different KGs may independently describe the same real-world entity, a crucial task when integrating knowledge from several KGs is to align their entity descriptions. \emph{Entity alignment (EA)}, also known as \emph{entity resolution}~\citep{DBLP:series/synthesis/2015Christophides,DBLP:journals/csur/ChristophidesEP21}, aims to identify pairs of descriptions from different KGs that refer to the same real-world entity, which we call matching pairs or simply, \emph{matches}.

Figure~\ref{fig:ea-example} shows an example of two KGs, each containing four entity descriptions, represented as nodes, with their properties, represented as edges. The entities described in those KGs can be aligned. For instance, node $v_1$ in $KG_1$ is a description of the director Stanley Kubrick, providing his name and birth year as attributes, and the facts that he directed and wrote the movie The Shining, as well as that he also directed an entity represented by $v_3$ as relations. Node $v_1$ should be aligned with node $v_5$ in $KG_2$, even if the ``name'' attribute is now called ``label'', and its value, ``S. Kubrick'', is slightly different than the name value, ``Stanley Kubrick'', used in $v_1$. The birth year attribute is also missing in $v_5$, compared to $v_1$, and its only relation to $v_6$ is ``directed'', missing the relation ``wrote'' that was also provided in $KG_1$ between $v_1$ and $v_2$. Similarly, the other entity alignments in the example of Figure~\ref{fig:ea-example} should be $v_2$ with $v_6$ (describing the movie ``The Shining''), $v_3$ with $v_7$ (describing the movie ``Barry Lyndon''), and $v_4$ with $v_8$ (describing the actor Philip Stone).

One way to implement entity alignment as a machine learning (ML) task, is to learn a vector-space representation of symbolic KGs, known as \emph{embeddings}.
{Numerical representations (embeddings) of KGs are preferred over symbolic ones in various ML tasks (link/node/subgraph prediction, matching, etc.), as they potentially mitigate the symbolic, linguistic and schematic heterogeneity of independently created KGs and thus aim to simplify knowledge reasoning.
The idea is to embed the nodes (entities) and edges (relations or attributes) of a KG into a low-dimensional vector space. Particularly, we would like similar entities in the original KG to be close to each other in the embedding space and dissimilar entities to be far from each other. In this respect, both \emph{positive} (i.e., actual KG edges) and \emph{negative} (i.e., synthetic edges, non-existing in the actual KG) samples of the KGs are used. This way, by measuring their embeddings distance in the vector space, we can decide whether two entities are matching or not. Any available ground truth regarding the alignment of entities, called \emph{seed alignment}, can be used for training and/or evaluating an embedding method.

\begin{figure} [t!]
    \centering
    \includegraphics[width=0.9\textwidth]{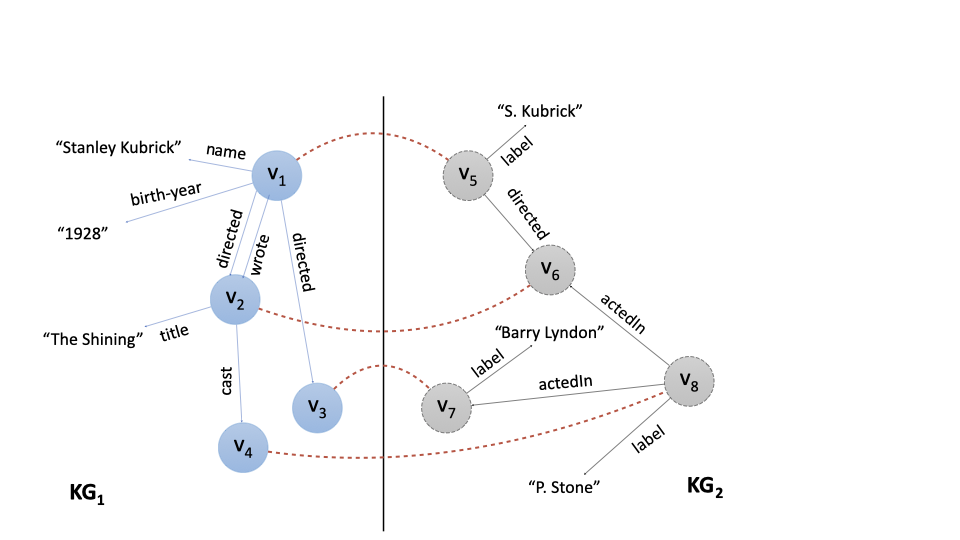}
    \caption{An example of two KGs whose entities can be aligned. The red dashed lines connect the aligned entities ($v_1$,$v_5$), ($v_2$,$v_6$), ($v_3$,$v_7$), and ($v_4$,$v_8$).}
    \label{fig:ea-example}
\vspace{-0.3cm}
\end{figure}

Learning low-dimensional representations of KGs in a way such that the semantic relatedness of entities is captured by the geometrical structures of an embedding space is a challenging task that gave birth to numerous methods. \emph{Embedding-based entity alignment} methods essentially exploit the \emph{relational} (i.e., entity structural neighborhood) and the \emph{factual} part (i.e., entity names/identities, attributes that represent literals) of descriptions. We refer to the former as \emph{relation-based} methods, e.g., MTransE~\citep{DBLP:conf/ijcai/ChenTYZ17}, MTransE+RotatE~\citep{DBLP:journals/pvldb/SunZHWCAL20}, RDGCN~\citep{DBLP:conf/ijcai/WuLF0Y019}, RREA~\citep{DBLP:conf/cikm/MaoWXWL20}, and to the latter as \emph{attribute-based} methods, e.g., MultiKE~\citep{DBLP:conf/ijcai/ZhangSHCGQ19}, AttrE~\citep{DBLP:conf/aaai/TrisedyaQZ19}, KDCoE~\citep{DBLP:conf/ijcai/ChenTCSZ18}, BERT\_INT~\citep{DBLP:conf/ijcai/Tang0C00L20}. Although this research direction is rapidly growing, there are still several open questions regarding the underlying assumptions of methods, as well as, the efficiency and effectiveness of entity alignment in realistic settings. In particular, in this work we address the following missing insights from the literature:

\begin{enumerate}
\item[\textbf{Q1.}] \textbf{Characteristics of Methods.} What are the critical factors that affect the effectiveness of relation-based (e.g., negative sampling, range of neighborhood) and attribute-based methods (e.g., usage of literals) and how sensitive are the methods to hyperparameters tuning?

\item[\textbf{Q2.}] \textbf{Families of Methods.} What is the improvement in the effectiveness of embedding-based entity alignment methods if we consider not only the structural relations of entities, but also their attribute values?

\item[\textbf{Q3.}] \textbf{Effectiveness vs Efficiency Tradeoff.} Is the runtime overhead of each method worth paying, with respect to the achieved effectiveness?

\item[\textbf{Q4.}] \textbf{Characteristics of Datasets.} To which characteristics of the datasets (e.g., sparsity, number of entity pairs in seed alignment, heterogeneity in terms of literals, predicate names and entity names) are supervised, semi-supervised and unsupervised methods sensitive?
\end{enumerate}

Although several recent works~\citep{zeng2021comprehensive,DBLP:journals/corr/abs-2107-07842,DBLP:journals/tkde/WangMWG17,DBLP:journals/pvldb/SunZHWCAL20,DBLP:conf/coling/ZhangLCCLXZ20,zhao2020experimental, DBLP:conf/dsc/JiangLG21, DBLP:conf/dsc/WangLG21, DBLP:journals/pvldb/LeoneHAGW22, DBLP:journals/vldb/ZhangTLJQ22, DBLP:journals/corr/abs-2205-08777} survey embedding-based entity alignment methods, only few of them~\citep{DBLP:journals/pvldb/SunZHWCAL20,DBLP:conf/coling/ZhangLCCLXZ20,zhao2020experimental, DBLP:journals/pvldb/LeoneHAGW22, DBLP:journals/vldb/ZhangTLJQ22, DBLP:journals/corr/abs-2205-08777} conduct an experimental evaluation to obtain useful insights. The conclusions drawn from~\cite{DBLP:conf/coling/ZhangLCCLXZ20} are limited, as it leaves out some representative methods in embedding-based alignment such as MTransE (no negative sampling), KDCoE (semi-supervised exploiting long textual descriptions\footnote{Although the term ``literals'' is inclusive of entity names and textual descriptions, in the EA literature we differentiate between them as follows: \emph{entity names} are the suffixes of entity identifiers (which are sometimes meaningful), \emph{textual descriptions} are the values of some pre-determined attributes that typically span several sentences, as opposed to the simple \emph{literals}, used to describe all other factual data, typically of much shorter length (1-2 words).}), RREA (semi-supervised exploiting structural information), AttrE (unsupervised), BERT\_INT (supervised exploiting both structural and factual information),  while neither RREA nor BERT\_INT were part of OpenEA~\citep{DBLP:journals/pvldb/SunZHWCAL20}, as both were published later. In addition, \cite{DBLP:journals/pvldb/LeoneHAGW22} does not include MTransE, MTransE+RotatE, AttrE, KDCoE, while RREA is not included in~\cite{DBLP:journals/pvldb/LeoneHAGW22,DBLP:journals/vldb/ZhangTLJQ22,DBLP:journals/corr/abs-2205-08777} either. Moreover, benchmarking efforts such as~\cite{DBLP:conf/coling/ZhangLCCLXZ20}, OpenEA~\citep{DBLP:journals/pvldb/SunZHWCAL20}, EAE~\citep{zhao2020experimental} and \cite{DBLP:journals/pvldb/LeoneHAGW22, DBLP:journals/vldb/ZhangTLJQ22,DBLP:journals/corr/abs-2205-08777}, do not shed light on questions Q1, Q2 and Q3, addressed in our work. Furthermore, only a subset of the dataset characteristics we study in our work, such as the density of the KGs, and the similarity of entity names, have been considered by previous works to answer question Q4.

To compare the effectiveness and efficiency of the methods in realistic settings, we have extended the testbed of datasets with pairs of KGs usually considered in related empirical studies. Specifically, OpenEA, EAE and \cite{DBLP:journals/pvldb/LeoneHAGW22, DBLP:journals/vldb/ZhangTLJQ22,DBLP:journals/corr/abs-2205-08777} employ only datasets with a low number of entities featuring descriptions and literal values. In our testbed, we have included five additional datasets whose unique characteristics allow us to draw new insights regarding the evaluated EA methods, which were not previously reported in~\cite{DBLP:journals/pvldb/SunZHWCAL20,DBLP:conf/coling/ZhangLCCLXZ20,zhao2020experimental, DBLP:journals/pvldb/LeoneHAGW22, DBLP:journals/vldb/ZhangTLJQ22,DBLP:journals/corr/abs-2205-08777}. More precisely, supervised methods like RDGCN exploiting KG relations, are outperformed by unsupervised (AttrE) and semi-supervised (KDCoE) methods that exploit the similarity of literals in datasets of decreasing density, but with rich factual information (i.e., attributes). 

Rather than simply reporting the raw experimental results, we conduct a meta-level analysis, aiming to find statistically significant correlations between the methods and the dataset characteristics (meta-features). Furthermore, we consider the non-parametric Friedman test~\citep{DBLP:journals/jmlr/Demsar06} and the post-hoc Nemenyi test~\citep{nemenyi1963distribution}, in order to perform a pairwise comparison of the methods and rank them based on their effectiveness across all datasets of our testbed. Finally, we are interested in the methods' training time curve and potential effectiveness vs efficiency trade-offs instead of simply reporting overall runtimes (as in OpenEA and EAE).

In a nutshell, the contributions of this work are the following:

\begin{itemize}

\item In Section~\ref{section:two}, we present a qualitative comparison of state-of-the-art embedding-based entity alignment methods that span from supervised, i.e.,  MTransE~\citep{DBLP:conf/ijcai/ChenTYZ17}, MTransE+RotatE~\citep{DBLP:journals/pvldb/SunZHWCAL20}, MultiKE~\citep{DBLP:conf/ijcai/ZhangSHCGQ19}, RDGCN~\citep{DBLP:conf/ijcai/WuLF0Y019}, RREA(basic)~\citep{DBLP:conf/cikm/MaoWXWL20}, BERT\_INT~\citep{DBLP:conf/ijcai/Tang0C00L20}, to unsupervised, i.e.,  AttrE~\citep{DBLP:conf/aaai/TrisedyaQZ19}, and semi-supervised, i.e., KDCoE~\citep{DBLP:conf/ijcai/ChenTCSZ18}, RREA(semi)~\citep{DBLP:conf/cikm/MaoWXWL20}, paradigms. They are representative methods of different embedding families covering both relation- and attribute-based, but also considering one-hop and multi-hop neighborhoods in KGs, as well as different negative sampling strategies.

\item In Section~\ref{section:three}, we describe our framework for a fair empirical comparison of the different methods. We detail the extended testbed of datasets that exhibit diverse characteristics (w.r.t. KG density, entity naming, textual descriptions, etc.) usually encountered in reality along with the corresponding pre-processing pipelines. We additionally introduce the evaluation protocol and metrics capturing different aspects of the methods' effectiveness.

\item In Section~\ref{section:four}, we report and analyze the results of a series of experiments, including a comparison to a state-of-the-art non-embedding-based method, PARIS~\citep{DBLP:journals/pvldb/SuchanekAS11}, conducted to answer the four open questions introduced previously, using a reliable, statistically sound methodology. First, we discover a statistical significant ranking of the methods according to their effectiveness across all real-world KGs of our testbed. Then, we study interesting trade-offs in terms of their effectiveness and efficiency. Last but not least, we extract statistically significant correlations between the methods' performance with various characteristics of our datasets (i.e., meta features).
\end{itemize}

Finally, the main conclusions drown from our experiments, as well as the plans for future work, are discussed in Section~\ref{section:five}.

\section{Entity Alignment With KG Embeddings}\label{section:two}

In this section, we first formally define the entity alignment problem on Knowledge Graphs (KGs), along with some related constraints. Then, we provide a qualitative comparison of entity alignment methods based on KG embedding.

\subsection{The Entity Alignment Problem}

Following the typical notation used in the literature~\citep{DBLP:conf/ijcai/ZhangSHCGQ19,DBLP:conf/emnlp/WangYY20}, we assume that entities (with the corresponding \emph{entity names})\footnote{\label{foot:name}The string suffix after the last slash of a URI of an entity or an attribute~\citep{DBLP:conf/ijcai/ZhangSHCGQ19}.}, are described in KGs by a collection of edges $\left<h,r,t\right>$, whose head $h$ is always an entity, and tail $t$ may be either another entity, in which case we call this edge a \emph{relation edge} and $r$ a \emph{relation}, or a literal (e.g., number, date, string), in which case we call this edge an \emph{attribute edge} and $r$ an \emph{attribute} with its corresponding \emph{attribute name}\footref{foot:name}. We represent a knowledge graph as $KG = (E, R, A, L, X, Y)$, where $E$ is a set of entities, $R$ is a set of relations, $A$ is a set of attributes, $L$ is a set of literals, $X \subseteq(E \times R \times E)$ and $Y \subseteq(E \times A \times L)$ are the sets of relation and attribute edges of the KG, respectively. Given a source $KG_1 = (E_1, R_1, A_1, L_1, X_1, Y_1)$ and a target $KG_2$ = $(E_2, R_2, A_2$, $L_2, X_2, Y_2)$, the task of entity alignment is to find pairs of matching entities $M = \{(e_i,e_j)\in E_1 \times E_2 \mid e_i \equiv e_j \}$, where ``$\equiv$'' denotes the equivalence relationship \citep{DBLP:conf/ijcai/ZhangSHCGQ19,DBLP:conf/emnlp/WangYY20}.
A subset $\delta \subseteq M$ of the matching pairs may be used as a seed alignment for training. For instance, in Figure~\ref{fig:ea-example}, the entities of the two KGs are $E_1 = \{v_1, v_2, v_3, v_4\}$ and $E_2 = \{v_5, v_6, v_7, v_8\}$. The relations are $R_1 = \{cast, directed, wrote\}$ and $R_2 = \{directed, actedIn\}$, while the attributes are $A_1 = \{name, birth$-$year, title\}$, $A_2=\{label\}$ and the literals are $L_1 = \{``Stanley Kubrick'', ``1928'', ``The Shining''\}$.  $L_2 = \{``S. Kubrick'', ``Barry Lyndon'', ``P. Stone''\}$. The relation edges are $X_1 = \{(v1,directed,v2), (v2,directed,v1), (v1,directed,v3), (v2, directed, v4)\}$, $X_2$ = $\{(v5,directed,v6)$, $(v8,actedIn,v6)$, $(v8,actedIn,v7)\}$, while the attribute edges are $Y_1$ = $\{(v1,name, ``Stanley Kubrick'')$, $(v1,birth$-$year,``1928'')$, $(v2,title,``The  Shining'')\}$ and $Y_2$ = $\{(v5,label, ``S. Kubrick'')$, $(v8,label,``P. Stone'')$, $(v7,label,``Barry Lyndon'')\}$. The task of entity alignment is to find the matches (denoted by dashed edges in Figure~\ref{fig:ea-example}) $M$ = $\{(v1,v5), (v2, v6), (v3, v7), (v4, v8)\}$.

In practice, all the evaluated entity alignment methods rely on a number of assumptions/constraints, as listed below: 
\begin{itemize}
    \item Every entity is assumed to be the head of at least one relation edge (so we do not consider entities that are not part of a connected component of the KG):\\
$\forall e \in E, \exists r \in R, t \in E: (e, r, t) \in X.$
    \item 1-to-1 constraint: Every entity in $E_1$ should be matched to exactly one entity in $E_2$: $\forall e_i \in E_1\;  \left(\exists e_j \in E_2: (e_i, e_j) \in M\right) \wedge \left(\nexists e_j' \in E_2: (e_i, e_j') \in M\right)$ and vice versa
    $\forall e_j \in E_2\;  \left(\exists e_i \in E_1: (e_i, e_j) \in M\right) \wedge \left(\nexists e_i' \in E_1: (e_i', e_j) \in M\right).$
    This also implies that $\mid$$M$$\mid = \mid$$E_1$$\mid = \mid$$E_2$$\mid$.
\end{itemize}

\subsection{Knowledge Graph Embeddings for Entity Alignment}\label{sec:kg_embedding_for_EA}

\begin{figure}
    \centering
\includegraphics[width=0.8\textwidth]{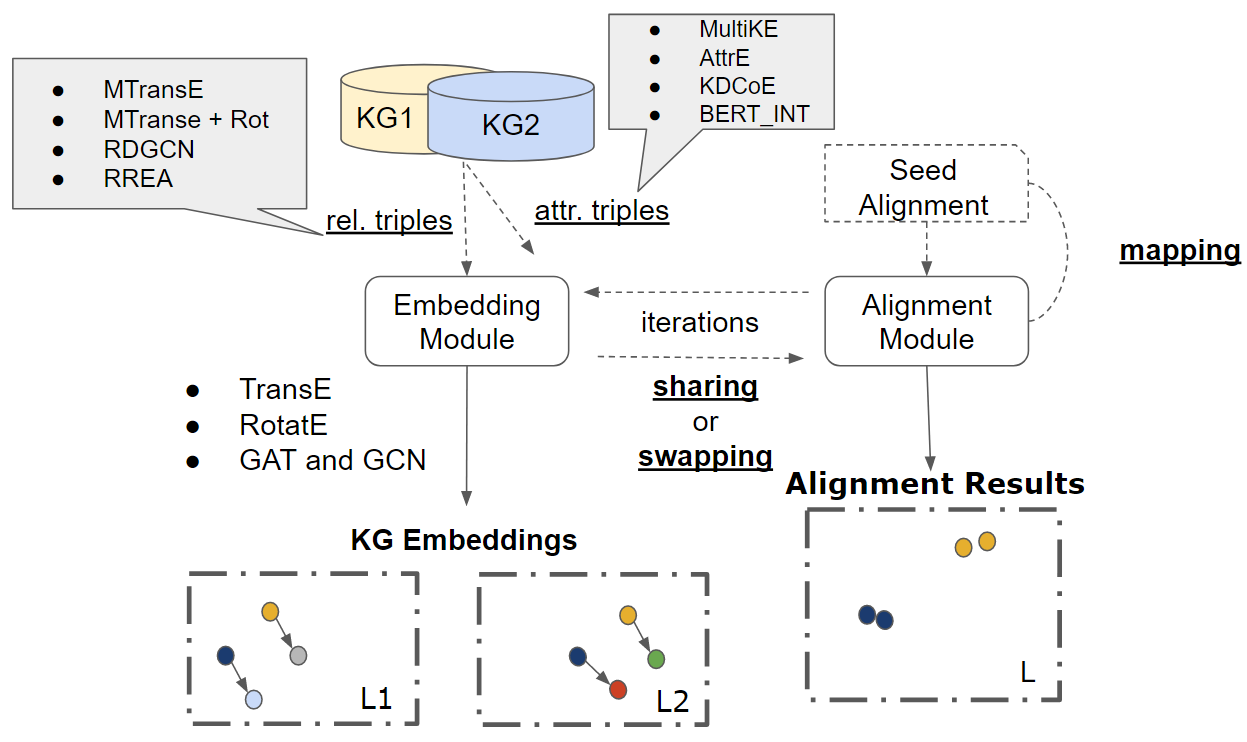}
    \caption{General framework of KG embeddings for entity alignment.}
    \label{fig:general_framework}
\vspace{-0.5cm}
\end{figure}

KG embedding methods aim to learn a low-dimensional vector-space representation of symbolic KGs, known as \emph{embeddings}. The idea is to embed the nodes (entities) and edges (relations or attributes) of a KG in an embedding space in a way that preserves their similarity in the original KG. Embedding methods have been proven to be effective in many machine learning tasks, such as node classification~\citep{DBLP:conf/iclr/KipfW17} that aims to assign entity types to KG nodes, or link prediction~\citep{DBLP:conf/nips/BordesUGWY13, DBLP:conf/iclr/SunDNT19} that aims to find missing relations between entities in a single KG. Lately, several embedding-based methods have been also proposed for entity alignment, exploiting either relation edges (\emph{relation-based methods}), such as MTransE~\citep{DBLP:conf/ijcai/ChenTYZ17}, MTransE + RotatE~\citep{DBLP:journals/pvldb/SunZHWCAL20}, RDGCN~\citep{DBLP:conf/ijcai/WuLF0Y019}, RREA(basic)~\citep{DBLP:conf/cikm/MaoWXWL20}, RREA(semi)~\citep{DBLP:conf/cikm/MaoWXWL20} or attribute edges (\emph{attribute-based methods}), such as MultiKE~\citep{DBLP:conf/ijcai/ZhangSHCGQ19}, AttrE~\citep{DBLP:conf/aaai/TrisedyaQZ19}, KDCoE~\citep{zhang2017knowledge}, and BERT\_INT~\citep{DBLP:conf/ijcai/Tang0C00L20}.

Figure~\ref{fig:general_framework} depicts the building blocks of embedding-based entity alignment methods: (i) The \emph{embedding module} $S_K$ that encodes the entities of each KG in an embedding space ($L1$ for $KG_1$ and $L2$ for $KG_2$) according to the \emph{relational} (i.e., entity structural neighborhood) and/or the \emph{factual} part (i.e., entity names/identities, literals/text) of descriptions.
(ii) The \emph{alignment module} $S_A$ that aligns the produced entity embeddings using the \emph{seed alignment} (supervised) or \emph{attribute-values} similarity (unsupervised), or \emph{both} (semi-supervised). It produces a common embedding space for the entities of two KGs, in order to generate the alignment result according to a distance metric (e.g., Euclidean), using three different techniques known as \emph{sharing}, \emph{swapping} and \emph{mapping}. Sharing and Swapping, update directly the entity embeddings produced by the embedding module according to the available similarity evidence of entities, while Mapping essentially learns a linear transformation between the two embedding spaces of aligned KGs. In the rest of this section, we will detail popular methods that implement those modules.

\subsubsection{Embedding Module}\label{embedding_module}

KG embedding methods proposed for the task of link prediction are used to implement the embedding module of entity alignment methods. There are several families of KG embedding methods for link prediction have been proposed in the literature, e.g.,~\cite{DBLP:journals/corr/YangYHGD14a, DBLP:conf/icml/TrouillonWRGB16,DBLP:conf/nips/NickelK17}. In this paper, we are interested in contrasting translational methods such as TransE~\citep{DBLP:conf/nips/BordesUGWY13} and RotatE~\citep{DBLP:conf/iclr/SunDNT19}, with Graph Neural Networks such as Graph Convolutional Networks (GCNs)~\citep{DBLP:conf/iclr/KipfW17} and Graph Attention Networks (GATs)~\citep{DBLP:journals/corr/abs-1710-10903}. 

\paragraph{Translational Methods}\label{translational}
Translational methods use distance-based scoring functions in order to optimize a margin-based loss function and learn the embeddings of entities in a KG. A distance-based scoring function is a function that measures the plausibility of a relation edge $\left<h,r,t\right>$ i.e., it measures the distance of the embedding of the head to the embedding of the tail entities, given the embedding of the relation. A margin-based loss function is a function that these methods aim to minimize, in order to minimize the distance of entity embeddings by a certain margin, computed by a distance-based scoring function. The key difference among all those translational methods is based on the degree that they are able to capture more complex graph structures such as cycles, by adopting the appropriate operator in the scoring function.

\textit{\textbf{TransE}}~\citep{DBLP:conf/nips/BordesUGWY13} is one of the most widely used translational KG embedding methods. In this method, both entities and relations are represented in the same vector space. The relation $r$ is equivalent to the translation of vectors from head entity $h$ to the tail entity $t$. If $\left<h,r,t\right> \in X$, then the embedding $\textbf{t}$ of $t$ should be close to the embedding $\textbf{h}$ of $h$, plus the vector $\textbf{r}$ of $r$ , i.e., $\textbf{h}+\textbf{r}\approx \textbf{t}$. Formally, TransE minimizes the margin-based loss function: 
\begin{equation}
\label{eq:TransE}
J_{S E}=\sum_{x \in X} \sum_{x' \in X'} \max \left(0, \gamma+f\left(\mathbf{x}\right)-f\left(\mathbf{x'}\right)\right),
\end{equation}
where $f\left(\mathbf{h},\mathbf{r},\mathbf{t}\right) =\; \mid$$\mathbf{h}+\mathbf{r}-\mathbf{t}$$\mid$ is the scoring function, $X$ is the set of \emph{positive relation edges} (relation edges that exists in the KG), $X'$ is the set of \emph{negative relation edges} (relation edges that do not exist in the KG), and $\gamma$ is the margin hyperparameter. Each negative edge $x' \in X'$ is created by replacing the head or the tail of a positive edge in $X$ with a random entity, ensuring that $x' \notin X$.

\textit{\textbf{RotatE}}~\citep{DBLP:conf/iclr/SunDNT19} is a translation-based embedding model that, unlike TransE, infers various relation patterns, such as symmetries. Specifically, RotatE maps the entities and relations to the complex vector space and defines each relation as a rotation from the head entity to the tail entity (Figure~\ref{fig:Rotate_example}). Given a relation edge $\left<h,r,t\right>$, we expect that $\textbf{t} = \textbf{h} \circ \textbf{r}$, where $\circ$ denotes the Hadamard \citep{million2007hadamard} (element-wise) product.

This model aims to minimize the margin-based loss function 
\begin{equation}
\label{eq:RotatE}
L=-\log \sigma\left(\gamma-d_{r}(\mathbf{h}, \mathbf{r}, \mathbf{t})\right)-\sum_{i=1}^{n} \frac{1}{k} \log \sigma\left(d_{r}\left(\mathbf{h}_{i}^{\prime}, \mathbf{r}, \mathbf{t}_{i}^{\prime}\right)-\gamma\right),
\end{equation}
by maximizing the scores of positive relation edges and minimizing the scores of negative relation edges, where $d_r\left(\left<h,r,t\right>\right) =\; \mid$$\mathbf{h}\circ\mathbf{r}-\mathbf{t}$$\mid$ is a scoring function, $\gamma$ is a fixed margin, $\sigma$ is the sigmoid function, and $(h'_i,r,t'_i)$ is the $i$-th negative edge. RotatE, like TransE, creates the negative relation edges by replacing the head and the tail of positive relation edges randomly.
\paragraph{Graph Neural Network Methods} \label{ssec:GCN}

Translational methods cannot deal with various complex graph structures. For example, TransE~\citep{DBLP:conf/nips/BordesUGWY13} cannot deal with triangular structures like the one in Figure~\ref{fig:triangular}, because it requires the three equations $v_1+r_a \approx v_2$, $v_2+r_a \approx v_3$ and $v_1+r_a \approx v_3$ to hold at the same time. This is impossible, because for satisfying the former two equations we would have $v_1+2 r_a \approx v_3$ which is contradictory to the equation $v_1+r_a \approx v_3$. 
 
In order to cope with that, Graph Neural Network (GNN) methods have been proposed. GNNs learn entity embeddings, by recursively aggregating the representations of neighboring nodes. They essentially rely on \emph{message passing}, according to which, each graph node recursively receives and aggregates features (node representations) from its neighbors in order to represent the local graph structure.   

There is a range of GNN variants, that implement different aggregation strategies. In this section, we focus on standard graph convolutional networks (GCNs)~\citep{DBLP:conf/iclr/KipfW17} and graph attention networks (GATs)~\citep{DBLP:journals/corr/abs-1710-10903}, since they are the core of both RDGCN~\citep{DBLP:conf/ijcai/WuLF0Y019} and RREA~\citep{DBLP:conf/cikm/MaoWXWL20}; two of the proposed methods that we evaluate in this study and describe in Section~\ref{ssec:knowledge_graph_embeddings_relations}.

\begin{figure}[ht]
\begin{minipage}{.5\textwidth}
    \centering
    \includegraphics[height=3cm]{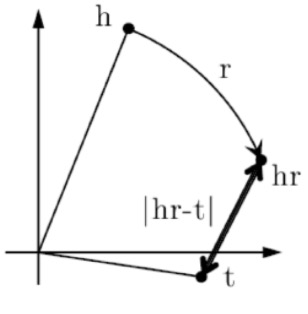}
    \caption{Example of RotatE.}
    \label{fig:Rotate_example}
\end{minipage}
\begin{minipage}{.5\textwidth}
    \centering
    \includegraphics[height=3cm]{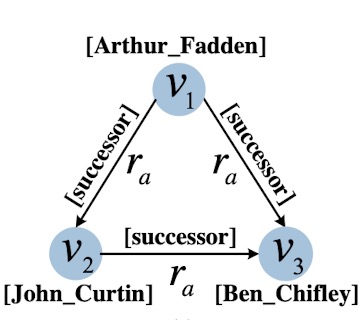}
    \caption{Triangular structure~\citep{DBLP:conf/ijcai/WuLF0Y019}.}
    \label{fig:triangular}
\end{minipage}
\vspace{-0.5cm}
\end{figure}

\textit{\textbf{GCN}}~\citep{DBLP:conf/iclr/KipfW17} takes as input the randomly initialized entity embeddings of the KG, which is treated as an undirected graph. Then, it learns a set of layer-specific weights, known as \emph{filters} or \emph{kernels}, that are multiplied with the input embeddings. In essence, it acts as a sliding window across the KG that learns entity features while preserving useful structural information from the neighborhoods. GCN uses the following function
\begin{equation}
\label{eq:gcn_layer}
H_{i}^{(l+1)}=\sigma\left(\sum_{j \in \mathcal{N}i} \frac{1}{c_{i j}} W^{(l)} H_{j}^{(l)}\right)
\end{equation}
to aggregate the entity embeddings of $l$ layers, where $\sigma$ is an activation function, $N_{i}$ is the set of the one-hop neighbors of the central entity $i$ (including itself by adding a self-loop), $c_{ij}$ is a normalization constant that defines isotropic average computation (each neighbor contributes equally to update the embedding of the central entity), $W^{(l)}$ is a trainable layer-specific weighted matrix for feature transformation and $H^{(l)}$ are the entity embeddings for layer~$l$. More precisely, in order to learn the final embedding of a central entity, GCN sums its embedding with the neighbors embeddings.

\textit{\textbf{GAT}}~\citep{DBLP:journals/corr/abs-1710-10903} expands the aggregation function of GCN, by an \emph{attention mechanism} that assigns different weights to each neighbor of a central entity. GAT uses the following aggregation function
\begin{equation}
H_{i}^{(l+1)}=\sigma\left(\sum_{j \in \mathcal{N}i} \alpha_{i j}^{(l)} z_{j}^{(l)}\right)
\end{equation}
that aggregates the entity embeddings of $l$ layers, where $\sigma$ is an activation function, $N_{i}$ is the set of the one-hop neighbors of the central entity $i$, $H^{(l)}$ are the entity embeddings for layer $l$, $z^{(l)}_{i}$ is a transformation operation  and $\alpha_{i j}^{(l)}$ is the normalized coefficient score. Particularly, $z^{(l)}_{i}$ and $\alpha_{i j}^{(l)}$ are calculated as:
\begin{equation}
z_{i}^{(l)}=W^{(l)} H_{i}^{(l)}
\end{equation} and
\begin{equation}
\alpha_{i j}^{(l)}=\frac{\exp \left(e_{i j}^{(l)}\right)}{\sum_{k \in \mathcal{N}(i)} \exp \left(e_{i k}^{(l)}\right)},
\end{equation}
where 
\begin{equation}
e_{i j}^{(l)}=\operatorname{LeakyReLU}\left(\vec{a}^{(l)^{T}}\left(z_{i}^{(l)} \| z_{j}^{(l)}\right)\right),
\end{equation}
$\vec{a}^{(l)^{T}}$ is a learnable weight vector, LeakyReLU a variant of the activation function ReLU \citep{DBLP:journals/eswa/ParisiNMC22} and $\mid\mid$ is the concatenation operation.

\subsubsection{Alignment Module}\label{sec:aligment}
For the alignment module $S_A$, there exist three techniques: sharing, swapping and mapping~\citep{DBLP:journals/pvldb/SunZHWCAL20}. We describe them below, while we extensively compare them in Section~\ref{compare_alignment_module}.

\paragraph{Sharing}\label{sharing}
Sharing aims to iteratively update the already produced entity embeddings, in order to minimize the embedding distance of each entity $e$ and its aligned entity $e'$ from the seed alignment $\delta$.
\begin{equation}
\label{eq:sharing}
S_{A}=\sum_{\left(e, e'\right) \in \delta\left(K G_{i}, K G_{j}\right)} {\mid\mid} \mathbf{e} - \mathbf{e'}{\mid\mid}.
\end{equation}

In Figure~\ref{fig:sharing}, we demonstrate the entity embeddings of $KG_1$ and $KG_2$ in $L1$ and $L2$ embedding spaces, respectively, while we also show the updates of the embeddings of the entities of seed alignment, in order to minimize their embedding distance. For simplification, we use a part of seed alignment, thus only the blue entities and the orange entities are considered as aligned. Therefore, by this technique, assuming the spatial similarity of aligned entities in two different KGs, we aim to adjust the axis of the two embedding spaces, so that entity vectors of the same entity in two KGs to overlap. It is worth mentioning that we started from two KGs encoded in two different embedding spaces (embeddings from the embedding module) and we ended up with two KGs encoded in an unified embedding space.

\begin{figure}
    \centering
\includegraphics[width=0.8\textwidth]{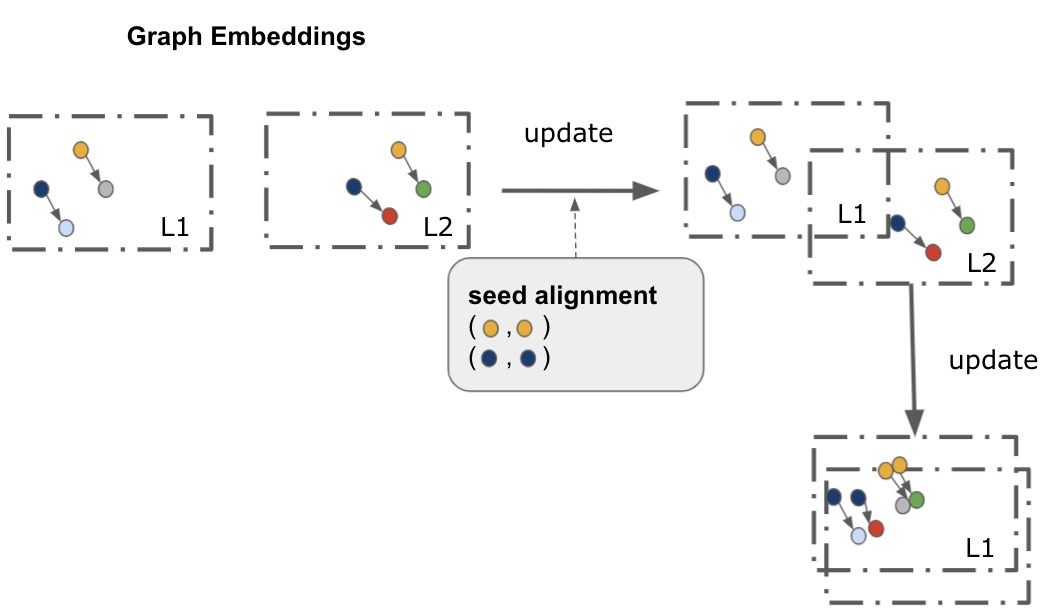}
    \caption{Sharing alignment technique.}
    \label{fig:sharing}
\vspace{-0.5cm}
\end{figure}

\paragraph{Swapping}\label{swapping}
Swapping is a variation of sharing that produces extra positive edges, preserving the same objective as sharing. For instance, given two aligned entity pairs $(h,h') \in \delta(KG_1,KG_2)$ and $(t,t') \in \delta(KG_1,KG_2)$ and a relation edge $(h,r,t)$ of $KG_1$, swapping produces two new positive edges $(h^{\prime},r,t)$ and $(h,r,t^{\prime})$ and feeds them in KG embedding models (embedding module) as positive relation edges, in order to increase the training data, benefiting the quality of the embeddings as we describe in Sections~\ref{relations} and \ref{negative}. Swapping does not introduce a new loss function.

\paragraph{Mapping}\label{mapping}
Mapping aims to learn a matrix $M$ as a linear transformation on entity vectors from $L_i$ to $L_j$, in order to minimize the embedding distance of each linearly transformed entity $e$ and its aligned entity $e'$ from the seed alignment $\delta$:
\begin{equation}
\label{eq:mapping}
S_{A}=\sum_{\left(e, e^{\prime}\right) \in \delta\left(K G_{i}, K G_{j}\right)} {\mid\mid}\mathbf{M}_{i j} \mathbf{e}-\mathbf{e'}{\mid\mid}.
\end{equation}

In Figure~\ref{fig:mapping}, we demonstrate the entity embeddings of $KG_1$ and $KG_2$ in $L1$ and $L2$ embedding spaces respectively, while we also show the process in which we learn the matrix $M_{i,j}$ that linearly transforms entities from $L1$ to $L2$. During this process, the linearly transformed entities of $KG_1$ should be close to their aligned entity of $KG_2$ according to the seed alignment. For simplification, we use a part of the seed alignment, thus only the blue entities and the orange entities are considered as aligned. Mapping, in contrary to sharing and swapping, aims to learn the mappings between the two embedding spaces (deducing the linear transformation from $L_1$ to $L_2$), without assuming the similarity of spatial emergence. More precisely, it does not force the entity vectors of aligned entities to overlap, instead, it treats the learned mappings as topological transformations (one-to-one correspondence) from $L_1$ to $L_2$, preserving the two KGs encoded in two different embedding spaces.
\begin{figure}
    \centering
\includegraphics[width=0.9\textwidth]{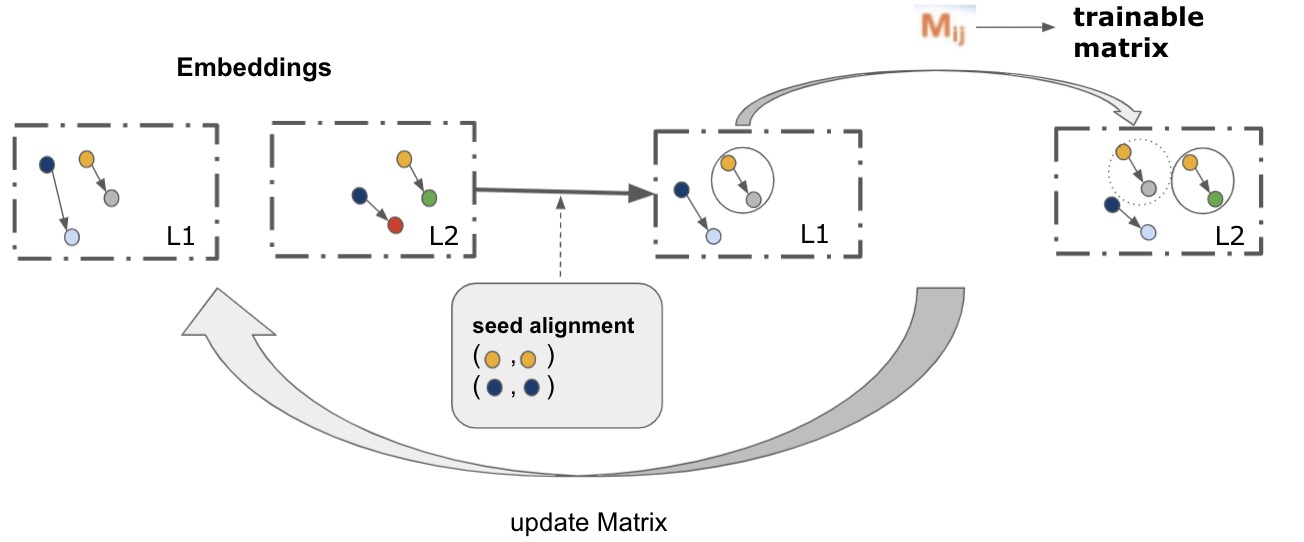}
    \caption{Mapping alignment technique.}
    \label{fig:mapping}
\vspace{-0.5cm}
\end{figure}

\subsection{Knowledge Graph Embeddings Using Relations}\label{ssec:knowledge_graph_embeddings_relations}
In this section, we discuss relation-based KG embedding methods, all of which are supervised. These methods use only the structural information (relation edges) for learning the entity embeddings.

\textit{\textbf{MTransE}}~\citep{DBLP:conf/ijcai/ChenTYZ17} is a translation-based model for multilingual KG embeddings, but it is also applicable to general-purpose KGs, capturing their structure. The objective is to minimize the loss function 
\begin{equation}\label{eq:mtranse_loss}
J = S_{K} + \alpha S_{A},
\end{equation}
where  $S_K$ is the loss function of the embedding module, $S_A$ is the loss function of the alignment module, and $\alpha$ is a factor that weights $S_K$ and $S_A$. As the loss function $S_K$ of the embedding module, MTransE utilizes a simplified version of TransE (Equation~\ref{eq:TransE}), in which no negative relation edges are considered, while as the loss function $S_A$ of the alignment module, it uses mapping (Equation~\ref{eq:mapping}). 

\textit{\textbf{MTransE+RotatE}}~\citep{DBLP:journals/pvldb/SunZHWCAL20} is a variation of MTransE that uses RotatE (Equation~\ref{eq:RotatE}) as $S_K$ in Equation~\ref{eq:mtranse_loss}, and sharing (Equation~\ref{eq:sharing}) as $S_A$, instead of TransE and mapping, respectively.

\textit{\textbf{RDGCN}}~\citep{DBLP:conf/ijcai/WuLF0Y019} leverages GCNs (described in ~\ref{ssec:GCN}) to incorporate structural information in the entity embeddings. Particularly, given $KG_1$ and $KG_2$, RDGCN constructs a primal (entity) graph $G^e$ by merging $KG_1$ and $KG_2$, and its dual (relation) graph $G^r$, by creating a node in $G^r$ for every relation type of $G^e$, and connecting two nodes in $G^r$ if the corresponding relations in $G^e$ share the same head or tail entities. 

Then, it uses a graph attention mechanism (a dual attention layer that assigns different importance to each neighbor's contribution) to make interactions between $G^e$ and $G^r$, in order the resulting entity representations in $G^e$ to capture the relation information, and then, to be fed to a GCN, capturing the structure of the neighborhood (Equation~\ref{eq:gcn_layer}). The resulting entity embeddings are refined using the mapping alignment technique (Section~\ref{mapping}). The loss function that RDGCN aims to minimize is
\begin{equation}
L=\sum_{(e_i, e_j) \in \delta, (e_i', e_j') \notin \delta} \text{max}\left(0, d(\mathbf{e_i}, \mathbf{e_j}) - d\left(\mathbf{e_i'}, \mathbf{e_j'}\right)+\gamma\right),
\end{equation}
where $(e_i, e_j)$ are entity pairs from the seed alignment $\delta$, $(e_i', e_j')$ are negative samples generated by replacing $e_i$ or $e_j$ with a random entity, 
and $d$ is the embedding distance function used in mapping (Section~\ref{mapping}).

\textit{\textbf{RREA}}~\citep{DBLP:conf/cikm/MaoWXWL20} integrates GCNs and GATs (described in Section \ref{ssec:GCN}) with a \emph{Relational Reflection Transformation}, in order to obtain relation-specific embeddings for KG entities. This transformation utilizes a matrix that, in contrary to standard GCN and GAT, is constrained to be orthogonal, in order to reflect entity embeddings across different relational hyperplanes. The orthogonal property of the aforementioned matrix keeps the norms and the relative distances of entities in the relational space unchanged.

More precisely, RREA stacks multiple GNN layers, in order to capture and aggregate multi-hop neighborhood information for each entity embedding. The output embedding of entity $e_i$ from the $l$-th layer is obtained as follows:
\begin{equation}
\boldsymbol{H}_{e_{i}}^{l+1}=\operatorname{ReLU}\left(\sum_{e_{j} \in \mathcal{N}_{e_{i}}^{e}} \sum_{r_{k} \in R_{i j}} \alpha_{i j k}^{l} \boldsymbol{M}_{r_{k}} \boldsymbol{h}_{e_{j}}^{l}\right),
\end{equation}
where ReLU~\citep{DBLP:journals/eswa/ParisiNMC22} is an activation function, $N^{e}_{e_i}$ are the neighboring entities of $e_i$, $R_{ij}$ denotes the relations between $e_i$ and $e_j$, $M_{r_{k}}$ the relational reflection matrix of $r_k$, and $\alpha_{i j k}^{l}$ is a weight coefficient of $M_{r_{k}}$ (similar to GAT). The final entity embedding comes from the concatenation of the embeddings of each layer. In addition, in order to include relational information around entities, RREA concatenates the summation of the relation embeddings with entity embeddings to get dual-aspect embeddings. The resulting entity embeddings are refined using the sharing alignment technique (Section~\ref{sharing}). The loss function that RREA aims to minimize is the following:
\begin{equation}
L=\sum_{\left(e_{i}, e_{j}\right) \in P} \max \left(\operatorname{dist}\left(e_{i}, e_{j}\right)-\operatorname{dist}\left(e_{i}^{\prime}, e_{j}^{\prime}\right)+\lambda, 0\right),
\end{equation}
where $e'_i$ and $e'_j$ represent the negative pair of $e_i$ and $e_j$, generated using \emph{truncated uniform negative sampling}~\citep{DBLP:conf/ijcai/SunHZQ18,DBLP:conf/ijcai/ZhuZ0TG19, DBLP:journals/corr/abs-1908-09898} and $dist$ is the embedding distance function used in sharing.

The methodology described above refers to the basic version of RREA, \emph{RREA(basic)}. RREA also comes with a semi-supervised version, \emph{RREA(semi)}, that proposes possibly aligned entity pairs in different iterations, in order to enrich the training set. According to~\cite{mao2020mraea}, the entity pair $(e_i,e_j)$ is proposed as aligned, if $e_i$ and $e_j$ are mutually nearest aligned. 

\subsection{Knowledge Graph Embeddings Using Attributes}\label{ssec:knowledge_graph_embeddings_attributes}

In this section, we focus on attribute-based KG embedding methods. These methods utilize not only the structural information of the KGs (relation edges) to learn the entity embeddings, but also the attribute values (literals). In addition, in many methods the attribute embeddings help to enrich the seed alignment or even in refining the entity embeddings. We categorize the attribute-based methods depending on the usage of seed alignment as \emph{supervised}, \emph{semi-supervised} and \emph{unsupervised}.

\subsubsection{Supervised} \label{sec:supervised}
\textit{\textbf{MultiKE}}~\citep{DBLP:conf/ijcai/ZhangSHCGQ19} first constructs the embeddings of each literal $l$
\begin{equation}
\phi(l)=\operatorname{encode}\left(\left[\operatorname{LP}\left(o_{1}\right) ; \operatorname{LP}\left(o_{2}\right) ; \ldots ; \mathrm{LP}\left(o_{n}\right)\right]\right),
\end{equation}
where $LP(o_n)$ is the pre-trained word embedding of word $o_n$, $encode(\cdot)$ is the encoder that does the compression of the embeddings, and $[;]$ is the concatenation operation. If $o_n$ is an out-of-vocabulary word (i.e., there is no pre-trained embedding for this word), then MultiKE builds it by using pre-trained character embeddings. 
Then, it learns entity embeddings by exploring three different views: the \emph{name view} $\Theta^{(1)}$, the \emph{relation view} $\Theta^{(2)}$ and the \emph{attribute view} $\Theta^{(3)}$. 

Given an entity $e$, its name view ($\Theta^{(1)}$) is defined as 
\begin{equation}
\mathbf{e}^{(1)}=\phi(\operatorname{name}(e)),
\end{equation}
where $name(\cdot)$ is the name of the entity.

For the relation view $\Theta^{(2)}$, it adopts TransE to learn the entity embeddings of the two KGs, minimizing the following loss function
\begin{equation}
\mathcal{L}\left(\Theta^{(2)}\right)=\sum_{(h, r, t) \in X^+ \cup X^-} \log \left(1+\exp \left(-\zeta_{(h, r, t)} f_{\mathrm{rel}}(\mathbf{h}, \mathbf{r}, \mathbf{t})\right)\right),
\end{equation}
where $X^+ = X_1 \cup X_2$ are the relation edges that exist in the two KGs, $X^-$ are relation edges that do not exist in the two KGs (negative relation edges, created as in TransE), $f_{\mathrm{rel}}$ is the scoring function of TransE (Equation~\ref{eq:TransE}), and $\zeta_{(h, r, t)} \in \{-1,1\}$ denotes whether $(h, r, t)$ is a positive or a negative edge.

For the attribute view $\Theta^{(3)}$, again it uses TransE to learn the embeddings exploiting the attributes and their values, aiming to minimize the loss function
\begin{equation}
\mathcal{L}\left(\Theta^{(3)}\right)=\sum_{(h, a, v) \in Y^+} \log \left(1+\exp \left(-f_{\text {attr }}\left(\mathbf{h}^{(3)}, \mathbf{a}, \mathbf{l}\right)\right)\right),
\end{equation}
where $Y^+ = Y_1 \cup Y_2$ are the attribute edges of the two KGs,  $f_{\text {attr}}\left(\mathbf{h}^{(3)}, \mathbf{a}, \mathbf{l}\right) = - \mid$$\mathbf{h}^{(3)} - \mathbf{CNN}(\langle\mathbf{a} ; \mathbf{l}\rangle)$$\mid$, which is using a Convolution Neural Network (CNN) representation of an attribute $a$ and its literal value $l$, as follows:
\begin{equation}
\mathbf{CNN}(\langle\mathbf{a} ; \mathbf{l}\rangle)=\sigma(\operatorname{vec}(\sigma(\langle\mathbf{a} ; \mathbf{l}\rangle * \Omega)) \mathbf{W}),
\end{equation}
where $[;]$ denotes the concatenation operation, $\mathbf{a}$ the embedding of an attribute $a$, $\mathbf{l}$ the embedding of a literal $l$, $\Omega$ the kernel of CNN, $\sigma$ the activation function, and $\mathbf{W}$ a trainable weighted matrix. 

For refining the entity embeddings of the relation and the attribute views, MultiKE minimizes the following two loss functions, respectively:
\begin{equation}
\begin{gathered}
\mathcal{L}_{\mathrm{CE}}\left(\Theta^{(2)}\right)=\sum_{(h, r, t) \in \mathcal{X}'} \log \left(1+\exp \left(-f_{\mathrm{rel}}\left(\hat{\mathbf{h}}^{(2)}, \mathbf{r}, \mathbf{t}^{(2)}\right)\right)\right) \\
+\sum_{(h, r, t) \in \mathcal{X}''} \log \left(1+\exp \left(-f_{\mathrm{rel}}\left(\mathbf{h}^{(2)}, \mathbf{r}, \hat{\mathbf{t}}^{(2)}\right)\right)\right)
\end{gathered}
\end{equation} and 
\begin{equation}
\mathcal{L}_{\mathrm{CE}}\left(\Theta^{(3)}\right)=\sum_{(h, a, l) \in \mathcal{Y}'} \log \left(1+\exp \left(-f_{\mathrm{attr}}\left(\hat{\mathbf{h}}^{(3)}, \mathbf{a}, \mathbf{l}\right)\right)\right),
\end{equation}
where $(h, \hat{h})$ and $(t, \hat{t})$ are entity pairs in the seed alignment, $\mathcal{X}'$ and $\mathcal{X}''$ are the relation edges whose head and tail entities are in seed alignment, respectively, and $\mathcal{Y}'$ are the attribute edges whose head entities are in seed alignment.

For the entity alignment (alignment module), MultiKE produces a set of aligned relations $S_{rel}$ and a set of aligned attributes $S_{attr}$ that are used to minimize the following loss function (cross-KG relation inference):
\begin{equation}
\mathcal{L}_{\mathrm{CRA}}\left(\Theta^{(2)}\right)=\sum_{(h, r, t) \in \mathcal{X}^{\prime \prime \prime}} \operatorname{sim}(r, \hat{r}) \log \left(1+\exp \left(-f_{\mathrm{rel}}\left(\mathbf{h}^{(2)}, \hat{\mathbf{r}}, \mathbf{t}^{(2)}\right)\right)\right),
\end{equation}
where $\mathcal{X}'''$ are the relation edges whose relations are in $S_{rel}$ and
\begin{equation}
\label{eq:sim_MultiKE}
\operatorname{sim}(r, \hat{r})=\alpha_{1} \text{ cosine} (\phi(\operatorname{name}(r)), \phi(\operatorname{name}(\hat{r})))+ (1-\alpha_{1}) \text{ cosine} (\mathbf{r}, \hat{\mathbf{r}})
\end{equation}
is the similarity measure that is used to align or not two relations, based on their name similarity (from literal embeddings) and their semantic similarity (from relation embeddings).

Finally, MultiKE jointly learns the final entity embeddings from the different views in a unified embedding space, by minimizing the loss function
\begin{equation}
\mathcal{L}_{\text {ITC }}(\tilde{\mathbf{H}}, \mathbf{H})=\sum_{i=1}^{D} \mid\mid \tilde{\mathbf{H}}-\mathbf{H}^{(i)}\mid\mid,
\end{equation}
where $\mathbf{\tilde{H}}=\bigcup_{i=1}^{D} \mathbf{H}^{(i)}$, $\mathbf{H}$ is the view-specific entity embedding, and $D$ is the number of views. 
For entity alignment, it uses swapping (Section~\ref{swapping}).

\textit{\textbf{BERT\_INT}}~\citep{DBLP:conf/ijcai/Tang0C00L20} utilizes a well-known language model,  BERT~\citep{DBLP:conf/naacl/DevlinCLT19}, in order to embed entities based on their factual information (e.g., descriptions, names) and an interaction model, in order to compute their interactions, instead of aggregating neighbors, which in many cases causes noisy matches. 

More precisely, for each entity $e$, it applies a pre-trained basic BERT unit that accepts the factual information as input, aiming to minimize the following loss function

\begin{equation}
\mathcal{L}=\sum_{\left(e, e^{\prime+}\right) \in \mathcal{D}} \max \left\{0, g\left(e, e^{\prime+}\right)-g\left(e, e^{\prime-}\right)+m\right)\}
\end{equation}
to fine-tune BERT. Here, $D$ is the seed alignment, $e^{\prime+}$ is the correctly aligned entity known from seed alignment, $e^{\prime-}$ is a randomly selected negative entity from the other KG - \emph{truncated uniform negative sampling}~\citep{DBLP:conf/ijcai/SunHZQ18}, $m$ the margin and $g$ is the $l1$ distance, used for measuring the similarity between the embeddings $C(e)$ and $C(e^{\prime})$.

Regarding the interaction model, it is divided into the name/description view, the neighbor-views and the attribute-view interactions. Firstly, as name/description interaction, it leverages the embeddings generated by BERT, calculating their cosine similarity. Then, neighbor-view interaction compares names/descriptions of each neighbor pair (considering also their neighboring relations and multi-hop neighbors), producing a similarity matrix. The similarity matrix is then processed by a dual aggregation function to extract the similarity vectors, i.e., the entity embeddings. Afterwards, rather than learning embeddings of entities by aggregating their attributes, it compares each attribute pair, learning similarly to neighbor-view the attribute similarity vectors. Finally, a unified dual aggregation function is applied to extract the features from the neighbor-view and attribute-view interactions and generate the final entity embeddings.

\subsubsection{Semi-Supervised}\label{semi-supervised}

\textit{\textbf{KDCoE}}~\citep{DBLP:conf/ijcai/ChenTCSZ18} leverages a weakly aligned KG for semi-supervised entity alignment using long (typically from a couple of sentences) textual descriptions of entities. It co-trains iteratively two embedding models, one on the structure of the KG (KGEM) and another on the textual descriptions of the entities (DEM), respectively, given a small seed alignment. During each iteration, each embedding model proposes a new set of aligned entity pairs alternately, in order to enrich the seed alignment. The process runs until one of the models has no entity pair to propose. 

KGEM is practically the same as MTransE (Equation~\ref{eq:mtranse_loss}), using the TransE (Equation~\ref{eq:TransE}) embedding module (including negative samples) and mapping (Equation~\ref{eq:mapping}). At the end of this model, if the embedding distance between an entity $e$ and its closest (based on the distance function of mapping) entity $\hat{e}'$ is lower than a threshold, then the pair $(e,\hat{e}^{\prime})$ is proposed as aligned. DEM utilizes an encoder to process textual description sequences of vectors $d_e$, that are produced by pre-trained word embeddings, and learn the description embeddings. The learning objective of DEM is to maximize the log likelihood of an entity $e$ and its counterpart $e'$ in terms of description embeddings, by minimizing the following loss function:

\begin{equation}
S_{D} =\sum_{\left(e, e^{\prime}\right) \in \delta}-L L_{1}-L L_{2} =\sum_{\left(e, e^{\prime}\right) \in \delta}-\log \left(P\left(e \mid e^{\prime}\right)\right)-\log \left(P\left(e^{\prime} \mid e\right)\right)
\end{equation}
where $\delta$ is the seed alignment and
\begin{equation}
L L_{1} =\log \sigma\left(\mathbf{d}_{e}^{\top} \mathbf{d}_{e^{\prime}}\right)+\sum_{k=1}^{\mid B_{d}\mid } \mathbb{E}_{e_{k} \sim U\left(e_{k} \in E_{L_{i}}\right)}\left[\log \sigma\left(-\mathbf{d}_{e_{k}}^{\top} \mathbf{d}_{e^{\prime}}\right)\right]
\end{equation}

\begin{equation}
L L_{2} =\log \sigma\left(\mathbf{d}_{e}^{\top} \mathbf{d}_{e^{\prime}}\right)+\sum_{k=1}^{\mid B_{d}\mid } \mathbb{E}_{e_{k} \sim \mathrm{U}\left(e_{k} \in E_{L_{j}}\right)}\left[\log \sigma\left(-\mathbf{d}_{e}^{\top} \mathbf{d}_{e_{k}}\right)\right],
\end{equation}
where $d_e$ and $d_{e'}$ are the embeddings of the textual descriptions of the two aligned entities $e$ and $e'$, unrelated entities $e_k$ are chosen randomly from a uniform distribution $U$, and $\mid$$B_{d}$$\mid$ is the batched sampling size. Intuitively, the encoder aims to maximize the dot product of descriptions of aligned entities and decrease the dot product of descriptions of unrelated entities. At the end of this model, if the embedding distance between $d_e$ and its closest entity $d_{e'}$ is lower than a threshold (different than the one used for KGEM), then the pair $(e, e')$ is proposed as aligned.

For generating entity embeddings from textual descriptions, KDCoE utilizes a self-attention Gated Recurrent Unit (GRU), in order to preserve the sequence of words in a textual description and remove information irrelevant for the prediction, while sharing information across the different descriptions. For more details on this, we refer the reader to~\cite{DBLP:conf/ijcai/ChenTCSZ18}.

\subsubsection{Unsupervised} \label{unsupervised}

\textit{\textbf{AttrE}}~\citep{DBLP:conf/aaai/TrisedyaQZ19} is an unsupervised method that leverages both structural embeddings and attribute character embeddings for entity alignment. Instead of relying on a seed alignment to refine the structural embeddings, it uses the factual information to minimize the embedding distance between entities that have similar attribute character embeddings.
 
\begin{figure}[t]
\centering
\includegraphics[width=\linewidth, height=7cm]{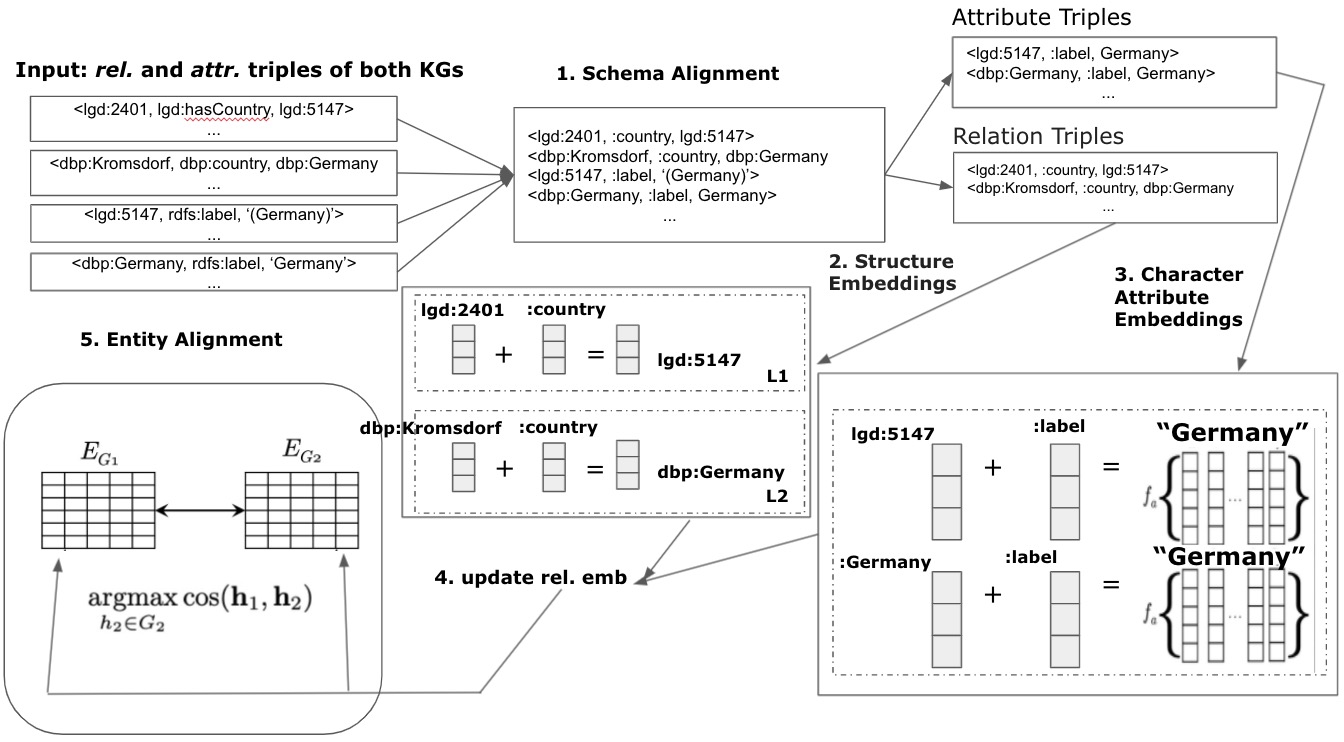}
\caption{Example of AttrE.}
\label{fig:AttrE_example}
\vspace{-0.5cm}
\end{figure}
 
As shown in Figure~\ref{fig:AttrE_example}, AttrE consists of four modules: the schema alignment module (preprocessing step), the structure embedding module $J_{S E}$, the attribute character embedding module $J_{C E}$, and the alignment module $J_{SIM}$
\begin{equation}
J=J_{S E}+J_{C E}+J_{SIM}.
\end{equation}
The predicate alignment module merges the two KGs and renames predicates with a similar predicate name from $KG_1$ and $KG_2$, with a unified naming schema. For example, lgd:hasCountry and dbp:country (with lgd and dbp being the prefixes of the Linked Geo Data and DBpedia namespaces, respectively) are converted to :country. To find the predicates with similar names, it computes the Levenshtein distance of the postfixes of the predicates' URIs. If this score is greater than a predefined threshold, then the two predicates are similar.

AttrE adopts TransE (Equation~\ref{eq:TransE}) to learn the structure embeddings by minimizing the $J_{SE}$ loss function, with $\alpha=\frac{\operatorname{count}(r)}{{\mid}X{\mid}}$,
where count(r) is the number of occurrences of relation r, and ${\mid}X{\mid}$ is the total number of relation edges in the merged KG.

To learn the attribute character embedding, AttrE minimizes the following objective function:
\begin{equation}
J_{C E}=\sum_{y \in Y} \sum_{y' \in Y'} \max \left(0,\left[\gamma+w\left(f\left(y\right)-f\left(y'\right)\right)\right]\right),
\end{equation}
where 
\begin{equation}
f\left(\left<h,a,l\right>\right) = {\mid}\mathbf{h}+\mathbf{r}-f_{a}(l){\mid}
\end{equation}
and $Y$ is the set of positive attribute edges, $Y'$ is the set of negative attribute edges (generated by replacing the head entity with a random entity), $w$ weights relation edges with aligned predicates, and $f_{a}(l)=\sum_{n=1}^{N}\left(\frac{\sum_{i=1}^{t} \sum_{j=i}^{n} \mathbf{c}_{\mathbf{j}}}{t-i-1}\right) \label{n-gram}$ is an n-gram based compositional function that encodes literals $l$ of attributes $a$ in the embedding space. At the end of this process, entities that have similar attribute embeddings should also have similar entity embeddings. In the example of Figure~\ref{fig:AttrE_example}, lgd:5147 has a similar embedding with :Germany, since their attributes label:Germany have similar embeddings.

Having learned the structure embeddings and the attribute embeddings, AttrE combines them to produce the final entity embeddings, by minimizing the loss function
\begin{equation}
J_{SIM}=\sum_{e \in E_{1} \cup E_{2}}\left[1-\cos \left(\mathbf{e}_{se}, \mathbf{e}_{ce}\right)\right],
\end{equation}
where $\mathbf{e_{se}}$ the structure embedding of $e$, and $\mathbf{e_{ce}}$ is the attribute character embedding of $e$. For example, the embedding of the entities lgd:5147 and :Germany will be updated in order their structured and attribute embeddings to be close. However, the entities lgd:2401 and dbp:Kromsdorf have tail entities with similar embeddings. At the end, the entities lgd:2401 and dbp:Kromsdorf will end up having similar embeddings too. For generating the alignment results, AttrE reports two entities as aligned, if the cosine similarity of their entity embeddings is greater than a pre-defined threshold.

\subsection{Qualitative Comparison of Embedding Methods}
In this section, we describe the general assumptions of the evaluated methods, while we also compare the embedding-based entity alignment methods from different perspectives. For this purpose, we summarize in Tables~\ref{tab:categories1} and \ref{tab:categories2} the basic characteristics and techniques of the methods, with respect to eight main categories: embedding module, literal size, alignment module, learning, schema alignment, embedding initialization, and negative sampling on relations and attributes, as described next.

\begin{table}
\centering
\caption{Method categories.}
\label{tab:categories1}
\resizebox{\textwidth}{!}{\begin{tabular}{|l|c|c|c|c|c|c|c|c|} 
\hline
\multicolumn{1}{|c|}{\textbf{Methods}} & \multicolumn{5}{c|}{\textbf{Embedding Module}}                                                                                                                                            & \textbf{Align. Module}                                 & \textbf{Learning}                                         & \textbf{Schema Align.}                                    \\ 
\hline
{\cellcolor[rgb]{0.655,0.573,0.573}}   & \textbf{Ent. Names} & \textbf{Relations} & \textbf{Attr. Names} & \textbf{Lit. Values}                                        & \textbf{\textbf{\textbf{\textbf{Lit. Size}}}} & \multicolumn{1}{l|}{{\cellcolor[rgb]{0.655,0.573,0.573}}} & \multicolumn{1}{l|}{{\cellcolor[rgb]{0.655,0.573,0.573}}} & \multicolumn{1}{l|}{{\cellcolor[rgb]{0.655,0.573,0.573}}}    \\ 
\hline
MTransE                                & \xmark                    & one-hop            & \xmark                       & \xmark                                                            & -                                    & mapping                                                   & supervised                                                & \xmark                                                           \\ 
\hline
MTransE+RotatE                       & \xmark                    & one-hop            & \xmark                       & \xmark                                                            & -                                    & sharing                                                   & supervised                                                & \xmark                                                           \\ 
\hline
RDGCN                                  & \cmark                    & multi-hop          & \xmark                       & \xmark                                                            & -                                    & mapping                                                   & supervised                                                & \xmark                                                           \\
\hline
RREA(basic)                                  & \xmark                    & multi-hop          & \xmark                       & \xmark                                                            & -                                    & sharing                                                   & supervised                                                & \xmark                                                           \\
\hline
RREA(semi)                                  & \xmark                    & multi-hop          & \xmark                       & \xmark                                                            & -                                    & sharing                                                   & semi-supervised                                                & \xmark                                                           \\
\hline
KDCoE                                  & \xmark                   & one-hop            & \xmark                       & descriptions                                                       & 4 words                                    & mapping                                                   & semi-supervised                                           & \xmark                                                           \\ 
\hline
MultiKE                                & \cmark                   & one-hop            & \cmark                      & \begin{tabular}[c]{@{}c@{}}word\\ and \\character\end{tabular} & 5 words                            & swapping                                                   & supervised                                                & \begin{tabular}[c]{@{}c@{}}name\\and\\semantic\end{tabular}  \\ 
\hline
AttrE                                  & \xmark                    & one-hop            & \xmark                       & character                                                     & 10 characters                               & sharing                                                   & unsupervised                                              & name                                                         \\ 
\hline
\colorrows{}BERT\_INT & \cmark                    & multi-hop            & \xmark                       & \begin{tabular}[c]{@{}c@{}}word\\and\\description\end{tabular}                                     & 128 words                               & sharing                                                   & supervised                                              & \xmark                                                         \\ 
\hline
\end{tabular}}
\end{table}

\subsubsection{Embedding Module}
The type of information that each method exploits in order to learn the embeddings reveals one of the most important assumptions that differentiates the methods. All embedding-based entity alignment methods utilize the relational structure of entities (relations), assuming that structural similarity is the key to entity alignment.
However, there are methods that also exploit the literal values of the attributes, as well as the entity and attribute names.

\paragraph{Entity Names}
Some embedding-based entity alignment methods learn the entity embeddings by encoding the entity names in the embedding space, so as entities with similar names to have similar embeddings. These methods work under the assumption that the same real-world entities have to follow the same or semantically similar naming. However, in many real-world KGs, this assumption is not holding as entity names are replaced with source-specific ids. For this reason, several EA methods either avoid to utilise the entity names for the entity embeddings, either use them jointly with the structural or factual information of KGs.

Among the evaluated methods, MultiKE, RDGCN and BERT\_INT are the only methods that use entity names in order to learn the final entity embeddings. Thus, for these methods, we expect an improved performance on KGs where there is some homogeneity in the naming of the entities.

\paragraph{Relations}\label{relations}
The structural information of the KGs constitutes the main type of information that all embedding-based entity alignment methods use in order to learn the entity embeddings. Entity embedding methods exploit the structural information either in \emph{one hop} or in \emph{multiple hops}. The former use the local structural information of the entities, ignoring the impact of more distant neighbors. For these methods, the more relations we have per entity, the better results we expect to get, because entities have the opportunity to minimize their embedding distance with multiple similar entities (neighbors), exploiting multiple features to learn the final embeddings (more information for learning). In addition, the more relation edges we have, the more negative samples we have per entity, and as a result, entities move away from dissimilar entities in the embedding space. 
The latter (multi-hop methods) use the subgraph structure, exploiting a larger amount of relations between entities. This way, they focus on the extended (multi-hop) neighborhood and aggregate the embeddings of the multi-hop neighbors in order to learn their own.

Among the methods evaluated in this study, all of them use the relations to learn the entity embeddings, with most  methods focusing on the one-hop neighborhood, except RDGCN, RREA(basic), RREA(semi), and BERT\_INT, which follow a multi-hop approach. This way, RDGCN and both versions of RREA increase the information that an entity can exploit and the final entity embeddings are more expressive. This benefit comes at the risk of incorporating more noisy information into the embeddings, due to considering some distant neighbors, irrelevant or less important for the entity alignment task. Thus, BERT\_INT, rather than aggregating the information of multi-hop neighbors, compares entity pairs from multi-hop neighbors, called interactions, based on factual information (e.g., textual descriptions, names).

\paragraph{Attribute Names}
Some embedding-based entity alignment methods use the attribute names for learning the entity embeddings. Usually, attribute names are exploited by the methods in order to enhance the similarity measure of entities and not as main source of information for learning the embeddings. That is because it requires KGs to have similar naming schemes, which is not usual in real-world KGs.

MultiKE is the only method that utilizes this type of information, requiring a homogeneous naming in terms of attributes. Thus, we expect this feature to perform better in datasets originating from the same or similar data sources (e.g., Wikipedia-based, or synthetic KGs). 

\paragraph{Literal Values and Literal Size}
Many embedding-based entity alignment methods use literal values as auxiliary information to learn the entity embeddings. Specifically, they learn the embeddings of the literal values and they use those embeddings for learning the entity embeddings, or for enriching the seed alignment, or even for alignment. In most methods, all literal values are used, but in some methods, only the literal values of the attributes with name ``description'' are used. In addition, there are two methods for learning the literal-value embeddings: character-based and word-based. The first one uses the characters of the literal to learn the final literal embedding, while in the second one, the literal is tokenized (split into words) and the final embedding is resulting by employing pre-trained word embeddings.

The character-based methods usually exploit a small part of the literals (e.g., the first few characters) and they are mostly used for short literals, such as dates. In addition, there are many methods to learn the character-based embeddings: the \emph{n-gram}-based that sums the n-gram combinations of the literals and the method that computes the average of the pre-trained character embeddings. 
The word-based methods typically exploit more information from the literal value (e.g., the first few words) and for this reason, they are mostly used for longer literals, such as names and descriptions. However, the word-based methods assume the existence of the word embeddings in the pre-trained set for all the words that appear in the literal values, which is very frequently not the case, resulting in out-of-vocabulary errors. For this reason, there are also hybrid methods, that use the word-based method first, but if a word does not exist in the pre-trained set (vocabulary), they use one of the aforementioned character-based method, in order to learn the specific word's embedding. Regarding the literal size, it is a very important hyperparameter that defines the size of the substring of the literal which will be used, including or excluding important words or characters for the final literal embeddings.

Among the evaluated methods, KDCoE, AttrE, MultiKE and BERT\_INT are the ones that use literal values for learning embeddings. KDCoE exploits the first 4 words of textual descriptions and pre-trained word embeddings, in order to learn the description embeddings and measure the entities similarity based on these embeddings. The objective is to enrich the training set (Section~\ref{ssec:preprocessing}) with extra aligned entity pairs based on this similarity. Thus, if the similarity of an entity pair (based on description embeddings) is above a pre-defined threshold (95\%), the entities are proposed as aligned, enriching the training set with an extra aligned entity pair that was not included in the training set before. AttrE, in contrary, exploits the first 10 characters of the literal values and follows an n-gram based method to learn the embeddings, for a combination of $n$ values ranging from 1 to 10. The literal embeddings are used in the alignment module in order entities of the KGs with similar literal embeddings to be close in the embedding space. Specifically, AttrE uses literal embeddings for the alignment, since, as an unsupervised method, it does not use the seed alignment during training process. Furthermore, MultiKE uses the first 5 words and pre-trained word embeddings for learning the literal embeddings. In case a word is not contained in the available pre-trained word embeddings, it averages the character embeddings of this word, i.e., it follows the hybrid approach described above. MultiKE uses the literal embeddings in many ways, as described in Section~\ref{sec:supervised}, e.g., for name view, attribute view and schema alignment. BERT\_INT utilizes long textual descriptions (limiting the words to 128), names, or literals for both BERT unit and interaction model, as described in Section~\ref{sec:supervised}. 

All these methods require homogeneity in terms of literal values, which is not usual in real-world KGs. Finally, the character-based method that MultiKE uses as alternative to word-based methods suffers in the literals that contain the same characters in different order. For example, the number ``5013'' and the number ``1350'' will have the same embeddings. AttrE does not have this issue, since it is following an n-gram based method for character embeddings. BERT\_INT, leveraging a variety of factual information and giving priorities depending on their availability and quality, deals with heterogeneity issues (e.g., it prioritizes descriptions over entity names, since descriptions contain richer information).

\subsubsection{Alignment Module} \label{compare_alignment_module}

The alignment module aims to produce a common embedding space for the entities of the two KGs, using three typical techniques described in Section \ref{sec:aligment}. MTransE+RotatE, AttrE, both versions of RREA, and BERT\_INT utilize the sharing technique to calibrate the axis of the embedding spaces of the two KGs, in order the aligned entities to overlap. Consequently, we expect methods that rely on this technique to work well on KGs with dense and similar neighborhoods. This setting helps them to achieve the desired overlapping and preserve the initial structure of the KGs in the embedding space. On the other hand, MTransE, KDCoE and RDGCN, that rely on the mapping technique, learn a transformation matrix that aligns entities in two separate embedding spaces. In contrary to sharing, this technique does not assume the spatial similarity of the KGs. Finally, methods such as MultiKE, that unify multiple embedding spaces (multi-view) and consume more relation edges for training, rely on the swapping alignment technique.

\subsubsection{Learning} \label{sec:learning}

We divide the evaluated methods into supervised, that require the seed alignment for the alignment module, semi-supervised, that use the seed for the alignment process, but also try to enrich it, and unsupervised, that do not use the seed alignment for the alignment process. All methods need seed alignment, either for training (alignment module) and testing (supervised and semi-supervised) or exclusively for testing (unsupervised). Supervised methods (MTransE, MTransE+RotatE, RDGCN, MultiKE RREA(basic) and BERT\_INT) assume the existence of seed alignment, which in many cases is hard to find (since it includes the matches of all the entities of the two KGs), even in widely used Wikipedia-based KGs, hindering the EA task especially when the KGs scale up in terms of content and density.

In contrary to the aforementioned category, semi-supervised methods (e.g., KDCoE and RREA(semi)) work sufficiently well when a small percentage of aligned entities is available, which is more realistic in real-world KGs. However, these methods usually require auxiliary information to support the alignment task that involves the enrichment of the seed alignment. For example, KDCoE does this by assuming the existence of multiple,  homogeneous textual descriptions. On the other hand,  RREA(semi) does not use auxiliary information; it utilizes specific rules (e.g., the two entities have to be mutually nearest aligned, in order be proposed as aligned). 

Finally, unsupervised methods carry out a more difficult task, since they use no seed alignment for the alignment process. Particularly, the lack of seed alignment makes AttrE leverage the literals similarity for alignment purposes, assuming their homogeneity. It is worth noting that unsupervised methods, in contrary to supervised and semi-supervised, do not assume that every entity of $KG_1$ should be matched to exactly one entity of $KG_2$.

\subsubsection{Schema Alignment} \label{schema} 
We refer to schema alignment as the process of finding similar relations and/or attributes. The similarity of the relations or the attributes can be measured by two different ways: measuring their name similarity (AttrE) or measuring both their name similarity and their structural (semantic) similarity (MultiKE) in the embedding space. All these methods can benefit if the initial KGs follow similar naming schemes, making the process of finding similar relations and attributes easier.

AttrE and MultiKE are two methods that perform schema alignment. MultiKE captures the schema similarity not only in terms of relation or attribute name similarity, but also exploits the semantic similarity of relations and attributes. In the case that we have KGs with heterogeneous naming at the schema level, MultiKE would have better performance than AttrE, that only exploits the name similarity for the schema alignment.

\subsubsection{Embedding Initialization}
For initializing the entity embeddings, there are two practices: the random one, that initializes them by picking random numbers from a normal distribution, and the one that initializes them with the entity name embeddings. The latter has better results~\citep{DBLP:conf/icon-nlp/KocmiB17}, because embeddings contain some information for the entities, but it works under the assumption that matching entities have similar entity names, which is not holding in many cases. RDGCN is the only method that initializes the embeddings with entity name embeddings.

\begin{table}
\centering
\caption{Initialization, hidden layers and negative sampling.}
\label{tab:categories2}
\resizebox{\textwidth}{!}{\begin{tabular}{|l|c|c|c|c|} 
\hline
\multicolumn{1}{|c|}{\begin{tabular}[c]{@{}c@{}}\\\textbf{Method}\end{tabular}} & \textbf{Embedding Initialization} & \textbf{Layers}     & \begin{tabular}[c]{@{}c@{}}\textbf{Negative Sampling on Relations}\\\textbf{\textbf{}\textbf{Negative per Positive}}\end{tabular} & \begin{tabular}[c]{@{}c@{}}\textbf{\textbf{Negative Sampling on Attributes}}\\\textbf{\textbf{Negative per Positive}}\end{tabular}  \\ 
\hline
MTransE                                                                         & random                            & 2                   & -                                                                                                                                 & -                                                                                                                                   \\ 
\hline
MTransE + RotatE                                                                & random                            & 2                   & 10 (uniform)                                                                                                                      & -                                                                                                                                   \\ 
\hline
KDCoE                                                                           & random                            & 2 + 2 GRU           & 1

~(uniform)                                                                                                                     & 1 (uniform)                                                                                                                         \\ 
\hline
AttrE                                                                           & random                            & 5                   & 1

~(uniform)                                                                                                                     & 1 (uniform)                                                                                                                         \\ 
\hline
MultiKE                                                                         & random                            & 2 lit.+ 4 + 3 conv. & 10

~(uniform)                                                                                                                    & -                                                                                                                                   \\ 
\hline
RDGCN                                                                           & name embeddings                   & 4 GAT + 2 GCN       & 125

~(uniform)                                                                                                                   & -                                                                                                                                   \\ 
\hline
RREA(semi)                                                                     & random                            & 2 GAT               & 1

~(truncated)                                                                                                                   & -                                                                                                                                   \\
\hline
RREA(basic)                                                                    & random                            & 2 GAT               & 1

~(truncated)                                                                                                                   & -                                                                                                                                   \\ 
\hline
\colorrows{}BERT\_INT                                                                     & random                            & 3 MLP + 4 RBF                & 2

~(truncated)                                                                                                                   & -                                                                                                                                   \\
\hline
\end{tabular}}
\end{table}

\subsubsection{Negative Sampling}\label{negative}
Negative sampling~\citep{DBLP:conf/icml/KamigaitoH22} is the process of generating $n$ (a hyperparameter) negative examples of edges that do not exist in the KG, by replacing either the head or the tail entities of each positive edge with another random entity - uniform negative sampling~\citep{DBLP:conf/nips/BordesUGWY13} - or with a highly similar neighbor - truncated negative sampling~\citep{DBLP:conf/ijcai/SunHZQ18,DBLP:conf/ijcai/ZhuZ0TG19, DBLP:journals/corr/abs-1908-09898}. Truncated negative sampling, in contrary to uniform negative sampling, ensures difficult negative samples that contribute more on learning process than the easy ones. For example, if we sample a negative triple (Titanic, capitalOf, Greece) from (Athens, capitalOf, Greece), using uniform sampling, we can find that Athens and Titanic have very low similarity and the generated negative sample contributes little in the learning process. If, instead, we generate the negative triple (Thessaloniki, capitalOf, Greece) from (Athens, capitalOf, Greece), using truncated sampling, this negative triple would contribute more in the learning process, since Thessaloniki (the second biggest city in Greece) and Athens, are two very similar entities. 

In general, negative sampling is widely used in KG embedding models on relations, as well as on attributes, in order to maximize the embedding distance of dissimilar entities. The more negative examples we have per positive example, the bigger the distance among dissimilar entities. In addition, an entity that participates in a high number of positive examples (high average relations per entity), will also participate in an increased number of negative examples. For instance, if an entity appears in the head or the tail of $x$ relation edges, then the number of the negative examples for this entity can be as high as $x*n$, where $n$ is the corresponding hyperparameter.

All the evaluated methods use negative sampling, with the exception of MTransE. The higher the ratio of negative per positive triples, the better the performance of the methods, since the embedding distance of dissimilar entities increases. Although requiring 10 or even 125 times bigger KGs (for 10 or 125 negative samples per positive, respectively) improves the performance, this comes at the cost of increased training time and less scalable methods. MultiKE further enriches the KGs by using the swapping alignment technique that generates additional positive examples, as described in Section~\ref{swapping}. This technique not only increases the available information for training, but it also increases the negative examples for the entities indirectly, as entities will appear in more relation edges.

\subsubsection{Neural Network Architectures}\label{architecture}
All evaluated methods are implemented on top of a neural network architecture with multiple hidden layers, each consisting of $d$ neurons, where $d$ are the dimensions of the embeddings (dim). More precisely, relation-based methods utilize either shallow neural networks (MTransE, MTransE+RotatE), or Graph Neural Network (RDGCN, RREA(basic), RREA(semi)) for learning the embeddings, while attribute-based methods use shallow neural networks for encoding relations (AttrE) or Graph Neural Networks for encoding literals (MultiKE) and textual descriptions (KDCoE) or Radial Basis Function kernel~\citep{DBLP:conf/sigir/XiongDCLP17} for the interactions (BERT\_INT).

Regarding relation-based methods, MTransE and MTransE+RotatE, utilize one layer for embedding entities and one layer for embedding entity relations. As shown in Table~\ref{tab:hyperparameters}, the dimensions of both embeddings are the same (100). On the other hand, RDGCN consists of four Graph Attention Networks (GATs)~\citep{DBLP:journals/corr/abs-1710-10903}, two primal layers and two dual layers. The final embeddings are output by two stacked GCN layers~\citep{DBLP:conf/iclr/KipfW17}. Both versions of RREA utilize two stacked GATs (Relational Reflection Aggregate Layer). It is worth mentioning that both RDGCN and both versions of RREA use variations of graph attention mechanisms, not standard GATs.

Attribute-based methods, as shown in Table~\ref{tab:categories2}, use two standard layers (for entity and relation embeddings) and some extra layers to encode additional information of entity descriptions. For example, KDCoE utilizes two layers for embedding entities and their relations and two GRU layers for encoding their textual descriptions. It is worth mentioning that while all other examined methods learn the structure and attribute embeddings jointly, KDCoE is the only method that trains two different models alternatively (co-training). AttrE consists of five layers: one layer for embedding entities of relation edges, one layer for embedding entities of attribute edges, one layer for relations, one layer for attributes, and one layer for character embeddings. All of these layers contain the same number of neurons (100). MultiKE utilizes two layers for the word embeddings, three layers for entity, relation and attribute embeddings, one convolution layer for encoding literals, and one layer that combines them. Finally, it uses two additional convolution layers for the cross-KG entity inference based on relations and attributes. BERT\_INT utilizes four Radial Basis Function kernels (one layer each one) for the interaction model and three Multilayer Perceptron for obtaining the entity embeddings.

It is worth mentioning that the simpler the architecture, the more scalable the method is (e.g., as shown in Table~\ref{tab:time}, the translation-based method MTransE is more scalable than the GNN-based method RDGCN).

\section{Experimental Setting}
\label{section:three}
In this section, we give an overview and statistics about the datasets, we describe the evaluation protocol and the metrics that we used. We also provide implementation details about this study. The source code of this work is publicly available\footnote{{\url{https://github.com/fanourakis/experimental-review-EA}}}.

\subsection{Datasets}\label{ssec:datasets}
For the experimental evaluation we utilized nine datasets, each consisting of a pair of KGs to be aligned, four from OpenEA~\citep{DBLP:journals/pvldb/SunZHWCAL20} and five datasets from~\cite{DBLP:conf/bigdataconf/EfthymiouSC15, DBLP:journals/corr/abs-2101-06126} in order to enrich the available data characteristics and ensure a trustworthy meta-level analysis.

For the datasets from OpenEA, three well-known KGs were used as sources: DBpedia (version 2016-10)~\citep{DBLP:journals/semweb/LehmannIJJKMHMK15},  Wikidata (version 20160801)~\citep{DBLP:journals/cacm/VrandecicK14} and  YAGO3~\citep{DBLP:conf/semweb/RebeleSHBKW16}. The seed alignment is constructed by the \emph{owl:sameAs} links, which were provided with those KGs. In addition, it is hard for embedding-based approaches to run on full KGs due to the large candidate space, hence the authors of OpenEA sampled real-world KGs, by their iterative degree-based sampling algorithm described in Section~\ref{ssec:preprocessing}. The resulting KGs have a size of 15K entities each, constituting the datasets D\_W\_15K\_V1 and D\_Y\_15K\_V1. In order to examine the behaviour of the methods with respect to the density of the KGs, the authors of OpenEA generated an additional dense version for each dataset (D\_W\_15K\_V2 and D\_Y\_15K\_V2). For generating the dense version, they deleted entities (nodes) with low ($\leq 5$) node degrees  and they performed again their iterative degree-based sampling algorithm, described in Section~\ref{ssec:preprocessing}. The above mentioned datasets are described as follows:

\begin{itemize}
  \item \textbf{D\_W\_15K\_V1} and \textbf{D\_W\_15K\_V2} are the sparse and the dense datasets, respectively, that were constructed from DBpedia and Wikidata KGs, describing actors, musicians, writers, films, songs, cities, football players and football teams. We chose them in order to examine the influence of KG density variations in the relation- and attribute-based methods. In addition, these datasets have both low entity name similarity and a low number of entities that have long textual descriptions, undermining the methods that use them. In these datasets, relation and attribute names have been replaced with special ids, which influences in a negative way methods that perform schema alignment using predicate names. 
  
  \item \textbf{D\_Y\_15K\_V1} and \textbf{D\_Y\_15K\_V2} are the sparse and the dense datasets, respectively, that were constructed from DBpedia and YAGO3. They contain the same entity types with D\_W\_15K but unlike D\_W\_15K, they have both high number of entities that have long textual descriptions and high entity name similarity.
\end{itemize}

Five additional datasets, from~\cite{DBLP:conf/bigdataconf/EfthymiouSC15, DBLP:journals/corr/abs-2101-06126}, were employed, that originate from the following eight KGs: the BTC2012 version of DBpedia\footnote{\url{http://km.aifb.kit.edu/projects/btc-2012/}}, BBCmusic (originating from Kasabi\footnote{\url{https://archive.org/details/kasabi}}), IMDb\footnote{\url{https://www.imdb.com/}}, TVDB\footnote{\url{https://www.thetvdb.com/}}, TMDb\footnote{\url{https://www.themoviedb.org/}}, and two restaurant KGs\footnote{\label{restaurants}\url{http://oaei.ontologymatching.org/2010/im/}}. None of these datasets originally cover the assumptions of embedding-based entity alignment methods, thus we pre-process them, as described in Section \ref{ssec:preprocessing}. The five extra datasets, built from those KGs, are the following:

\begin{itemize}
  \item \textbf{BBC-DB}~\citep{DBLP:conf/bigdataconf/EfthymiouSC15} is a sparse dataset constructed by BBCmusic and the BTC2012 version of DBpedia. It contains various entity types such as musicians, their birth places and bands, while also it has the highest number of average attributes per entity and the highest number of entities that have textual descriptions among other datasets. In addition, this dataset is better suited for the methods that perform schema alignment, since these KGs follow similar predicate naming conventions.
  
   \item \textbf{imdb-tmdb}, \textbf{imdb-tvdb}, and \textbf{tmdb-tvdb}~\citep{DBLP:journals/corr/abs-2101-06126} were constructed from IMDb, TMDb, and TVDB that contain descriptions about people, movies, tv series, episodes and companies. 
   Those datasets have a low number of entities, which makes them interesting for examining the performance of supervised vs unsupervised methods with a small training dataset.
   In addition, they are much sparser than the previous datasets, with low average attributes per entity, while they exhibit high homogeneity in terms of literals and predicate names. 
   tmdb-tvdb features higher homogeneity in terms of literals and lower similarity of long-text descriptions, compared to imdb-tvdb and tmdb-tvdb.

\item \textbf{Restaurants}\footref{restaurants} contains descriptions of real restaurants and their addresses from two different KGs. It is the smallest dataset in terms of number of entities, relation types and attribute types, but its main challenge is that it does not cover the assumptions of all the embedding-based entity alignment methods. Specifically, not all entities of KG1 are matched with all the entities KG2 and vice versa. 
We chose this dataset to examine whether the evaluated methods can be generalized beyond datasets that satisfy the 1-to-1 assumption.

\end{itemize}

\subsection{Statistics and Meta-Features}\label{ssec:statistics}
In order to quantify useful information about the knowledge graphs, we calculated some basic dataset statistics regrading the type and the number of relations and attributes that are presented in Table~\ref{tab:basic_statistics} and some additional dataset statistics that are shown in Table~\ref{tab:statistics}, categorized in the three following categories: 
    seed alignment size,
    density and
    heterogeneity
that were used for constructing the meta-features of Table~\ref{metafeatures}.

\emph{Seed alignment size} refers to the number of entity pairs in the seed alignment. 
\emph{Density} consists of the average relations per entity and the average attributes per entity, which indicate the density of the knowledge graphs on structural and attribute level, respectively. In addition, it consists of the proportion of sole and hyper relation types of the two knowledge graphs, where a sole relation type is a relation type that does not co-occur with another relation type in any entity pair~\citep{zhang2017knowledge}, otherwise, a relation is called a hyper relation~\citep{zhang2017knowledge}. 
Finally, in order to measure the \emph{heterogeneity} of the two knowledge graphs, we calculate the textual description similarity (Descr\_Sim) in the embedding space, of entities that are included in the test and validation sets, the number of entities of the two KGs that have textual descriptions (\#Ents\_Descr), the similarity of entity names (Ent\_Name\_Sim), the similarity of literals (Lit\_Sim) and the similarity of predicate names (Pred\_Name\_Sim). Specifically, for Descr\_Sim, Ent\_Name\_Sim and Lit\_Sim, we report the average of MAX similarities of each description, entity name and literal, respectively. As for the Pred\_Name\_Sim, we measure the name similarity of relation and attribute types (predicates) using Levenshtein Distance, as suggested in AttrE~\citep{DBLP:conf/aaai/TrisedyaQZ19}. Specifically, we report the average of MAX similarities of each predicate (relation or attribute).

The meta-features of Table~\ref{metafeatures} were constructed by applying the aggregation functions of Table~\ref{aggregation} on the statistics of Table~\ref{tab:statistics}, in order to construct features for each individual dataset instead of each individual KG of the dataset and use them in Section \ref{ssec:meta-learning}. More precisely, for Avg\_Rels\_per\_Entity, Avg\_Attrs\_per\_Entity, Sole\_Rels, Hyper\_Rels and \#Ents\_Descr (based on KDCoE that proposes possibly aligned entities of the entities of the first KG) were used aggregation functions, while \#Entity\_Pairs, Descr\_Sim, Ents\_Name\_Sim, Lit\_Sim and Pred\_Name\_Sim were used as meta-features without the application of any aggregation function.

\begin{table}
\centering
\caption{Statistics per dataset.}
\label{tab:basic_statistics}
\resizebox{\textwidth}{!}{\begin{tabular}{|c|c|c|c|c|c|} 
\hline
\textbf{Datasets}            & \textbf{\#Rel\_Triples} & \textbf{\#Attr\_Triples} & \textbf{\#Rel\_Types} & \textbf{\#Attr\_Types} & \textbf{\textbf{Symmetric\_Rels}}  \\ 
\hline
\begin{tabular}[c]{@{}c@{}}D\_W\_V1\\KG1\end{tabular}    & 38,265                  & 68,258                   & 248                   & 342                    & 7                                  \\ 
\hline
\begin{tabular}[c]{@{}c@{}}D\_W\_V1\\KG2\end{tabular}    & 42,746                  & 138,246                  & 169                   & 649                    & 2                                  \\ 
\hline
\begin{tabular}[c]{@{}c@{}}D\_W\_V2\\KG1\end{tabular}    & 73,983                  & 66,813                   & 167                   & 175                    & 1                                  \\ 
\hline
\begin{tabular}[c]{@{}c@{}}D\_W\_V2\\KG2\end{tabular}    & 83,365                  & 175,686                  & 121                   & 457                    & 0                                  \\ 
\hline
\begin{tabular}[c]{@{}c@{}}D\_Y\_V1\\KG1\end{tabular}    & 30,291                  & 71,716                   & 165                   & 257                    & 4                                  \\ 
\hline
\begin{tabular}[c]{@{}c@{}}D\_Y\_V1\\KG2\end{tabular}    & 26,638                  & 132,114                  & 28                    & 35                     & 2                                  \\ 
\hline
\begin{tabular}[c]{@{}c@{}}D\_Y\_V2\\KG1\end{tabular}    & 68,063                  & 65,100                   & 72                    & 90                     & 0                                  \\ 
\hline
\begin{tabular}[c]{@{}c@{}}D\_Y\_V2\\KG2\end{tabular}    & 60,970                  & 131,151                  & 21                    & 20                     & 0                                  \\ 
\hline
\begin{tabular}[c]{@{}c@{}}BBC-DB\\KG1\end{tabular}      & 15,478                  & 18,165                   & 9                     & 4                      & 3                                  \\ 
\hline
\begin{tabular}[c]{@{}c@{}}BBC-DB\\KG2\end{tabular}      & 45,561                  & 149,720                  & 98                    & 723                    & 1                                  \\ 
\hline
\begin{tabular}[c]{@{}c@{}}imdb-tmdb\\KG1\end{tabular}   & 4,747                   & 10,279                   & 2                     & 13                     & 0                                  \\ 
\hline
\begin{tabular}[c]{@{}c@{}}imdb-tmdb\\KG2\end{tabular}   & 4,857                   & 8,947                    & 2                     & 29                     & 0                                  \\ 
\hline
\begin{tabular}[c]{@{}c@{}}imdb-tvdb\\KG1\end{tabular}   & 3,389                   & 5,648                    & 2                     & 13                     & 0                                  \\ 
\hline
\begin{tabular}[c]{@{}c@{}}imdb-tvdb\\KG2\end{tabular}   & 1,051                   & 3,535                    & 2                     & 8                      & 0                                  \\ 
\hline
\begin{tabular}[c]{@{}c@{}}tmdb-tvdb\\KG1\end{tabular}   & 3,396                   & 5,326                    & 2                     & 24                     & 0                                  \\ 
\hline
\begin{tabular}[c]{@{}c@{}}tmdb-tvdb\\KG2\end{tabular}   & 938                     & 4,096                    & 2                     & 8                      & 0                                  \\ 
\hline
\begin{tabular}[c]{@{}c@{}}Restaurants\\KG1\end{tabular} & 226                     & 1,504                    & 2                     & 4                      & 0                                  \\ 
\hline
\begin{tabular}[c]{@{}c@{}}Restaurants\\KG2\end{tabular} & 1,504                   & 3,760                    & 2                     & 4                      & 0                                  \\
\hline
\end{tabular}}
\end{table}

\begin{table}
\vspace{-.5cm}
\centering
\caption{Statistics per dataset for the Meta-features.}
\label{tab:statistics}
\resizebox{\textwidth}{!}{\begin{tabular}{|c|c|c|c|c|c|c|c|c|c|c|} 
\hline
                                              & \begin{tabular}[c]{@{}c@{}}\textbf{\textbf{Seed }}\\\textbf{\textbf{Alignment Size}}\end{tabular}                                                      & \multicolumn{4}{c|}{\textbf{\textbf{Density}}}                                                                                                                                                                                                                                                                                                                                                                                                                                                                                                                                                                                            & \multicolumn{5}{c|}{\textbf{\textbf{Heterogeneity }}}                                                                                                                                                                                                                                                                                                                                                                                                                                                                                                                                                           \\ 
\hline
{\cellcolor[rgb]{0.969,1,1}}\textbf{Datasets} & {\cellcolor[rgb]{0.969,1,1}}\begin{tabular}[c]{@{}>{\cellcolor[rgb]{0.969,1,1}}c@{}}\textbf{\#Entity\_}\\\textbf{Pairs}\\\textbf{(KG1/KG2)}\end{tabular} & {\cellcolor[rgb]{0.969,1,1}}\begin{tabular}[c]{@{}>{\cellcolor[rgb]{0.969,1,1}}c@{}}\textbf{\#Avg\_Rels\_}\\\textbf{per\_ Entity}\\\textbf{(KG1/KG2)}\end{tabular} & {\cellcolor[rgb]{0.969,1,1}}\begin{tabular}[c]{@{}>{\cellcolor[rgb]{0.969,1,1}}c@{}}\textbf{\#Avg\_Attrs\_}\\\textbf{per\_ Entity}\\\textbf{(KG1/KG2)}\end{tabular} & {\cellcolor[rgb]{0.969,1,1}}\begin{tabular}[c]{@{}>{\cellcolor[rgb]{0.969,1,1}}c@{}}\textbf{Sole\_}\\\textbf{Rels}\\\textbf{(KG1/KG2)}\end{tabular} & {\cellcolor[rgb]{0.969,1,1}}\begin{tabular}[c]{@{}>{\cellcolor[rgb]{0.969,1,1}}c@{}}\textbf{Hyper\_}\\\textbf{Rels}\\\textbf{(KG1/KG2)}\end{tabular} & {\cellcolor[rgb]{0.969,1,1}}\begin{tabular}[c]{@{}>{\cellcolor[rgb]{0.969,1,1}}c@{}}\textbf{\#Ents\_}\\\textbf{Descr}\end{tabular} & {\cellcolor[rgb]{0.969,1,1}}\begin{tabular}[c]{@{}>{\cellcolor[rgb]{0.969,1,1}}c@{}}\textbf{Descr\_}\\\textbf{Sim}\end{tabular} & {\cellcolor[rgb]{0.969,1,1}}\begin{tabular}[c]{@{}>{\cellcolor[rgb]{0.969,1,1}}c@{}}\textbf{Ent\_Name\_}\\\textbf{Sim}\end{tabular} & \begin{tabular}[c]{@{}>{\cellcolor[rgb]{0.969,1,1}}c@{}}\textbf{Lit\_}\\\textbf{Sim}\end{tabular} & {\cellcolor[rgb]{0.969,1,1}}\begin{tabular}[c]{@{}>{\cellcolor[rgb]{0.969,1,1}}c@{}}\textbf{Pred\_Name\_}\\\textbf{Sim}\end{tabular}  \\ 
\hline
D\_W\_V1                                      & 15K / 15K                                                                                                                                              & 2.86 / 3.15                                                                                                                                                   & 4.69 / 9.61                                                                                                                                                    & 91 / 74                                                                                                                                            & 157 / 95                                                                                                                                            & 1,181 / 9,032                                                                                                                    & 0.69                                                                                                                          & 0.72                                                                                                                            & 0.81                                                               & 0                                                                                                                                   \\ 
\hline
D\_W\_V2                                      & 15K / 15K                                                                                                                                              & 5.39 / 5.77                                                                                                                                                   & 4.52 / 11.97                                                                                                                                                   & 53 / 44                                                                                                                                            & 114 / 77                                                                                                                                            & 278 / 3,579                                                                                                                      & 0.73                                                                                                                          & 0.78                                                                                                                            & 0.83                                                               & 0                                                                                                                                   \\ 
\hline
D\_Y\_V1                                      & 15K / 15K                                                                                                                                              & 2.8 / 3.58                                                                                                                                                    & 4.89 / 8.8                                                                                                                                                     & 53 / 5                                                                                                                                             & 112 / 23                                                                                                                                            & 3,455 / 12,000                                                                                                                   & 0.88                                                                                                                          & 0.99                                                                                                                             & 0.78                                                               & 47.37                                                                                                                               \\ 
\hline
D\_Y\_V2                                      & 15K / 15K                                                                                                                                              & 6.12 / 10.98                                                                                                                                                  & 4.38 / 8.74                                                                                                                                                    & 17 / 6                                                                                                                                             & 55 / 15                                                                                                                                             & 2,945 / 12,000                                                                                                                   & 0.87                                                                                                                          & 0.91                                                                                                                             & 0.79                                                               & 47.99                                                                                                                               \\ 
\hline
BBC-DB                                        & 9,396 / 9,396                                                                                                                                          & 1.72 / 6.03                                                                                                                                                   & 1.94 / 20.15                                                                                                                                                   & 9 / 20                                                                                                                                             & 0 / 78                                                                                                                                              & 2,430 / 2,133                                                                                                                    & 0.62                                                                                                                          & 0.34                                                                                                                             & 0.87                                                               & 80.93                                                                                                                               \\ 
\hline
imdb-tmdb                                     & 1,933 / 1,933                                                                                                                                          & 2.65 / 2.71                                                                                                                                                   & 5.31 / 4.62                                                                                                                                                    & 2 /2                                                                                                                                               & 0 / 0                                                                                                                                               & 753 / 753                                                                                                                        & 0.35                                                                                                                          & 0.48                                                                                                                             & 0.92                                                               & 88.49                                                                                                                               \\ 
\hline
imdb-tvdb                                     & 1,076 / 1,076                                                                                                                                          & 3.22 / 1                                                                                                                                                      & 5.25 / 3.28                                                                                                                                                    & 2 / 2                                                                                                                                              & 0 / 0                                                                                                                                               & 472 / 473                                                                                                                        & 0.38                                                                                                                          & 0.43                                                                                                                             & 0.91                                                               & 79.96                                                                                                                               \\ 
\hline
tmdb-tvdb                                     & 982 / 982                                                                                                                                              & 3.62 / 1                                                                                                                                                      & 5.42 / 4.17                                                                                                                                                    & 2 / 2                                                                                                                                              & 0 / 0                                                                                                                                               & 180 / 180                                                                                                                        & 0.22                                                                                                                          & 0.84                                                                                                                             & 0.94                                                               & 80.33                                                                                                                               \\ 
\hline
Restaurants                                   & 112 / 112                                                                                                                                              & 1 / 2                                                                                                                                                         & 1.66 / 1.66                                                                                                                                                    & 2 / 2                                                                                                                                              & 0 / 0                                                                                                                                               & 90 / 73                                                                                                                          & 0                                                                                                                             & 0.95                                                                                                                             & 0.96                                                               & 80.11                                                                                                                               \\
\hline
\end{tabular}}
\end{table}

\begin{table}
\centering
\caption{Meta-features of the datasets.}
\label{metafeatures}
\resizebox{\textwidth}{!}{\begin{tabular}{|c|c|c|c|c|c|c|c|c|l|c|} 
\hline
                                              & \begin{tabular}[c]{@{}c@{}}\textbf{Seed}\\\textbf{Alignment Size}\end{tabular}                                                     & \multicolumn{4}{c|}{\textbf{Density}}                                                                                                                                                                                                                                                                                                                                                                                                                                                                                                                & \multicolumn{5}{c|}{\textbf{Heterogeneity }}                                                                                                                                                                                                                                                                                                                                                                                                                                                                                                                                                                                         \\ 
\hline
{\cellcolor[rgb]{0.969,1,1}}\textbf{Datasets} & {\cellcolor[rgb]{0.969,1,1}}\begin{tabular}[c]{@{}>{\cellcolor[rgb]{0.969,1,1}}c@{}}\textbf{\#Entity\_Pairs of}\\\textbf{Seed Alignment}\end{tabular} & {\cellcolor[rgb]{0.969,1,1}}\begin{tabular}[c]{@{}>{\cellcolor[rgb]{0.969,1,1}}c@{}}\textbf{Avg\_Rels\_}\\\textbf{per\_Entity}\end{tabular} & {\cellcolor[rgb]{0.969,1,1}}\begin{tabular}[c]{@{}>{\cellcolor[rgb]{0.969,1,1}}c@{}}\textbf{Avg\_Attrs\_}\\\textbf{per\_ Entity}\end{tabular} & {\cellcolor[rgb]{0.969,1,1}}\begin{tabular}[c]{@{}>{\cellcolor[rgb]{0.969,1,1}}c@{}}\textbf{Sole\_}\\\textbf{Rels}\end{tabular} & {\cellcolor[rgb]{0.969,1,1}}\begin{tabular}[c]{@{}>{\cellcolor[rgb]{0.969,1,1}}c@{}}\textbf{Hyper\_}\\\textbf{Rels}\end{tabular} & {\cellcolor[rgb]{0.969,1,1}}\begin{tabular}[c]{@{}>{\cellcolor[rgb]{0.969,1,1}}c@{}}\textbf{\#Ents\_}\\\textbf{Descr}\end{tabular} & {\cellcolor[rgb]{0.969,1,1}}\begin{tabular}[c]{@{}>{\cellcolor[rgb]{0.969,1,1}}c@{}}\textbf{Descr}\\\textbf{Sim}\end{tabular} & {\cellcolor[rgb]{0.969,1,1}}\begin{tabular}[c]{@{}>{\cellcolor[rgb]{0.969,1,1}}c@{}}\textbf{Ent\_Name\_}\\\textbf{Sim}\end{tabular} & \multicolumn{1}{c|}{\begin{tabular}[c]{@{}>{\cellcolor[rgb]{0.969,1,1}}c@{}}\textbf{Lit\_}\\\textbf{Sim}\end{tabular}} & {\cellcolor[rgb]{0.969,1,1}}\begin{tabular}[c]{@{}>{\cellcolor[rgb]{0.969,1,1}}c@{}}\textbf{Pred\_Name\_}\\\textbf{Sim}\end{tabular}  \\ 
\hline
D\_W\_V1                                      & 15,000                                                                                                                              & 3.00                                                                                                                                    & 7.15                                                                                                                                      & 39.56                                                                                                                         & 60.43                                                                                                                          & 1,181                                                                                                                             & 0.69                                                                                                                          & 0.72                                                                                                                            & 0.81                                                                                    & 0                                                                                                                                   \\ 
\hline
D\_W\_V2                                      & 15,000                                                                                                                              & 5.58                                                                                                                                    & 8.24                                                                                                                                      & 33.68                                                                                                                         & 66.31                                                                                                                          & 278                                                                                                                              & 0.73                                                                                                                          & 0.78                                                                                                                            & 0.83                                                                                    & 0                                                                                                                                   \\ 
\hline
D\_Y\_V1                                      & 15,000                                                                                                                              & 3.19                                                                                                                                    & 6.84                                                                                                                                      & 30.05                                                                                                                         & 69.94                                                                                                                          & 3,455                                                                                                                             & 0.88                                                                                                                          & 0.99                                                                                                                             & 0.79                                                                                    & 47.37                                                                                                                               \\ 
\hline
D\_Y\_V2                                      & 15,000                                                                                                                              & 8.55                                                                                                                                    & 6.56                                                                                                                                      & 24.73                                                                                                                         & 75.26                                                                                                                          & 2,945                                                                                                                             & 0.87                                                                                                                          & 0.91                                                                                                                             & 0.78                                                                                    & 47.99                                                                                                                               \\ 
\hline
BBC-DB                                        & 9,396                                                                                                                               & 3.87                                                                                                                                    & 11.04                                                                                                                                     & 27.10                                                                                                                         & 72.89                                                                                                                          & 2,430                                                                                                                             & 0.62                                                                                                                          & 0.34                                                                                                                             & 0.87                                                                                   & 80.93                                                                                                                               \\ 
\hline
imdb-tmdb                                     & 1,933                                                                                                                               & 2.68                                                                                                                                    & 4.96                                                                                                                                      & 100                                                                                                                           & 0                                                                                                                              & 753                                                                                                                              & 0.35                                                                                                                          & 0.48                                                                                                                             & 0.92                                                                                    & 88.49                                                                                                                               \\ 
\hline
imdb-tvdb                                     & 1,076                                                                                                                               & 2.11                                                                                                                                    & 4.26                                                                                                                                      & 100                                                                                                                           & 0                                                                                                                              & 472                                                                                                                              & 0.38                                                                                                                          & 0.43                                                                                                                             & 0.91                                                                                    & 79.96                                                                                                                               \\ 
\hline
tmdb-tvdb                                     & 982                                                                                                                                & 2.31                                                                                                                                    & 4.79                                                                                                                                      & 100                                                                                                                           & 0                                                                                                                              & 180                                                                                                                              & 0.22                                                                                                                          & 0.84                                                                                                                             & 0.94                                                                                    & 80.33                                                                                                                               \\ 
\hline
Restaurants                                   & 112                                                                                                                                & 1.50                                                                                                                                    & 1.66                                                                                                                                      & 100                                                                                                                           & 0                                                                                                                              & 90                                                                                                                               & 0                                                                                                                             & 0.95                                                                                                                             & 0.96                                                                                    & 80.11                                                                                                                               \\
\hline
\end{tabular}}
\end{table}

\begin{table}
\vspace{-.5cm}
\arrayrulecolor{black}
\caption{Aggregation functions.}
\resizebox{\textwidth}{!}{\begin{tabular}{!{\color{black}\vrule}l!{\color{black}\vrule}l!{\color{black}\vrule}} 
\hline
\textbf{Statistics}       & \textbf{Aggregation Function}s                                   \\ 
\hline
Avg\_Rels\_per\_Entity & (avg\_rel\_per\_entity\_KG1 + avg\_rel\_per\_entity\_KG2) / 2    \\ 
\hline
Avg\_Attrs\_per\_Entity & (avg\_attr\_per\_entity\_KG1 + avg\_attr\_per\_entity\_KG2) / 2  \\ 
\hline
Sole\_Rels       & ((sole\_KG1 + sole\_KG2) / rels.)*100                            \\ 
\hline
Hyper\_Rels      & ((hyper\_KG1 + hyper\_KG2) / rels.)*100                          \\ 
\hline
\#Ents\_Descr             & ents\_with\_descr\_KG1                                           \\
\hline
\end{tabular}}
\arrayrulecolor{black}
\label{aggregation}
\end{table}

\subsection{Evaluation Protocol and Metrics}\label{ssec:measures}
In this section, we describe the evaluation protocol that embedding-based entity alignment methods use, the metrics used for their evaluation, while we also compare those metrics.

\subsubsection{Evaluation Protocol}\label{ssec:protocol}
EA methods have been traditionally evaluated using classification-based metrics (e.g., precision and recall), comparing the matches proposed by those methods to the correct matches of a given ground truth~\citep{DBLP:journals/pvldb/LeoneHAGW22}. On the contrary, KG embedding-based EA methods, adopted rank-based evaluation metrics (see Section~\ref{metrics}), possibly influenced by the recent literature on the embedding-based methods for link prediction~\citep{DBLP:conf/nips/BordesUGWY13,DBLP:conf/iclr/SunDNT19}. In order to compare embedding-based EA methods on an equal basis and re-use the available open-source code as much as possible, we adopt the latter evaluation metrics and acknowledge that the criticism of~\cite{DBLP:journals/pvldb/LeoneHAGW22} is valid.

During the evaluation, the embedding-based EA methods use two embedding matrices, one for each KG, with the entity embeddings in their rows, and the test set\footnote{\label{split_footnote}The seed alignment is split into train (20\%), validation (10\%) and test (70\%), as in~\citep{DBLP:journals/pvldb/SunZHWCAL20}.} of the seed alignment (Figure~\ref{fig:evaluation}).
The alignment methods calculate the similarity of every entity of the first matrix with every entity of the second matrix, in the embedding space using similarity measures such as \emph{Euclidean distance}, \emph{cosine similarity}, etc. The result of this process is a similarity list, that we sort in descending order and aim to find the index (ranking) of the aligned entity in the similarity list. 

In the example of Figure~\ref{fig:evaluation}, $e_1'$ is the aligned entity of entity $e_1$, known as \emph{true entity}, according to seed alignment, and it is in the first position in the similarity list. In the last similarity list, there is a tie in the scores of the two candidate entities $e_3'$ and $e_2'$, that is decisive for some evaluation metrics. According to~\cite{berrendorf2020ambiguity}, there are different behaviors for this phenomenon that are categorized as \emph{optimistic}, \emph{pessimistic}, \emph{non-deterministic} and \emph{realistic}. 
In optimistic behavior, it is assumed that the true entity is ranked first among other entities with the same score, in pessimistic, it is assumed that it is ranked last, and realistic is the mean of optimistic and pessimistic.
In our case, all of the evaluated methods are non-deterministic, which means that handling ties depends on the inner working of the sorting algorithm that the methods use to sort each similarity list. 

In the context of ensuring a reliable and unbiased experimental process, we conducted the experiments with 5-fold cross-validation~\citep{DBLP:journals/pvldb/SunZHWCAL20}, where each fold is split into train, validation and test\footref{split_footnote}. Finally, the performance of each method, per dataset, for a specific metric is calculated by the average scores that the method achieves in the 5 folds for this metric.

\begin{figure}
    \centering
    \includegraphics[width=0.9\textwidth]{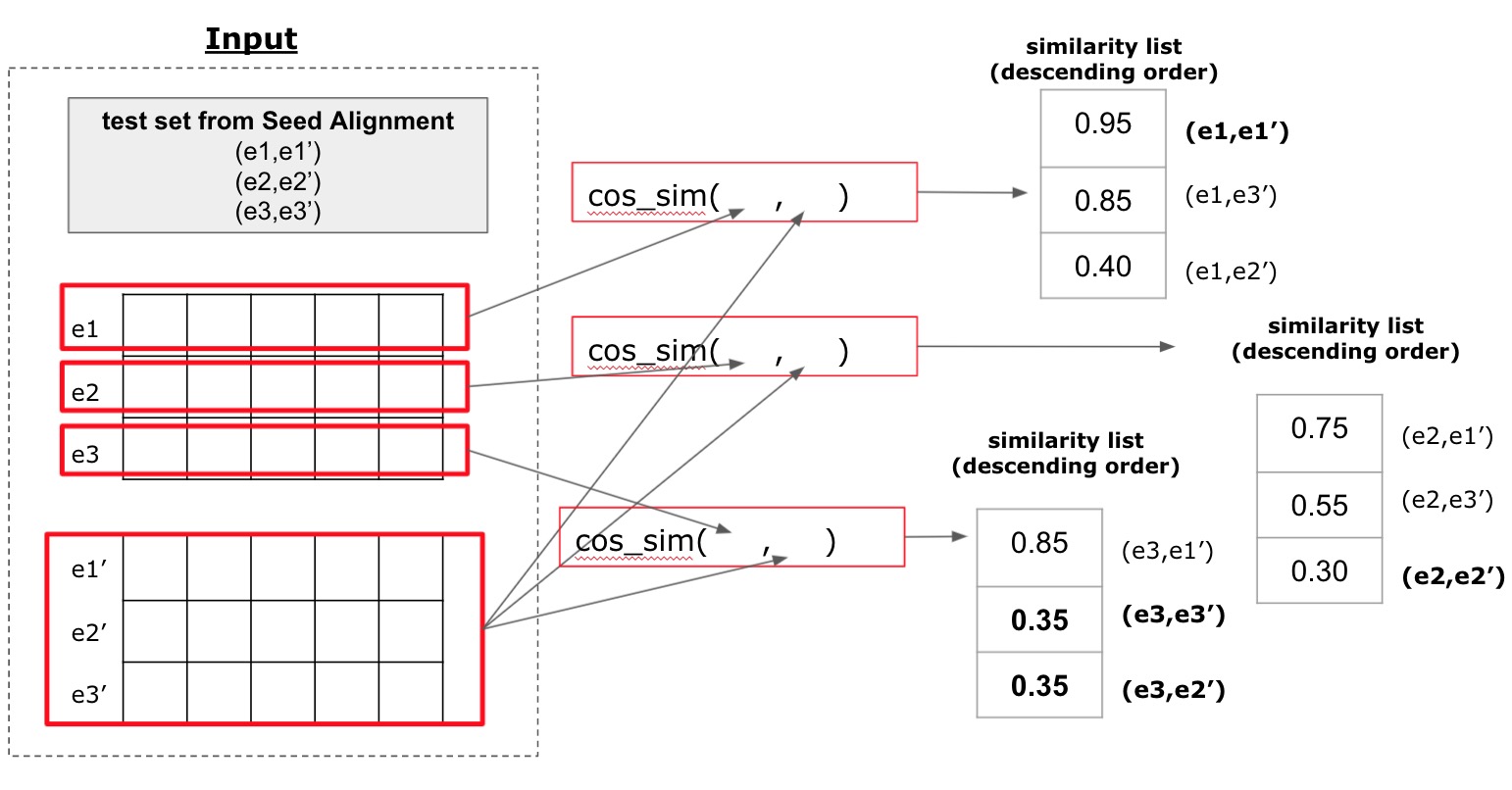}
    \caption{Evaluation Process.}
    \label{fig:evaluation}
\vspace{-0.5cm}
\end{figure}

\subsubsection{Evaluation Metrics}\label{metrics}
For the evaluation of the alignment methods, we use the following metrics, which depend on the individual ranks $\mathcal{I}$ generated by each method, as described in the previous section.

\textit{\textbf{Hits@k}} describes the fraction of hits (true entities) that appear in the first $k$ ranks of the sorted similarity lists:

\begin{equation}
Hits@k = \frac{\mid\{r \in \mathcal{I} \mid r \leq k\}\mid}{\mid\mathcal{I}\mid},
\end{equation}
where $Hits@k \in [0, 1]$. In the experimental results, we report Hits@k values as percentages (i.e., $Hits@k \cdot 100\%$).
Specifically, this metric measures the accuracy of the methods with adjustable error rate. For instance, Hits@10 allows a small error rate in contrary to Hits@1 that does not allow any errors. The weakness of this metric is that it
considers only the first $k$ position of the similarity list, and as a result, all the other positions have no effect to the final score. An advantage of this metric is its easy interpretation. In the example of Figure~\ref{fig:evaluation}, $Hits@1 = 1/3 = 0.33$, since among the three similarity lists ($\mid$$\mathcal{I}$$\mid$$ = 3$), only in one similarity list, a true entity pair is observed in the first rank.

\textit{\textbf{Mean Rank (MR)}}
computes the average ranks of the true entity pairs:
\begin{equation}
MR = \frac{1}{\mid\mathcal{I}\mid} \sum_{r \in \mathcal{I}} r  ,
\end{equation}
where $MR \in [1,$$\mid$$E$$\mid$$]$ (the lower the better). 
An advantage of MR is its sensitivity to any model performance changes, since it reflects the average performance and not only the fist $k$ ranks. However, it is also sensitive to outliers, because a very high or very low rank of a true entity pair affects the mean of the ranks. Finally, MR is an easily interpretable metric, but we should keep in mind the size of the candidate entities, because MR = 10 is better for 1,000,000 entities than for 20 entities. In the example of Figure~\ref{fig:evaluation}, assuming that in the third similarity list the true pair is in the second rank, $MR = (1 + 3 + 2) / 3 = 2$.

\textit{\textbf{Mean Reciprocal Rank (MRR)}} is the inverse of harmonic mean rank:
\begin{equation}
MRR = \frac{1}{\mid\mathcal{I}\mid} \sum_{r \in \mathcal{I}} \frac{1}{r} ,
\end{equation}
where $MRR \in (0,1]$. This metric is affected more by the top-ranked values rather than the bottom ones. For instance, the change is larger when moving from 1 to 2 compared to moving from 10 to 1,000. Thus, MRR is less sensitive to outliers. In the example of Figure~\ref{fig:evaluation}, assuming that in the third similarity list the true pair is in the second rank, $MRR = (1/1 + 1/3 + 1/2)/3 = 0.61$.

\subsection{Pre-processing Pipelines}\label{ssec:preprocessing}

The embedding-based entity alignment methods are working under some strict assumptions that make them inapplicable to the five new datasets. However, for a reliable experimental evaluation of the methods under the same conditions and data characteristics, we pre-process those datasets in order to conform to the assumptions described in Section~\ref{section:two}. 

\textbf{1-to-1 and structural-information assumptions.}
Firstly, we processed the seed alignment, in order to satisfy the 1-to-1 mapping assumption, as follows. If there are two entity pairs $(e_i,e_j)$ and $(e_i,e_j')$ in the seed alignment, we need to keep one of them. In the dilemma of choosing which entity pair to remove, we examined two factors. The first factor is to keep the entity pair in which the true entity, $e_j$ or $e_j'$ in this example, is involved in at least one relation edge (one of the other assumptions of these methods). The second factor is to keep the entity pairs in which the true entity has the highest node degree (number of incoming and outgoing edges), in the effort to reduce as little as possible the initial KGs. In addition, we processed the initial KGs, by removing relation edges $\left(h,r,t\right)$, whose head ($h$) or tail ($t$) entities did not appear in the seed alignment. Then, we repeated the same process for the attribute edges, 
removing the edges whose head entity did not appear in the seed alignment, and also removing the attribute edges whose head entity was not involved in at least one relation edge.

\textbf{Pre-aligned predicates assumption.} In addition to these assumptions, AttrE requires predicate alignment, as described in Section~\ref{ssec:knowledge_graph_embeddings_attributes} and merging the two KGs into one KG. The predicate alignment step is performed by manual inspection of the data, as suggested by the authors of AttrE\footnote{\label{AttrEcode}\url{https://bitbucket.org/bayudt/kba/src/35c67d56a8f0e2bcc05e8d56fb98c4374e3ff542/}}. For example, given the attributes \emph{:name} and \emph{:hasName}, the second one is replaced by \emph{:name}.

\textbf{Sampling.} For sampling real-world KGs, the authors of OpenEA proposed an iterative degree-based sampling (IDS) algorithm which simultaneously removes entities of the two KGs using seed alignment. The objective of this algorithm is to generate a dataset of a specific size from real-world KGs, such that the difference of the entity degree distribution, between the original KG and the sampled KG, does not exceed a certain value. In order to determine which entities to remove without greatly affecting the entity degree distribution, they delete entities with low PageRank scores,
while assessing the divergence of the two degree distributions of the KGs using the Jensen-Shannon (JS)~\citep{DBLP:journals/tit/Lin91} measure. The algorithm stops when the size of the datasets is the desired one (15K) and the divergence is $\leq 5\%$.

\subsection{Implementation Details}\label{ssec:details}
All experiments were performed in a server with 16 cores (AMD EPYC 7232P @ 3.1 GHz), 64 GB RAM, one RTX-4090 GPU (24 GB) and Ubuntu 18.04.5 LTS. We used the code of OpenEA\footnote{\url{https://github.com/nju-websoft/OpenEA/tree/2a6e0b03ec8cdcad4920704d1c38547a3ad72abe}} (2020-07-28) for the experiments of methods MTransE, MTransE + RotatE, KDCoE, MultiKE and RDGCN. For AttrE, we noticed that the implementation of OpenEA utilizes the seed alignment for training, which contradicts the unsupervised learning nature of AttrE~\citep{DBLP:conf/aaai/TrisedyaQZ19}. For this reason, we used the code provided by the authors of AttrE\footref{AttrEcode} (2019-01-11). Finally, for RREA (basic and semi), BERT\_INT and PARIS, we used the code provided by the authors\footnote{\url{https://github.com/MaoXinn/RREA}}\textsuperscript{,}\footnote{\url{https://github.com/kosugi11037/bert-int}}\textsuperscript{,}\footnote{\url{https://github.com/epfl-dlab/entity-matchers}}.

For conducting a fair experimental process, we used the same hyperparameter values as described in the works of OpenEA, AttrE, RREA and BERT\_INT, and are shown in Table~\ref{tab:hyperparameters}. \emph{Dim} is the dimensions of the embeddings (length of the vector representations). The higher dimensions an embedding has, the more features (expressiveness) it has. \emph{Batch size} is the number of relation edges that we used in each iteration during training. All the examined methods adopt a batch size of 5,000, except AttrE which adopts 100 and BERT\_INT 128. \emph{Max epochs} refers to the times the learning algorithm will work in the entire dataset. In our experiments, there are two ways to define the number of epochs. The first one is to give a fixed number (50, 205, 1,200 or 6,000 epochs), as AttrE, BERT\_INT, RREA(basic) and RREA(semi) do. The second one, \emph{early stopping}, is to terminate the training process when Hits@1 starts dropping, based on the validation set, checked every 10 epochs. The latter ensures that the model does not overfit. It is worth noting RREA(semi) requires a fixed number of iterations (5), while KDCoE decides the number of iterations dynamically. RREA(basic) is the first iteration of RREA(semi).

\begin{table}
\centering
\caption{Hyperparameters of methods.}
\label{tab:hyperparameters}
{\begin{tabular}{|c|c|c|c|c|c|} 
\hline
\textbf{Methods} & \textbf{Dim} & \textbf{Batch Size} & \textbf{Max Epochs}  \\ 
\hline
MTransE          & 100          & 5,000                                             & 2,000                                     \\ 
\hline
MTransE+RotatE   & 100          & 5,000                                            & 2,000                                     \\ 
\hline
RDGCN            & 300          & 5,000                                          & 2,000                                     \\ 
\hline
RREA(basic)       & 100          & 5,000                                       & 1,200 (fix)                                \\ 
\hline
RREA(semi)        & 100          & 5,000                                       & 6,000 (fix)                                \\ 
\hline
KDCoE            & 100          & 5,000                                            & 2,000                               \\ 
\hline
MultiKE          & 100          & 5,000                                           & 2,000                               \\ 
\hline
AttrE            & 100          & 100                                            & 50 (fix)                      \\
\hline
\colorrows{}BERT\_INT            & 300          & 128                                            & 205 (fix)                      \\
\hline
\end{tabular}}
\end{table}

\section{Analysis of Experimental Results}
\label{section:four}

In this section, we report and analyze the results of a series of experiments we conducted to answer the open questions Q1-Q4 posed in Section~\ref{section:intro}, regarding embedding-based entity alignment methods. More precisely, to shed light regarding Q1 (characteristics  of  methods) and Q2 (families  of  methods), we analyze the performance of the methods described in Section~\ref{section:two} across all datasets (Section~\ref{sec:evaluation}). To answer Q3 (effectiveness-efficiency trade-off), we rank the methods in a statistically significant way, while we analyze their corresponding execution time over all datasets of our testbed, and compare the time they require to reach a relative effectiveness threshold (Section~\ref{sec:effectivenessVSefficiency}). Finally, in Section~\ref{ssec:meta-learning}, we address Q4 (characteristics of datasets), by conducting a meta-level analysis to identify correlations between the methods and the various meta-features extracted from the KGs of our testbed.

\begin{table}
\centering
\caption{Performance of methods in different datasets and metrics.}
\label{tab:scores}
\arrayrulecolor{black}
\resizebox{\textwidth}{!}{\begin{tabular}{lccccccccccc}
\multicolumn{1}{l}{}                                               & \multicolumn{1}{l}{}                                                & \multicolumn{1}{l}{} & \multicolumn{5}{c}{\textbf{Relation-Based Methods}} 
& \multicolumn{4}{c}{\textbf{Attribute-Based Methods }}\\\cmidrule(l){4-8}\cmidrule(l){9-12}                                        \\ 

                                                                   & {\cellcolor[rgb]{0.937,0.937,0.937}}\textbf{Dataset}                &                      & \textbf{MTransE} & \textbf{MTransE+RotatE} & \textbf{\textbf{RDGCN}} & \textbf{RREA(basic)} & \textbf{RREA(semi)} & \textbf{KDCoE} & \textbf{MultiKE} & \textbf{AttrE} & \multicolumn{1}{l}{\textbf{BERT\_INT}}  \\ 
\hhline{~-----------|}
\multirow{4}{*}{}                                                  & {\cellcolor[rgb]{0.937,0.937,0.937}}                                & H@1                  & 26.07            & 27.27                   & 53.02                   & 65.57                & \textbf{71.82}      & 23.22          & 40.49            & N/A            & 44.08                                    \\ 

                                                                   & {\cellcolor[rgb]{0.937,0.937,0.937}}                                & H@10                 & 54.05            & 56.98                   & 72.96                   & 88.49                & \textbf{90.03}      & 46.29          & 56.71            & N/A            & 48.96                                    \\ 

                                                                   & {\cellcolor[rgb]{0.937,0.937,0.937}}                                & MR                   & 352.92           & 501.58                  & 494.69                  & 46.32~      & 57.71               & 868.85         & 320.29           & N/A            & \textbf{13.27\tnote{*}}                                   \\ 

                                                                   & \multirow{-4}{*}{{\cellcolor[rgb]{0.937,0.937,0.937}}D\_W\_15K\_V1} & MRR                  & 0.35             & 0.36                    & 0.59                    & 0.74                 & \textbf{0.79}       & 0.30           & 0.45             & N/A            & 0.45                                     \\ 
\hhline{~-----------|}
\multirow{4}{*}{\rotcell{\textbf{Datasets}}}                       & {\cellcolor[rgb]{0.937,0.937,0.937}}                                & H@1                  & 26.27            & 49.22                   & 63.71                   & 87.86                & \textbf{93.72}      & 33.78          & 49.54            & N/A            & 42.63                                    \\ 

                                                                   & {\cellcolor[rgb]{0.937,0.937,0.937}}                                & H@10                 & 57.45            & 81.69                   & 82.33                   & 98.66                & \textbf{99.18~}     & 64.38          & 72.76            & N/A            & 48.50                                    \\ 

                                                                   & {\cellcolor[rgb]{0.937,0.937,0.937}}                                & MR                   & 145.53           & 58.53                   & 210.94                  & \textbf{2.64}        & 2.74                & 169.46         & 37.44            & N/A            & 10.38\tnote{*}                                    \\ 

                                                                   & \multirow{-4}{*}{{\cellcolor[rgb]{0.937,0.937,0.937}}D\_W\_15K\_V2} & MRR                  & 0.36             & 0.59                    & 0.69                    & 0.92                 & \textbf{0.96}       & 0.44           & 0.56             & N/A            & 0.44                                     \\ 
\hhline{~-----------|}
\multirow{4}{*}{}                         & {\cellcolor[rgb]{0.937,0.937,0.937}}                                & H@1                  & 45.54            & 46.25                   & 94.10                   & 76.19                & 81.79               & 69.57          & 90.18            & 0.03           & \textbf{99.23}                           \\ 

                                                                   & {\cellcolor[rgb]{0.937,0.937,0.937}}                                & H@10                 & 72.78            & 72.94                   & 97.68                   & 91.31                & 91.95               & 88.02          & 95.03            & 0.2            & \textbf{99.28}                           \\ 

                                                                   & {\cellcolor[rgb]{0.937,0.937,0.937}}                                & MR                   & 245.02           & 574.13                  & 14.87                   & 29.79                & 73.34               & 143.78         & 19.69            & 5077.59        & \textbf{1.0004\tnote{*}}                          \\ 

                                                                   & \multirow{-4}{*}{{\cellcolor[rgb]{0.937,0.937,0.937}}D\_Y\_15K\_V1} & MRR                  & 0.55             & 0.55                    & 0.95                    & 0.82                 & 0.86                & 0.75           & 0.91             & 0.001          & \textbf{0.99}                            \\ 
\hhline{~-----------|}
\multirow{4}{*}{\rotcell{\mbox{\textbf{OpenEA}}}}                                        & {\cellcolor[rgb]{0.937,0.937,0.937}}                                & H@1                  & 44.72            & 93.89                   & 94.15                   & 96.56                & 97.53               & 86.72          & 85.64            & 0.04           & \textbf{99.37}                           \\ 

                                                                   & {\cellcolor[rgb]{0.937,0.937,0.937}}                                & H@10                 & 71.21            & 98.93                   & 97.73                   & 99.46                & \textbf{99.55}      & 98.06          & 92.70            & 0.3            & 99.37                                    \\ 

                                                                   & {\cellcolor[rgb]{0.937,0.937,0.937}}                                & MR                   & 79.47            & 3.46                    & 11.50                   & 1.32                 & 1.39                & 2.28           & 9.86             & 4900.23        & \textbf{1\tnote{*}}                               \\ 

                                                                   & \multirow{-4}{*}{{\cellcolor[rgb]{0.937,0.937,0.937}}D\_Y\_15K\_V2} & MRR                  & 0.53             & 0.95                    & 0.97                    & 0.98                 & 0.98                & 0.91           & 0.88             & 0.002          & \textbf{0.99}                            \\ 
\hhline{~===========|}
\multirow{4}{*}{}                                                  & {\cellcolor[rgb]{0.937,0.937,0.937}}                                & H@1                  & 24.98            & 18.41                   & 6.38                    & 43.62~               & 46.87               & 32.69          & 15.22            & 17.31          & \textbf{92.50}                           \\ 

                                                                   & {\cellcolor[rgb]{0.937,0.937,0.937}}                                & H@10                 & 50.28            & 41.66                   & 11.01                   & 65.28                & 65.19               & 52.88          & 33.47            & 46.65          & \textbf{93.78}                           \\ 

                                                                   & {\cellcolor[rgb]{0.937,0.937,0.937}}                                & MR                   & 672.17           & 1343.30                 & 2656.5                  & 318.03               & 631.69~             & 643.35         & 1299.81          & 760.98         & \textbf{1.38\tnote{*}}                           \\ 

                                                                   & \multirow{-4}{*}{{\cellcolor[rgb]{0.937,0.937,0.937}}BBC-DB}        & MRR                  & 0.33             & 0.26                    & 0.08                    & 0.52                 & 0.54                & 0.38           & 0.21             & 0.27           & \textbf{0.93}                            \\ 
\hhline{~-----------|}
\multicolumn{1}{c}{\multirow{4}{*}{\rotcell{\textbf{}}}} & {\cellcolor[rgb]{0.937,0.937,0.937}}                                & H@1                  & 15.00            & 9.30                    & 0.07                    & 27.06~               & 29.56~              & 20.02          & 15.28            & 87.01          & \textbf{99.70}                          \\ 

\multicolumn{1}{c}{}                                              & {\cellcolor[rgb]{0.937,0.937,0.937}}                                & H@10                 & 54.04            & 37.82                   & 0.73                    & 74.86                & 73.65               & 29.43          & 54.60            & 98.06          & \textbf{100}                           \\ 

\multicolumn{1}{c}{}                                              & {\cellcolor[rgb]{0.937,0.937,0.937}}                                & MR                   & 52.99            & 131.41                  & 677.50                  & 16.25                & 61.83~              & 199.59         & 59.72            & 5.96           & \textbf{1.002}                           \\ 

\multicolumn{1}{c}{}                                              & \multirow{-4}{*}{{\cellcolor[rgb]{0.937,0.937,0.937}}imdb-tmdb}     & MRR                  & 0.27             & 0.18                    & 0.005                   & 0.42                 & 0.44                & 0.23           & 0.27             & 0.90           & \textbf{0.99}                            \\ 
\hhline{~-----------|}
\multicolumn{1}{c}{\multirow{4}{*}{\rotcell{\textbf{Datasets}}}}       & {\cellcolor[rgb]{0.937,0.937,0.937}}                                & H@1                  & 13.63            & 7.18                    & 0.13                    & 20.53~               & 19.41               & 15.80          & 8.61             & 84.54          & \textbf{99.86}                           \\ 

\multicolumn{1}{c}{}                                              & {\cellcolor[rgb]{0.937,0.937,0.937}}                                & H@10                 & 43.82            & 24.16                   & 1.32                    & 52.76~               & 46.82               & 24.11          & 33.21            & 98.99          & \textbf{100}                             \\ 

\multicolumn{1}{c}{}                                              & {\cellcolor[rgb]{0.937,0.937,0.937}}                                & MR                   & 40.41            & 149.27                  & 377.50                  & 25.86                & 78.26               & 149.64         & 94.45            & 1.89           & \textbf{1.001}                           \\ 

\multicolumn{1}{c}{}                                              & \multirow{-4}{*}{{\cellcolor[rgb]{0.937,0.937,0.937}}imdb-tvdb}     & MRR                  & 0.23             & 0.13                    & 0.009                   & 0.32                 & 0.29                & 0.18           & 0.16             & 0.89           & \textbf{0.99}                            \\ 
\hhline{~-----------|}
\multirow{4}{*}{\rotcell{\textbf{New}}}                                                  & {\cellcolor[rgb]{0.937,0.937,0.937}}                                & H@1                  & 10.63            & 5.60                    & 0.14                    & 14.13~               & 14.51               & 20.00          & 12.26            & 34.41          & \textbf{100}                             \\ 

                                                                   & {\cellcolor[rgb]{0.937,0.937,0.937}}                                & H@10                 & 49.68            & 29.73                   & 1.45                    & 58.31                & 54.27               & 27.09          & 52.23            & 65.42          & \textbf{100}                             \\ 

                                                                   & {\cellcolor[rgb]{0.937,0.937,0.937}}                                & MR                   & 22.89            & 74.29                   & 344.50                  & 14.59~               & 31.79               & 133.18         & 23.94            & 28.11          & \textbf{1}                               \\ 

                                                                   & \multirow{-4}{*}{{\cellcolor[rgb]{0.937,0.937,0.937}}tmdb-tvdb}     & MRR                  & 0.22             & 0.13                    & 0.01                    & 0.27~                & 0.26                & 0.22           & 0.24             & 0.44           & \textbf{1}                               \\ 
\hhline{~-----------|}
\multirow{4}{*}{}                                                  & {\cellcolor[rgb]{0.937,0.937,0.937}}                                & H@1                  & N/A              & N/A                     & N/A                     & N/A                  & N/A                 & N/A            & N/A              & \textbf{94.6}  & N/A                                      \\ 

                                                                   & {\cellcolor[rgb]{0.937,0.937,0.937}}                                & H@10                 & N/A              & N/A                     & N/A                     & N/A                  & N/A                 & N/A            & N/A              & \textbf{98.19} & N/A                                      \\ 

                                                                   & {\cellcolor[rgb]{0.937,0.937,0.937}}                                & MR                   & N/A              & N/A                     & N/A                     & N/A                  & N/A                 & N/A            & N/A              & \textbf{1.68}  & N/A                                      \\ 

                                                                   & \multirow{-4}{*}{{\cellcolor[rgb]{0.937,0.937,0.937}}Restaurants}   & MRR                  & N/A              & N/A                     & N/A                     & N/A                  & N/A                 & N/A            & N/A              & \textbf{0.95}  & N/A                                      \\
\end{tabular}}
 \begin{tablenotes}
  \item[*] For memory reasons, the source code of BERT\_INT restricts the evaluation computations to the top-$k$ ranks, which may yield better results in some cases (default $k$=50). We set $k$=1000, which is the largest $k$ value for which we did not run into  memory issues.
  \end{tablenotes}
\end{table}

\subsection{Effectiveness of EA Methods Using  Different Metrics}\label{sec:evaluation}

To answer Q1 and Q2, we experimentally compare five relation-based (MTransE, MTransE+RotatE, RDGCN, RREA(basic) and RREA(semi)) and three attribute-based methods (KDCoE, MultiKE and AttrE), using the protocol and metrics described in Section~\ref{ssec:measures}. Table~\ref{tab:scores} reports the different evaluation metrics per method over all datasets of our testbed.

\subsubsection{Relation-based EA Methods}

\textbf{MTransE} exploits one-hop entity neighborhoods and does not use negative sampling. As we can see in Table~\ref{tab:scores}, the performance of MTransE remains relatively the same between the sparse version D\_W\_15K\_V1 with 3 average relations per entity\footnote{A KG is considered ``dense'' in~\cite{DBLP:journals/pvldb/SunZHWCAL20} if it has more than an average of 5 relations per entity.} (Table~\ref{metafeatures}), and the dense one D\_W\_15K\_V2 with 5.58 average relations per entity, as well as between the sparse D\_Y\_15K\_V1 with 3.19 average relations per entity, and the dense D\_Y\_15K\_V2 with 8.55 average relations per entity. In general, entities in dense KGs are more likely to minimize their embedding distance with multiple similar entities (neighbors), exploiting richer semantic information to learn the embeddings. For this reason, dense KGs are the best use cases for relation-based methods. However, the density of KGs is also reflected in the embedding space. In case that the embedding space has limited dimensions and the KGs are dense, dissimilar entities end up close in the embedding space. The lack of negative sampling (that distances dissimilar entities in the embedding space) is the reason why MTransE is not improved in dense datasets.

In the new datasets (BBC-DB, imdb-tmdb, imdb-tvdb and tmdb-tvdb) that are all sparse, fewer entity pairs in seed alignment and fewer relation edges are used for training, resulting in worse performance, too, for this method (when comparing BBC-DB to tmdb-tvdb). For example, Hits@1 for BBC-DB with 61,039 relation edges is 24.98, while Hits@1 for imdb-tmdb with 9,604 relation edges is 15. On the contrary, MR is improved as can been seen in Table~\ref{tab:scores}. This is attributed to the dependency of MR on the size of the similarity lists (see the evaluation protocol in Section~\ref{ssec:protocol}). Specifically, as the entity pairs of seed alignment are reduced, the size of similarity lists is reduced too. The smaller the similarity lists, the lower the ranks for true entity pairs in the test set, resulting to an improved MR metric.

\textbf{MTransE+RotatE} is a variation of MTransE that exploits RotatE as embedding module, as well as negative sampling, and was originally proposed for better handling symmetrical relations in KGs~\citep{DBLP:journals/pvldb/SunZHWCAL20}. Since such relations are not encountered in datasets imdb\_tmdb, imdb\_tvdb, and tmdb\_tvdb (see Table~\ref{tab:basic_statistics}), we investigate only the effect of negative sampling on the performance of the method. MTransE+RotatE exhibits a similar performance to MTransE in the sparse versions of D\_W\_15K\_V1 and  D\_Y\_15K\_V1 (e.g., in dataset  D\_W\_15K\_V1, MTransE Hits@10 is 72.78 and MTransE+RotatE Hits@10 is 72.94). However, the negative sampling used by MTransE+RotatE significantly improves its performance in dense datasets compared to MTransE. For example, for D\_Y\_15K\_V2, the MR of MTransE is 79.47 and that of MTransE+RotatE is 3.46, and for   D\_W\_15K\_V2, MTransE Hits@1 is 26.27 and MTransE+RotatE Hits@1 is 49.22. This means that MTransE+RotatE represents similar entities closer in the embedding space and dissimilar entities farther, exploiting the rich semantic information that dense KGs offer.

In the new datasets, MTransE+RotatE is outperformed by MTransE (e.g., for tmdb-tvdb Hits@1 of MTransE is 10.63, while Hits@1 of MTransE+RotatE drops to 5.60). This is due to the small size of the new datasets in terms of seed entity pairs (e.g., 982 in tmdb-tvdb) and relation edges (e.g., 3,396 for $KG_1$ vs only 938 for $KG_2$ in tmdb-tvdb) that, in conjunction with the high dimensionality of the embedding space, place similar entities far away. Negative sampling seems to amplify this behavior.

\textbf{RDGCN} exploits multi-hop entity neighborhoods, as well as a higher number of negative samples and dimensions in the embedding space than the other methods (see Table \ref{tab:hyperparameters}). As shown in Table~\ref{tab:scores}, RDGCN has the highest Hits@1 and Hits@10, compared to MTransE and MTransE+RotatE in datasets D\_W\_15K and D\_Y\_15K, both for sparse (V1) and dense (V2) versions (e.g., for D\_W\_15K\_V2 and  D\_Y\_15K\_V1, RDGCN Hits@1 is respectively 63.71 and 94.10, while MTransE Hits@1 is 26.27 and 45.54 and MTransE+RotatE Hits@1 is 49.22 and 46.25). This behavior can be attributed to the fact that RDGCN exploits an extended neighborhood of entities (multi-hop) for constructing the embeddings. Of course, this richer semantic context incurs the risk of incorporating noisy information into the embeddings, due to potentially irrelevant or less important distant neighbors for aligning entities. For these reasons, RDGCN requires both high-dimensional embeddings to capture more features, as well as increased negative samples per positive edge that place irrelevant neighbors far away in the embedding space. It is worth mentioning that RDGCN outperforms the other methods in all OpenEA datasets, demonstrating the added value of multi-hop neighbors in embeddings. On the other hand, RDGCN is outperformed by MTransE+RotatE in datasets BBC-DB, imdb-tmdb, imdb-tvdb and tmdb-tvdb (e.g., in BBC-DB Hits@1 drops from 18.41 to 6.38). This behavior stems from the same problem that negative sampling causes to MTransE+RotatE in these datasets. As RDGCN uses more negative samples and higher dimensions in the embedding space than MTransE+RotatE, it aggravates the problem.

\textbf{RREA(basic)}, like RDGCN, exploits multi-hop entity neighborhoods, having the same issue about irrelevant or less important neighbors. Contrary to RDGCN, that needs higher number of negative samples and dimensions to solve the aforementioned issue, RREA(basic) utilizes an attention mechanism that assigns high weights to relevant and important neighbors, outperforming RDGCN, but also MTransE and MTransE+RotatE in all datasets, regardless of the metric (e.g., in D\_W\_15K\_V2, RREA(basic)'s Hits@1 is 87.86 and for MTransE it is 26.27, in D\_W\_15K\_V1, RREA(basic)'s Hits@10 is 88.49, while for MTransE+RotatE it is 56.98, and in D\_Y\_15K\_V2, RREA(basic)'s MR is 1.32, while the corresponding MR for RDGCN is 11.50). As we can see from D\_Y\_15K\_V1, RREA(basic) fails to differentiate the importance of relevant and irrelevant neighbors, and RDGCN achieves higher performance in all metrics (e.g., Hits@1 for RDGCN is 94.10, while for RREA(basic) it is 76.19). The fact that RREA(basic) solves the issue of noisy information and irrelevant neighbors by utilizing the attention mechanism, without requiring higher number of negative samples, leads to outperforming MTransE, MTransE+RotatE and RDGCN in all new datasets (e.g., RREA(basic)'s Hits@1 for imdb\_tmdb is 27.06, while for MTransE it is 15, RREA(basic)'s Hits@10 for bbc\_db is 65.28, while for MTransE+RotatE it is 41.66 and RREA(basic)'s MR for imdb\_tvdb is 25.86, while for RDGCN it is 377.50).

\textbf{RREA(semi)} runs in 5 iterations, where the first iteration is RREA(basic), which means that the 4 extra iterations, increase the performance of RREA(semi) compared to RREA(basic) and consequently to MTransE, MTransE+RotatE and RDGCN in all datasets, in Hits@1, Hits@10 and MRR (e.g., RREA(semi)'s Hits@1 for D\_Y\_15K\_V2 is 97.53, while for RREA(basic) it is 96.56, RREA(semi)'s Hits@10 for imdb\_tmdb is 73.65, while for RDGCN it is 0.73, and RREA(semi)'s MRR for imdb\_tvdb is 0.29, while for MTransE it is 0.23). RREA(semi), like RREA(basic) fails to distinguish relevant and irrelevant neighbors in D\_Y\_15K\_V1, where RDGCN achieves better performance in all metrics (e.g., Hits@1 for RDGCN is 94.10, while for RREA(semi) it is 81.79). The fact that RREA(semi) enriches the seed alignment with new entity pairs leads to increasing the candidate-space (denominator of MR) and also the probability of wrong alignment (compared to RREA(basic)), making RREA(semi) exhibit worse MR in all datasets (e.g., in D\_Y\_15K\_V2, RREA(semi)'s MR is 1.39, while for RREA(basic) it is 1.32).

\subsubsection{Attribute-based EA Methods}

\textbf{KDCoE} in addition to one-hop neighborhood and negative sampling, exploits the textual descriptions of entities, when available. Negative sampling also improves the performance of KDCoE in dense versions of the datasets, compared to their sparse versions (e.g., in dataset  D\_W\_15K\_V1 Hits@10 is 46.29, while in  D\_W\_15K\_V2 Hits@10 is 64.38, and in dataset  D\_Y\_15K\_V1 MR is 143.78, while in  D\_Y\_15K\_V2 MR is 2.28). In dense entity graphs, negative sampling helps KDCoE to increase the distance of dissimilar entities in an embedding space with limited dimensions. The use of textual descriptions as an additional source of similarity evidence favors KDCoE to outperform the other methods in datasets that are rich in terms of textual descriptions. As shown in Table~\ref{tab:scores}, this is the case of BBC\_DB, which features the highest number of entities with textual descriptions (2,430), compared to the size of the seed alignment (9,396), and where KDCoE exhibits better performance compared to MTransE, MTransE+RotatE and RDGCN. The higher ratio of entities with similar textual descriptions in a dataset boosts the performance of KDCoE. It is worth mentioning that although KDCoE outperforms the aforementioned methods in BBC\_DB, both versions of RREA exhibit higher performance in all metrics, due to the attention mechanism that both utilize and the rule that RREA(semi) uses. In case of lower ratio of entities with similar textual descriptions in a dataset, KDCoE's performance drops as the training set is flooded with falsely similar entity pairs in terms of textual descriptions. This behavior is observed in datasets imdb\_tmdb, imdb\_tvdb, tmdb\_tvdb, where textual descriptions of entities exhibit the minimum similarity among all datasets (e.g., for imdb-tmdb descriptions similarity is 0.35). Statistically meaningful correlations of the method with the specific data characteristics (\#Ents\_Descr and Descr\_Sim) are reported in Section~\ref{ssec:meta-learning}. Note that the disfavored effect of negative sampling in these datasets still persists (as in MTransE+RotatE and RDGCN), but with an attenuated strength, due the smaller number of negative samples this method uses.

\textbf{MultiKE} exploits multiple sources of similarity evidence of entities (i.e., from attribute names, literals and relations) along with a high number of negative samples. In this respect, MultiKE embeddings are less sensitive to the number of entity relations than in the previous methods. For this reason, MultiKE, in contrary to KDCoE, is not improving in all dense versions of the datasets, even if it uses negative sampling (see Table~\ref{tab:scores}). Specifically, in D\_W\_15K, when comparing the dense to the sparse version, Hits@1 increases from 40.49 to 49.54 as both the number (average attributes per entity is 3 for sparse and 5.58 for dense) and the similarity score of literals (i.e., 81.24 for sparse and 83.79 for dense) increase. In contrary, from the sparse D\_Y\_15K\_V1 to the dense  D\_Y\_15K\_V2, there is a small decrease ($\approx8\%$) in both entity names similarity (from 0.99 to 0.91), causing Hits@1 to drop from 90.18 to 85.64. Finally, in datasets imdb\_tmdb, imdb\_tvdb and tmdb\_tvdb, MultiKE has a lower Hits@1 than KDCoE (i.e., for imdb\_tmdb Hits@1 of KDCoE is 20.02, while for MultiKE it is 15.28), due to the increased number of negative samples. The correlation of MultiKE with the similarity of entity names will be further investigated in Section~\ref{hom_het}.

\textbf{AttrE} is an unsupervised method that consumes fewer negative samples (like KDCoE) and exploits the similarity of literal values. On the other hand, it requires the alignment of predicates in KGs as a pre-processing step. The highest the number of aligned predicates, the better the performance of the method. In datasets without aligned predicates, AttrE is unable to run. Specifically, AttrE is not applicable to datasets D\_W\_15K\_V1 and D\_W\_15K\_V2, while on D\_Y\_15K\_V1 and  D\_Y\_15K\_V2, it exhibits a very low performance (e.g., Hits@1 is 0.0003 for D\_Y\_15K\_V1) due to the low predicate similarity (e.g., 0 in  D\_W\_15K\_V1 and 47.37\% in  D\_Y\_15K\_V1). In datasets like BBC\_DB, where both the similarity of predicates (80.93) and the similarity of literals (0.87) are high, AttrE's performance is significantly improved. In datasets imdb-tmdb, imdb-tvdb and tmdb-tvdb, exhibiting the highest similarity of predicates and literals, AttrE outperforms all other methods. The correlation of the method with specific data characteristics (Pred\_Sim and Lit\_Sim) will be presented in Section~\ref{ssec:meta-learning}. Finally, AttrE is the only method that is able to run on the Restaurants datasets, where not all the entities of $KG_1$ are aligned with entities of $KG_2$ and vice versa.

\textbf{BERT\_INT} is an attribute-based method that exploits multiple sources of similarity evidence between entities (i.e., long-text descriptions, literals, names, and relations). More precisely, BERT\_INT, rather than aggregating the information of multi-hop neighbors that in some cases introduces noise (as in RDGCN), even by the usage of attention mechanisms (as in RREA), it employs interactions between the different views (i.e., description/name-view, neighbor-view and attribute-views), considering multi-hops and consuming very few negative samples compared to RDGCN. Regarding the factual information, it prioritizes  textual descriptions over entity names, in order to leverage as much textual information as possible (textual descriptions are typically longer than entity names). BERT\_INT has the advantage, compared to MultiKE, that it also exploits multiple resources but without prioritizing them, since the noise of irrelevant factual information is not incorporated in the embeddings. BERT\_INT also has an advantage over AttrE and KDCoE, that they exclusively exploit either longer textual descriptions or other literals.

Table~\ref{tab:scores} shows that BERT\_INT outperforms all methods in all datasets and measures, except the two versions of D\_W\_15K. For example, in D\_Y\_15K\_V1, BERT\_INT achieves 99.23 H@1 when RDGCN and RREA(semi) achieve 94.10 and 81.79. In D\_Y\_15K\_V2, we observe similar behavior with BERT\_INT exhibiting 99.37 when RDGCN and RREA(semi) achieve 94.15 and 97.53. This happens, because in both versions of D\_Y\_15K the number of the entities that have textual descriptions, the descriptions similarity and the entity names are high (e.g., in D\_Y\_15K\_V1 3,455 entities have textual descriptions with 0.88 similarity and 0.99 similarity in names). The opposite behavior is observed in both versions of D\_W\_15K, where RREA outperforms BERT\_INT with H@10 90.03 instead of 48.96 that BERT\_INT achieves, since in these datasets the number of the entities that have descriptions is low (only 278 in D\_W\_15K\_V2), with low similarity (0.73) and low similarity in entity names (0.78). In new datasets, the similarity of long-text descriptions and entity names is very low, the literal similarity is very high (0.94 in tmdb-tvdb); as a result, the performance of BERT\_INT is even better, outperforming even AttrE (e.g., in tmdb-tvdb, BERT\_INT finds all the matches correctly, with an impressive MRR 1, while the corresponding MRR of the second best method, AttrE, is only 0.44).
\vspace{-.3cm}
\subsubsection{Conventional EA Methods}

\begin{table}
\centering
\caption{Performance of PARIS and of the best embedding-based method per dataset.}
\label{tab:PARIS}
\arrayrulecolor{black}
\resizebox{0.6\textwidth}{!}{\begin{tabular}{l|c|c|c|c|} 
\hhline{~----|}
                                                                  & {\cellcolor[rgb]{0.937,0.937,0.937}}\textbf{Dataset}                &           & \textbf{PARIS} & \textbf{Best Method}  \\ 
\hhline{~----|}
\multirow{3}{*}{}                                                 & {\cellcolor[rgb]{0.937,0.937,0.937}}                                & Precision & \textbf{96.51} & 71.82                 \\ 
\hhline{~>{\arrayrulecolor[rgb]{0.937,0.937,0.937}}->{\arrayrulecolor{black}}---|}
                                                                  & {\cellcolor[rgb]{0.937,0.937,0.937}}                                & Recall    & \textbf{74.55} & 71.82                 \\ 
\hhline{~>{\arrayrulecolor[rgb]{0.937,0.937,0.937}}->{\arrayrulecolor{black}}---|}
                                                                  & \multirow{-3}{*}{{\cellcolor[rgb]{0.937,0.937,0.937}}D\_W\_15K\_V1} & F1-score  & \textbf{84.12} & 71.82                 \\ 
\hhline{~----|}
\multirow{3}{*}{\rotcell{\textbf{Datasets}}}                      & {\cellcolor[rgb]{0.937,0.937,0.937}}                                & Precision & \textbf{97.59} & 93.72                 \\ 
\hhline{~>{\arrayrulecolor[rgb]{0.937,0.937,0.937}}->{\arrayrulecolor{black}}---|}
                                                                  & {\cellcolor[rgb]{0.937,0.937,0.937}}                                & Recall    & 90.32          & \textbf{93.72}        \\ 
\hhline{~>{\arrayrulecolor[rgb]{0.937,0.937,0.937}}->{\arrayrulecolor{black}}---|}
                                                                  & \multirow{-3}{*}{{\cellcolor[rgb]{0.937,0.937,0.937}}D\_W\_15K\_V2} & F1-score  & \textbf{93.81} & 93.72                 \\ 
\hhline{~----|}
\multirow{3}{*}{}                                                 & {\cellcolor[rgb]{0.937,0.937,0.937}}                                & Precision & \textbf{99.72} & 99.23                 \\ 
\hhline{~>{\arrayrulecolor[rgb]{0.937,0.937,0.937}}->{\arrayrulecolor{black}}---|}
                                                                  & {\cellcolor[rgb]{0.937,0.937,0.937}}                                & Recall    & 96.24          & \textbf{99.23}        \\ 
\hhline{~>{\arrayrulecolor[rgb]{0.937,0.937,0.937}}->{\arrayrulecolor{black}}---|}
                                                                  & \multirow{-3}{*}{{\cellcolor[rgb]{0.937,0.937,0.937}}D\_Y\_15K\_V1} & F1-score  & 97.95          & \textbf{99.23}        \\ 
\hhline{~----|}
\multirow{3}{*}{\rotcell{\mbox{\textbf{OpenEA}}}}                        & {\cellcolor[rgb]{0.937,0.937,0.937}}                                & Precision & \textbf{99.85} & 99.37                 \\ 
\hhline{~>{\arrayrulecolor[rgb]{0.937,0.937,0.937}}->{\arrayrulecolor{black}}---|}
                                                                  & {\cellcolor[rgb]{0.937,0.937,0.937}}                                & Recall    & 98.37          & \textbf{99.37}        \\ 
\hhline{~>{\arrayrulecolor[rgb]{0.937,0.937,0.937}}->{\arrayrulecolor{black}}---|}
                                                                  & \multirow{-3}{*}{{\cellcolor[rgb]{0.937,0.937,0.937}}D\_Y\_15K\_V2} & F1-score  & 99.10          & \textbf{99.37}        \\ 
\hhline{~====|}
\multirow{3}{*}{}                                                 & {\cellcolor[rgb]{0.937,0.937,0.937}}                                & Precision & 78.43          & \textbf{92.50}        \\ 
\hhline{~>{\arrayrulecolor[rgb]{0.937,0.937,0.937}}->{\arrayrulecolor{black}}---|}
                                                                  & {\cellcolor[rgb]{0.937,0.937,0.937}}                                & Recall    & 26.38          & \textbf{92.50}        \\ 
\hhline{~>{\arrayrulecolor[rgb]{0.937,0.937,0.937}}->{\arrayrulecolor{black}}---|}
                                                                  & \multirow{-3}{*}{{\cellcolor[rgb]{0.937,0.937,0.937}}BBC-DB}        & F1-score  & 38.72          & \textbf{92.50}        \\ 
\hhline{~----|}
\multicolumn{1}{c|}{\multirow{3}{*}{\rotcell{\textbf{ }}}}        & {\cellcolor[rgb]{0.937,0.937,0.937}}                                & Precision & 56.84          & \textbf{99.70}        \\ 
\hhline{~>{\arrayrulecolor[rgb]{0.937,0.937,0.937}}->{\arrayrulecolor{black}}---|}
\multicolumn{1}{c|}{}                                             & {\cellcolor[rgb]{0.937,0.937,0.937}}                                & Recall    & 8.99           & \textbf{99.70}        \\ 
\hhline{~>{\arrayrulecolor[rgb]{0.937,0.937,0.937}}->{\arrayrulecolor{black}}---|}
\multicolumn{1}{c|}{}                                             & \multirow{-3}{*}{{\cellcolor[rgb]{0.937,0.937,0.937}}imdb-tmdb}     & F1-score  & 15.53          & \textbf{99.70}        \\ 
\hhline{~----|}
\multicolumn{1}{c|}{\multirow{3}{*}{\rotcell{\textbf{Datasets}}}} & {\cellcolor[rgb]{0.937,0.937,0.937}}                                & Precision & 58.37          & \textbf{99.86}        \\ 
\hhline{~>{\arrayrulecolor[rgb]{0.937,0.937,0.937}}->{\arrayrulecolor{black}}---|}
\multicolumn{1}{c|}{}                                             & {\cellcolor[rgb]{0.937,0.937,0.937}}                                & Recall    & 9.17           & \textbf{99.86}        \\ 
\hhline{~>{\arrayrulecolor[rgb]{0.937,0.937,0.937}}->{\arrayrulecolor{black}}---|}
\multicolumn{1}{c|}{}                                             & \multirow{-3}{*}{{\cellcolor[rgb]{0.937,0.937,0.937}}imdb-tvdb}     & F1-score  & 15.84          & \textbf{99.86}        \\ 
\hhline{~----|}
\multirow{3}{*}{\rotcell{\textbf{New}}}                           & {\cellcolor[rgb]{0.937,0.937,0.937}}                                & Precision & 99.76          & \textbf{100}          \\ 
\hhline{~>{\arrayrulecolor[rgb]{0.937,0.937,0.937}}->{\arrayrulecolor{black}}---|}
                                                                  & {\cellcolor[rgb]{0.937,0.937,0.937}}                                & Recall    & 97.50          & \textbf{100}          \\ 
\hhline{~>{\arrayrulecolor[rgb]{0.937,0.937,0.937}}->{\arrayrulecolor{black}}---|}
                                                                  & \multirow{-3}{*}{{\cellcolor[rgb]{0.937,0.937,0.937}}tmdb-tvdb}     & F1-score  & 98.62          & \textbf{100}          \\ 
\hhline{~----|}
\multirow{3}{*}{}                                                 & {\cellcolor[rgb]{0.937,0.937,0.937}}                                & Precision & 28.79          & \textbf{94.60}        \\ 
\hhline{~>{\arrayrulecolor[rgb]{0.937,0.937,0.937}}->{\arrayrulecolor{black}}---|}
                                                                  & {\cellcolor[rgb]{0.937,0.937,0.937}}                                & Recall    & 69.76          & \textbf{94.60}        \\ 
\hhline{~>{\arrayrulecolor[rgb]{0.937,0.937,0.937}}->{\arrayrulecolor{black}}---|}
                                                                  & \multirow{-3}{*}{{\cellcolor[rgb]{0.937,0.937,0.937}}Restaurants}   & F1-score  & 40.18          & \textbf{94.60}        \\
\hhline{~----|}
\end{tabular}}
\end{table}

PARIS~\citep{DBLP:journals/pvldb/SuchanekAS11} is a conventional, probabilistic, holistic approach, i.e., it performs both instance and schema matching, by estimating probabilities of matching (i.e., equivalence), without learning KG embeddings. Such probability estimation relies on the knowledge of quasi-functional relations, i.e., relations that for a given head entity, the expected number of tail entities is close to 1. Given, for example, two relation triples $<$$h, r, t$$>$ and $<$$h', r, t'$$>$, with $r$ being a functional relation (e.g., born\_in), the equivalence probability $h \equiv h'$ depends recursively on other equivalence probabilities ($t \equiv t'$). Initial equivalences are computed between literals, using a certain string distance (their similarity is inverse proportional to their edit distance).

Table~\ref{tab:PARIS} depicts the performance of PARIS along with the best embedding-based method for each dataset according to Table~\ref{tab:scores}. As in~\cite{DBLP:journals/pvldb/SunZHWCAL20}, for PARIS, we report the classification-based metrics (Precision, Recall and F1-score), while for the embedding-based method, we used the H@1. This is because, in the test phase, each source entity gets a list of candidates; as a result, precision, recall and F1-score can be considered equal to Hits@1. Regarding the OpenEA datasets, we observe that in both versions of D\_W\_15K, where the literal similarity is high (for D\_W\_15K\_V1, it is 0.81), PARIS outperforms the best embedding-based method (in this case RREA(semi)). On the other hand, in both versions of D\_Y\_15K, the best method (BERT\_INT) exhibits higher Recall and F1-score (in D\_Y\_15K\_V2 Recall is 99.37, instead of 98.37 of PARIS and F1-score is 99.37, instead of 99.10 of PARIS), while the precision is lower (99.37) but still close to PARIS (99.85). This happens because of the low literal similarity that these datasets exhibit (affecting the Precision) and the high textual descriptions and entity names similarity (improving Recall and F1-score). Finally, in the new datasets, even if the literal similarity is high, the lack of functional relations affect negatively the performance of PARIS, and as a result, BERT\_INT is the clear winner in these datasets (e.g., in imdb-tmdb F1-score is 15.53 for PARIS, while BERT\_INT achieves 99.70).

\subsection{Effectiveness vs Efficiency Trade-offs}\label{sec:effectivenessVSefficiency}

In this section, we investigate whether it is worth paying the time overhead of each method, with respect to its effectiveness (Q3). In this respect, in Section~\ref{ssec:ranking}, we first rank the methods according to their effectiveness across all datasets of our testbed. For that, we rely on a statistically sound methodology for ranking the different EA methods across all datasets of our testbed, that exploits the non-parametric Friedman test~\citep{DBLP:journals/jmlr/Demsar06} and the post-hoc Nemenyi test~\citep{nemenyi1963distribution}. Then, in Section~\ref{ssec:efficiency}, we report the execution time of each method (Table~\ref{tab:time}) and analyze their trade-offs in terms of effectiveness and efficiency (Figure~\ref{fig:trade}), demonstrating also the training curves of the methods and the time that a method needs to reach 90\% of its highest MRR (Figure~\ref{fig:TTA}).
\vspace{-.2cm}
\subsubsection{Effectiveness-based Ranking of EA Methods}\label{ssec:ranking} 
In this section, for each of the metrics used in Table~\ref{tab:scores}, we first calculate the ranks (rank 1 corresponds to the best performance) of all methods per dataset and average their ranks for each metric. The resulting ranks are reported in Tables~\ref{tab:average_ranks_hits1}, \ref{tab:average_ranks_hits10}, \ref{tab:average_ranks_MR}, and \ref{tab:average_ranks_MRR}. 

\begin{table}
\centering
\caption{Ranking of methods per dataset according to Hits@1.}
\label{tab:average_ranks_hits1}
\resizebox{\textwidth}{!}{\begin{tabular}{|l|c|c|c|c|c|c|c|c|c|} 
\hline
\multicolumn{1}{|c|}{\textbf{Dataset}}              & \textbf{MTransE} & \textbf{MTransE+RotatE} & \textbf{\textbf{RDGCN}} & \textbf{RREA(basic)} & \textbf{RREA(semi)} & \textbf{KDCoE} & \textbf{\textbf{MultiKE}} & \textbf{AttrE} & \textbf{BERT\_INT}  \\ 
\hline
D\_W\_15K\_V1                                       & 7                & 6                       & 3                       & 2                    & 1                   & 8              & 5                         & 9              & 4                   \\ 
\hline
D\_W\_15K\_V2                                       & 8                & 5                       & 3                       & 2                    & 1                   & 7              & 4                         & 9              & 6                   \\ 
\hline
D\_Y\_15K\_V1                                       & 8                & 7                       & 2                       & 5                    & 4                   & 6              & 3                         & 9              & 1                   \\ 
\hline
D\_Y\_15K\_V2                                       & 8                & 5                       & 4                       & 3                    & 2                   & 6              & 7                         & 9              & 1                   \\ 
\hline
BBC-DB                                              & 5                & 6                       & 9                       & 3                    & 2                   & 4              & 8                         & 7              & 1                   \\ 
\hline
imdb-tmdb                                           & 7                & 8                       & 9                       & 4                    & 3                   & 5              & 6                         & 2              & 1                   \\ 
\hline
imdb-tvdb                                           & 6                & 8                       & 9                       & 3                    & 4                   & 5              & 7                         & 2              & 1                   \\ 
\hline
tmdb-tvdb                                           & 7                & 8                       & 9                       & 5                    & 4                   & 3              & 56                        & 2              & 1                   \\ 
\hline
\rowcolor[rgb]{0.937,0.937,0.937} \textbf{Avg rank} & 7                & 6.62                   & 6                       & 3.37                & 2.62               & 5.5            & 5.75                      & 6.12          & 2                   \\
\hline
\end{tabular}}
\end{table}

\begin{table}
\vspace{-.5cm}
\centering
\caption{Ranking of methods per dataset according to Hits@10.}
\label{tab:average_ranks_hits10}
\resizebox{\textwidth}{!}{\begin{tabular}{|l|c|c|c|c|c|c|c|c|c|} 
\hline
\multicolumn{1}{|c|}{\textbf{Dataset}}              & \textbf{MTransE} & \textbf{MTransE+RotatE} & \textbf{\textbf{RDGCN}} & \textbf{RREA(basic)} & \textbf{RREA(semi)} & \textbf{KDCoE} & \textbf{\textbf{MultiKE}} & \textbf{AttrE} & \textbf{BERT\_INT}  \\ 
\hline
D\_W\_15K\_V1                                       & 6                & 4                       & 3                       & 2                    & 1                   & 8              & 5                         & 9              & 7                   \\ 
\hline
D\_W\_15K\_V2                                       & 7                & 4                       & 3                       & 2                    & 1                   & 6              & 5                         & 9              & 8                   \\ 
\hline
D\_Y\_15K\_V1                                       & 8                & 7                       & 2                       & 5                    & 4                   & 6              & 3                         & 9              & 1                   \\ 
\hline
D\_Y\_15K\_V2                                       & 8                & 4                       & 6                       & 2                    & 1                   & 5              & 7                         & 9              & 3                   \\ 
\hline
BBC-DB                                              & 5                & 7                       & 9                       & 2                    & 3                   & 4              & 8                         & 6              & 1                   \\ 
\hline
imdb-tmdb                                           & 6                & 7                       & 9                       & 3                    & 4                   & 8              & 5                         & 2              & 1                   \\ 
\hline
imdb-tvdb                                           & 5                & 7                       & 9                       & 3                    & 4                   & 8              & 6                         & 2              & 1                   \\ 
\hline
tmdb-tvdb                                           & 6                & 7                       & 9                       & 3                    & 4                   & 8              & 5                         & 2              & 1                   \\ 
\hline
\rowcolor[rgb]{0.937,0.937,0.937} \textbf{Avg rank} & 6.37            & 5.87                   & 6.25                    & 2.75                 & 2.75                & 6.62          & 5.5                       & 6              & 2.87               \\
\hline
\end{tabular}}
\end{table}

\begin{table}
\centering
\caption{Ranking of methods per dataset according to MR.}
\label{tab:average_ranks_MR}
\resizebox{\textwidth}{!}{\begin{tabular}{|l|c|c|c|c|c|c|c|c|c|} 
\hline
\multicolumn{1}{|c|}{\textbf{Dataset}}              & \textbf{MTransE} & \textbf{MTransE+RotatE} & \textbf{\textbf{RDGCN}} & \textbf{RREA(basic)} & \textbf{RREA(semi)} & \textbf{KDCoE} & \textbf{\textbf{MultiKE}} & \textbf{AttrE} & \textbf{BERT\_INT}  \\ 
\hline
D\_W\_15K\_V1                                       & 5                & 7                       & 6                       & 2                    & 3                   & 8              & 4                         & 9              & 1                   \\ 
\hline
D\_W\_15K\_V2                                       & 6                & 5                       & 8                       & 1                    & 2                   & 8              & 4                         & 9              & 3                   \\ 
\hline
D\_Y\_15K\_V1                                       & 7                & 8                       & 2                       & 4                    & 5                   & 6              & 3                         & 9              & 1                   \\ 
\hline
D\_Y\_15K\_V2                                       & 8                & 5                       & 7                       & 2                    & 3                   & 4              & 6                         & 9              & 1                   \\ 
\hline
BBC-DB                                              & 5                & 8                       & 9                       & 2                    & 3                   & 4              & 7                         & 6              & 1                   \\ 
\hline
imdb-tmdb                                           & 4                & 7                       & 9                       & 3                    & 6                   & 8              & 5                         & 2              & 1                   \\ 
\hline
imdb-tvdb                                           & 4                & 7                       & 9                       & 3                    & 5                   & 8              & 6                         & 2              & 1                   \\ 
\hline
tmdb-tvdb                                           & 3                & 7                       & 9                       & 2                    & 6                   & 8              & 4                         & 5              & 1                   \\ 
\hline
\rowcolor[rgb]{0.937,0.937,0.937} \textbf{Avg rank} & 5.25             & 6.75                    & 7.37                   & 2.37                & 4.12               & 6.62          & 4.87                     & 6.37          & 1.25                \\
\hline
\end{tabular}}
\end{table}

\begin{table}
\vspace{-.5cm}
\centering
\caption{Ranking of methods per dataset according to MRR.}
\label{tab:average_ranks_MRR}
\resizebox{\textwidth}{!}{\begin{tabular}{|l|c|c|c|c|c|c|c|c|c|} 
\hline
\multicolumn{1}{|c|}{\textbf{Dataset}}              & \textbf{MTransE} & \textbf{MTransE+RotatE} & \textbf{\textbf{RDGCN}} & \textbf{RREA(basic)} & \textbf{RREA(semi)} & \textbf{KDCoE} & \textbf{\textbf{MultiKE}} & \textbf{AttrE} & \textbf{BERT\_INT}  \\ 
\hline
D\_W\_15K\_V1                                       & 7                & 6                       & 3                       & 2                    & 1                   & 8              & 4                         & 9              & 5                   \\ 
\hline
D\_W\_15K\_V2                                       & 8                & 4                       & 3                       & 2                    & 1                   & 6              & 5                         & 9              & 7                   \\ 
\hline
D\_Y\_15K\_V1                                       & 8                & 7                       & 2                       & 5                    & 4                   & 6              & 3                         & 9              & 1                   \\ 
\hline
D\_Y\_15K\_V2                                       & 8                & 5                       & 4                       & 2.5                  & 2.5                 & 6              & 7                         & 9              & 1                   \\ 
\hline
BBC-DB                                              & 5                & 7                       & 9                       & 3                    & 2                   & 4              & 8                         & 6              & 1                   \\ 
\hline
imdb-tmdb                                           & 6                & 8                       & 9                       & 4                    & 3                   & 7              & 5                         & 2              & 1                   \\ 
\hline
imdb-tvdb                                           & 5                & 8                       & 9                       & 3                    & 4                   & 6              & 7                         & 2              & 1                   \\ 
\hline
tmdb-tvdb                                           & 7                & 8                       & 9                       & 3                    & 4                   & 6              & 5                         & 2              & 1                   \\ 
\hline
\rowcolor[rgb]{0.937,0.937,0.937} \textbf{Avg rank} & 6.75             & 6.62                    & 6                       & 3.06                 & 2.68                & 6.12           & 5.50                      & 6              & 2.25                \\
\hline
\end{tabular}}
\end{table}

Then, to infer a statistically significant ranking of the methods, we rely on the non-parametric Friedman test~\citep{DBLP:journals/jmlr/Demsar06}. The null hypothesis $H_0$ of the Friedman test is that ``The mean performance for each method is equal'', while the alternative hypothesis $(H_a)$ states exactly the opposite. With p-values 0.004, 0.003, 0.001 and 0.003 of Friedman test for Hits@1, Hits@10, MR and MRR, respectively, we can reject the null hypothesis $H_0$ at a $10\%$ confidence-level $(\alpha)$. In the sequel, we conduct the Nemenyi post-hoc test to compare the methods pairwise. This test reports as significance the average ranks of two methods if they differ by a critical distance (CD) given by $q_{\alpha} \sqrt{\frac{k(k+1)}{6 N}}$, where $N$ is the number of the datasets, $q_{\alpha}$ is a constant based on $\alpha$, and $k$ is the number of methods in total. For 9 EA methods, 8 datasets and $\alpha=0.1$, the value of $CD$ is 3.90. 

From Figure~\ref{fig:neme}, we can see that BERT\_INT is a clear winner, since it outperforms the other methods with statistically significant difference (bigger difference than CD). 
Figure~\ref{fig:neme_hits@1}, which captures accuracy with 0 error tolerance (Hits@1), shows that BERT\_INT has statistically significant performance difference to MTransE, MTransE+RotatE, AttrE and RDGCN, while RREA(semi) only to MTransE and MTransE+RotatE. 
In case we accept a 10\% error tolerance (Hits@10 in Figure~\ref{fig:neme_hits@10}), none of the examined methods has significant performance difference to the other methods, so the choice should depend on different factors, e.g., the training time (see Section~\ref{ssec:efficiency}) or dataset characteristics (see Section~\ref{ssec:meta-learning}). 
In addition, BERT\_INT that exploits multiple sources of similarity evidence of entities (e.g., textual descriptions, entity names) and not only structural information, as relation-based methods such as RREA(basic) do, improves the general performance, improving the ranks even of lower ranked entities in the similarities lists (see Section~\ref{ssec:protocol}). 
The result is that BERT\_INT has significant performance difference with MTransE, RDGCN, MTransE+RotatE, KDCoE and AttrE, while RREA(basic) has significant performance difference with only RDGCN, MTransE+RotatE, KDCoE and AttrE, for MR. 
In contrast, MRR  (Figure~\ref{fig:neme_MRR}) which is affected by the first ranks (see Section~\ref{metrics}) and is less sensitive to outliers, shows that BERT\_INT is the only method that has statistically significant performance difference with other methods (MTransE and MTransE+RotatE).

\begin{figure}[ht]
\begin{subfigure}{.5\textwidth}
  \centering
  \includegraphics[width=1\linewidth]{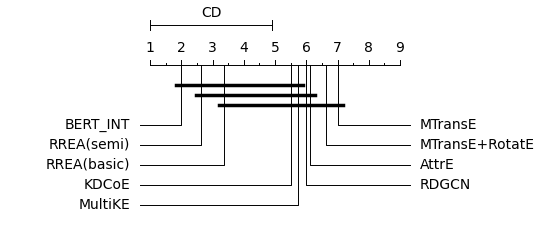}  
  \caption{Hits@1}
  \label{fig:neme_hits@1}
\end{subfigure}
\begin{subfigure}{.5\textwidth}
  \centering
  \includegraphics[width=1\linewidth]{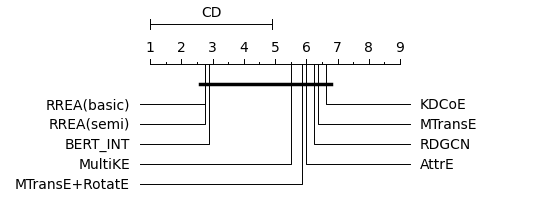}  
  \caption{Hits@10}
  \label{fig:neme_hits@10}
\end{subfigure}
\begin{subfigure}{.5\textwidth}
  \centering
  \includegraphics[width=1\linewidth]{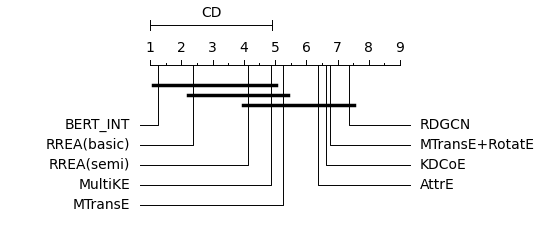}  
  \caption{MR}
  \label{fig:neme_MR}
\end{subfigure}
\begin{subfigure}{.5\textwidth}
  \centering
  \includegraphics[width=1\linewidth]{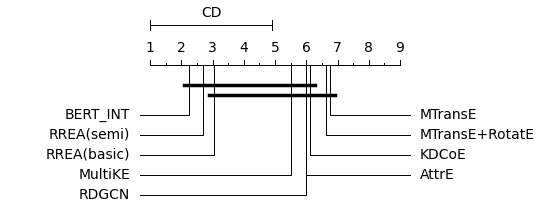}  
  \caption{MRR}
  \label{fig:neme_MRR}
\end{subfigure}
\caption{Statistical significance rankings of methods according to all metrics.}
\label{fig:neme}
\vspace{-0.5cm}
\end{figure}

\subsubsection{Efficiency-based Ranking of EA Methods}\label{ssec:efficiency}

In this subsection, we compare the efficiency of the evaluated EA methods. Particularly, in Table~\ref{tab:time}, we are measuring the execution time of both training and testing. Figure~\ref{fig:trade} depicts the trade-off between effectiveness and efficiency of EA methods using BERT\_INT as baseline. In addition, Figure \ref{fig:TTA} demonstrates the training curves of the methods, as well as the time or the epochs that a method needs to achieve 90\% of its own highest MRR.

As MTransE has the most lightweight implementation among all methods~\citep{DBLP:journals/pvldb/SunZHWCAL20} (only two layers of entity and relation embeddings), it is the fastest method (Table~\ref{tab:time}). Although MTransE+RotatE also uses only two layers, MTransE is faster because it employs a simpler scoring function than RotatE (see Section~\ref{translational}). BERT\_INT, that achieves the best performance in most datasets (excluding the two versions of D\_W\_15K), and MultiKE are competing for the second most efficient method. Specifically, as both BERT\_INT and MultiKE trains and combines multi-view embeddings, they require more training time (long-lasting epochs) than MTransE, despite the fact that the latter needs more epochs according to Table~\ref{tab:epochs} (short-lasting epochs). On the other hand, BERT\_INT and MultiKE are faster than MTransE+RotatE and KDCoE, since BERT\_INT and MultiKE converge faster (see Table~\ref{tab:epochs}). In addition, BERT\_INT and MultiKE are faster than RREA(basic) and RREA(semi), since the last two do not utilize early stopping and they need multiple epochs to be trained (1,200 and 5$\times$1,200=6,000, respectively). Finally, BERT\_INT and MultiKE are faster than the unsupervised method AttrE, even if the latter needs fewer epochs for training (it does not utilize early stopping). This is due to the nature of AttrE refining embeddings in multiple iterations, by exploiting all the possible literal values and not the seed alignment.

\begin{table}
\centering
\caption{Training + testing time (execution time in seconds) of methods in each dataset. The dash `-' means that the method could not run in this dataset.}
\label{tab:time}
\resizebox{\textwidth}{!}{\begin{tabular}{|l|c|c|c|c|c|c|c|c|c|} 
\hline
\multicolumn{1}{|c|}{\textbf{Dataset}} & \textbf{MTransE}       & \textbf{MTransE+RotatE} & \textbf{RDGCN}   & \textbf{RREA(basic)} & \textbf{RREA(semi)} & \textbf{KDCoE}  & \textbf{MultiKE} & \textbf{AttrE}   & \textbf{BERT\_INT}  \\ 
\hline
D\_W\_15K\_V1                          & \textbf{53.54 + 3.67}  & 236.76 + 6.85           & 853.80 + 26      & 442.68 + 8.69        & 1,158.73 + 8.22     & 784.75 + 5.67   & 403.53 + 4.38    & -                & 333.34 + 30.09      \\ 
\hline
D\_W\_15K\_V2                          & \textbf{44.41 + 3.62}  & 1,028.26 + 6.94         & 1,520.21 + 26.86 & 593.32 + 8.33        & 1,543.96 + 9        & 1,052.56 + 4.50 & 372.29 + 5.37    & -                & 333.61 + 29.80      \\ 
\hline
D\_Y\_15K\_V1                          & \textbf{101.76 + 3.63} & 405.33 + 6.92           & 356.50 + 27.07   & 387.43 + 8.12        & 1,027.93 + 7.56     & 795.42 + 5.34   & 348.55 + 4.40    & 1,472.26 + 54    & 331.73 + 29.95      \\ 
\hline
D\_Y\_15K\_V2                          & \textbf{88.85 + 3.62}  & 470.96 + 6.78           & 1,185.88 + 27.80 & 570.38 + 9           & 1,468 + 8.25        & 1,585.22 + 4.71 & 154.12 + 4.52    & 1,416.78 + 20.59 & 331.40 + 29.75      \\ 
\hline
BBC-DB                                 & \textbf{44.88 + 2.21}  & 209.26 + 2.84           & 221.97 + 10.38   & 230.33 + 5.27        & 595.32 + 5.56       & 690.21 + 2.95   & 400.93 + 2.05    & 938.05 + 7.48    & 206.69 + 17.13      \\ 
\hline
imdb-tmdb                              & \textbf{11.94 + 0.30}  & 52.32 + 0.30            & 12.09 + 1.44     & 71.41 + 3.74         & 195.12 + 3.90       & 57.23 + 1.12    & 41.90 + 0.64     & 107.47 + 0.32    & 42.62 + 4.04        \\ 
\hline
imdb-tvdb                              & \textbf{5.16 + 0.24}   & 37.93 + 0.19            & 9.31 + 1.03      & 42.77 + 3.48         & 117.37 + 3.48       & 32.90 + 1.22    & 47.57 + 0.55     & 53.16 + 0.08     & 23.80 + 1.16        \\ 
\hline
tmdb-tvdb                              & \textbf{6.03 + 0.24}   & 37.62 + 0.17            & 5.61 + 1.02      & 41.07 + 3.53         & 110.67 + 3.53       & 25.93 + 1.22    & 34.96 + 0.54     & 53.79 + 0.08     & 22.06 + 0.91        \\
\hline
\end{tabular}}

\end{table}

\begin{table}
\vspace{-.5cm}
\centering
\caption{Number of epochs employed for training.}
\label{tab:epochs}
\resizebox{\textwidth}{!}{\begin{tabular}{|l|c|c|c|c|c|c|c|c|c|} 
\hline
\multicolumn{1}{|c|}{\textbf{Dataset}} & \begin{tabular}[c]{@{}c@{}}\textbf{MTransE}\\rel\_emb\end{tabular} & \begin{tabular}[c]{@{}c@{}}\textbf{MTransE+RotatE}\\rel\_emb\end{tabular} & \begin{tabular}[c]{@{}c@{}}\textbf{RDGCN}\\rel\_emb\end{tabular} & \begin{tabular}[c]{@{}c@{}}\textbf{RREA(basic)}\\rel\_emb\end{tabular} & \begin{tabular}[c]{@{}c@{}}\textbf{RREA(semi)}\\rel\_emb\end{tabular} & \begin{tabular}[c]{@{}c@{}}\textbf{KDCoE}\\rel\_emb + desc\_emb\end{tabular} & \begin{tabular}[c]{@{}c@{}}\textbf{MultiKE}\\multi-view\_emb\end{tabular} & \begin{tabular}[c]{@{}c@{}}\textbf{AttrE}\\rel\_emb\end{tabular} & \begin{tabular}[c]{@{}c@{}}\textbf{BERT\_INT}\\BERT + interactions\end{tabular}  \\ 
\hline
D\_W\_15K\_V1                          & 224                                                                & 322                                                                       & 220                                                              & 1200 (fix)                                                             & 6000 (fix)                                                            & 256/iter. + 568/iter.                                                        & 114                                                                       & -                                                                & 5 + 200 (fix)                                                                          \\ 
\hline
D\_W\_15K\_V2                          & 144                                                                & 300                                                                       & 224                                                              & 1200 (fix)                                                             & 6000 (fix)                                                            & 304/iter. + 640/iter.                                                        & 154                                                                       & -                                                                & 5 + 200 (fix)                                                                          \\ 
\hline
D\_Y\_15K\_V1                          & 472                                                                & 396                                                                       & 70                                                               & 1200 (fix)                                                             & 6000 (fix)                                                            & 204/iter. + 490/iter.                                                        & 90                                                                        & 50 (fix)                                                         & 5 + 200 (fix)                                                                          \\ 
\hline
D\_Y\_15K\_V2                          & 318                                                                & 332                                                                       & 140                                                              & 1200 (fix)                                                             & 6000 (fix)                                                            & 268/iter. + 568/iter.                                                        & 86                                                                        & 50 (fix)                                                         & 5 + 200 (fix)                                                                          \\ 
\hline
BBC-DB                                 & 236                                                                & 234                                                                       & 126                                                              & 1200 (\textbar{}fix)                                                   & 6000 (fix)                                                            & 230/iter. + 344/iter.                                                        & 116                                                                       & 50 (fix)                                                         & 5 + 200 (fix)                                                                          \\ 
\hline
imdb-tmdb                              & 196                                                                & 178                                                                       & 50                                                               & 1200 (fix)                                                             & 6000 (fix)                                                            & 220/iter. + 184/iter.                                                        & 74                                                                        & 50 (fix)                                                         & 5 + 200 (fix)                                                                          \\ 
\hline
imdb-tvdb                              & 130                                                                & 224                                                                       & 50                                                               & 1200 (fix)                                                             & 6000 (fix)                                                            & 122/iter. + 144/iter.                                                        & 56                                                                        & 50 (fix)                                                         & 5 + 200 (fix)                                                                          \\ 
\hline
tmdb-tvdb                              & 120                                                                & 228                                                                       & 50                                                               & 1200 (fix)                                                             & 6000 (fix)                                                            & 74/iter. + 42/iter.                                                          & 72                                                                        & 50 (fix)                                                         & 5 + 200 (fix)                                                                          \\
\hline
\end{tabular}}
\end{table}

\begin{figure}[ht]
  \centering    
\includegraphics[width=1\textwidth]{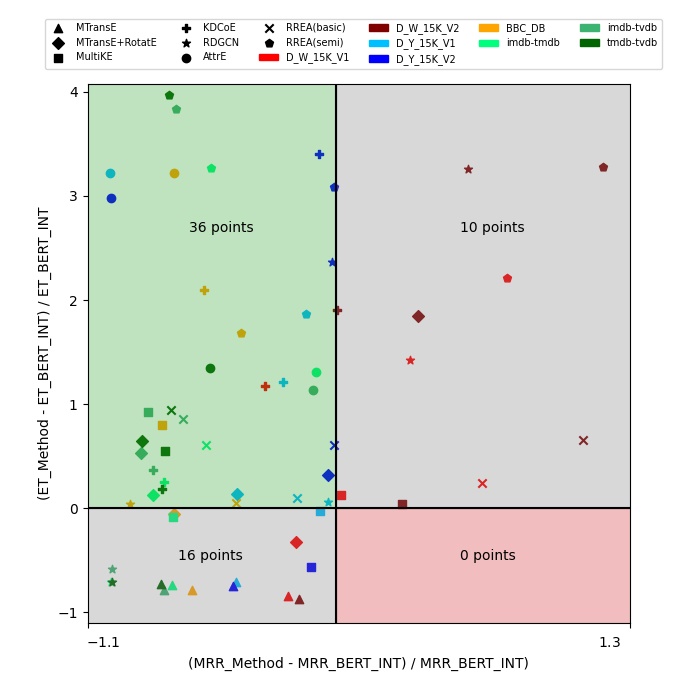}
\caption{Trade-off between efficiency and effectiveness (MRR) of methods across all datasets, using BERT\_INT as baseline. The x-axis is log-scaled using symmetric log function.}
\label{fig:trade}
\vspace{-0.5cm}
\end{figure}

Figure~\ref{fig:trade} illustrates the trade-off between effectiveness and efficiency of all methods, compared to a baseline method (BERT\_INT).
The x-axis represents the relative difference of the methods in effectiveness, measured by MRR, the metric that is the least sensitive to outliers (see Section \ref{metrics}), and the y-axis represents the relative difference in execution time, with respect to the baseline. In particular, the red quadrant (bottom right) indicates methods that both run faster and have a better effectiveness than the baseline (i.e., dominate the baseline) in a specific dataset, and the green quadrant (top left) exactly the opposite (i.e., methods dominated by the baseline). The bottom-left grey quadrant indicates methods that run faster with lower performance than the baseline and the top-right grey quadrant indicates methods that have better performance, but are slower than the baseline method.

The choice of BERT\_INT as baseline is because it is the method that dominates most methods (highest number of methods on the top-left quadrant), and it is not dominated by any other method (0 methods in the bottom-right quadrant). As shown in Figure~\ref{fig:neme_MRR}, BERT\_INT is the method with the best MRR in both OpenEA and new datasets. Indicatively, the corresponding values for the fastest method, MTransE, when chosen as a baseline, are: 16 (top left), 46 (top right), 0 (bottom left), and 0 (bottom right).

As BERT\_INT outperforms all methods in all datasets (except of the two versions of D\_W\_15K) in terms of MRR, with low a execution time, the top-left quadrant consists of 36 out of 62 (58.06\%) points. The bottom-left grey quadrant contains 16 out of 62 (25.80\%) points, as BERT\_INT is slower than the fastest method MTransE in all datasets, as well as slower than the competing method MultiKE in some datasets. More precisely, MTransE is on average 394.14\% faster than BERT\_INT, achieving 138.73\% lower MRR, while MultiKE is 43.79\% faster than BERT\_INT, at a cost of 10.61\% lower MRR in the two versions of D\_Y\_15K. Finally, since BERT\_INT is negatively affected by the low attribute (e.g., textual descriptions and entity names) similarity in dense KGs, in terms of effectiveness, the top-right quadrant contains 10 out of 62 (16.12\%) points. In KGs with such characteristics, RREA(basic) and RDGCN outperform BERT\_INT, achieving 88.63\% and 45.45\% better effectiveness with an overhead of 226.62\% and 42.53\% longer execution time, respectively.

In Figure~\ref{fig:TTA}, we demonstrate the training curves of the methods for each of the OpenEA datasets, while we also point out the training time that a method needs, in order to achieve 90\% of its highest MRR (represented as a dot in each curve), similar to the \emph{time-to-accuracy} (TTA) metric in the literature~\citep{coleman2017dawnbench, DBLP:journals/sigops/ColemanKNNZZBOR19}, measured here in seconds (x-axis labels) and epochs (first element of the points coordinates). As suggested in~\cite{coleman2017dawnbench, DBLP:journals/sigops/ColemanKNNZZBOR19}, we run the methods in only one fold of the datasets. We exclude the AttrE method, since it cannot run with D\_W\_15K\_V1 and D\_W\_15K\_V2, while for D\_Y\_15K\_V1 and D\_Y\_15K\_V2 it yields MRR close to zero. In addition, it is worth mentioning that the training curve of RREA(basic) (Figure~\ref{fig:TTA}) overlaps with part of the training curve of RREA(semi), since the former is the first iteration (out of 5) of the latter. Last but not least, in Table~\ref{tab:coefficient_variation} we demonstrate the coefficient of variation (CV) of methods training epochs per dataset. More precisely, CV measures the relative variability of the methods epochs and is given by $\sigma/\mu$, where $\sigma$ is the standard deviation and $\mu$ is the mean. The higher the values, the greater the degree of relative variability. 

From Figure~\ref{fig:TTA}, we firstly conclude that the increased training time of RREA(semi) is not worth spending, since the improvement of its MRR is very low, compared to RREA(basic), and it reaches 90\% of its highest MRR in its first iteration. In addition, we conclude that in most cases, both versions of RREA, not only achieve the highest MRR, but also reach every threshold faster (steeper slope) in only few seconds or epochs, e.g., in D\_W\_15K\_V1, RREA(basic) needs only 9.43s (7 epochs) for 0.69 MRR. RDGCN achieves quite lower MRR in reasonable time e.g., in D\_W\_15K\_V2 it needs only 107.19s (19 epochs) for 0.62 MRR. Although BERT\_INT is the fastest or the second fastest method (see Table~\ref{tab:time}), as it needs few short-lasting epochs for training, yet, it takes more time to fine-tune the BERT model and launch the training. This explains why BERT\_INT appears on right extreme of the x-axis in Figure~\ref{fig:TTA}, but it still manages to finish before all other methods. It is worth mentioning that it takes only 1 or 2 epochs to achieve an MRR that is even better than RREA or RDGCN. E.g., in D\_Y\_15K\_V1 it needs 281.97s (1 epoch) for 0.99 MRR. Last but not least, as we can see from Table~\ref{tab:coefficient_variation}, KDCoE, RDGCN and BERT\_INT have CVs greater than 1, which means that the time of each epoch varies a lot during training. In contrary to those methods, MTransE, MTransE+RotatE, RREA(basic), RREA(semi), and MultiKE exhibit low variability in epochs duration.

\begin{figure}[ht]
\begin{subfigure}{.5\textwidth}
  \centering
  \includegraphics[width=\textwidth]{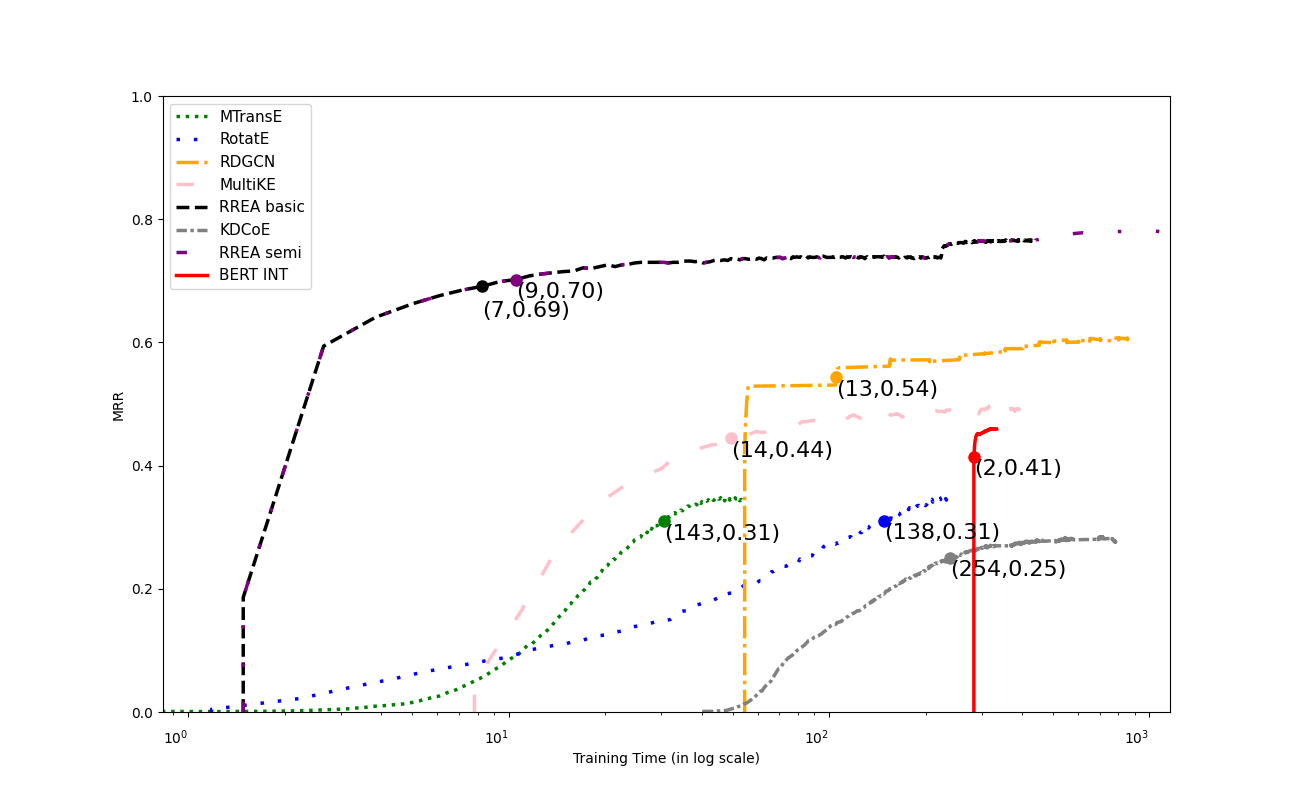}
  \caption{D\_W\_15K\_V1}  
  \label{fig:TTA_D_W_15K_V1}
\end{subfigure}
\begin{subfigure}{.5\textwidth}
  \centering
  \includegraphics[width=\textwidth]{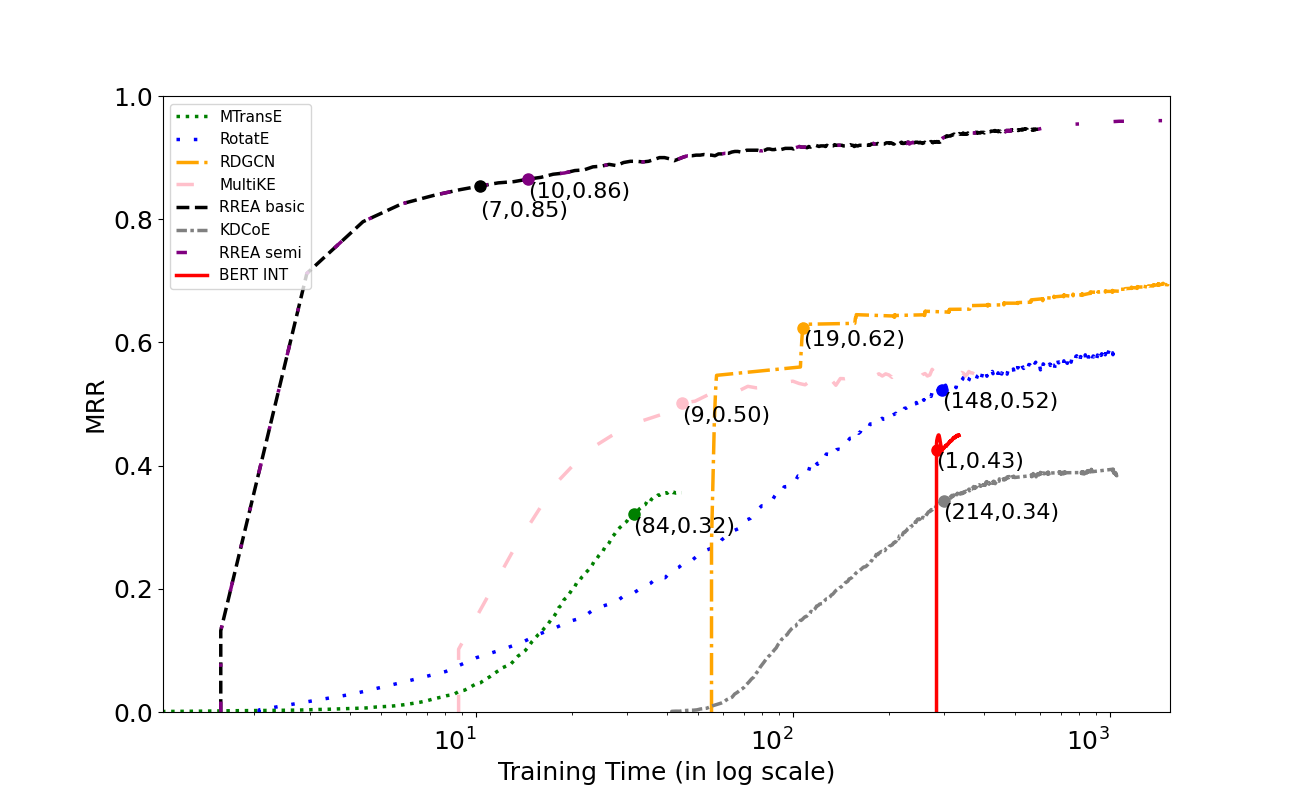}
  \caption{D\_W\_15K\_V2}  
  \label{fig:TTA_D_W_15K_V2}
\end{subfigure}
\begin{subfigure}{.5\textwidth}
  \centering
  \includegraphics[width=\textwidth]{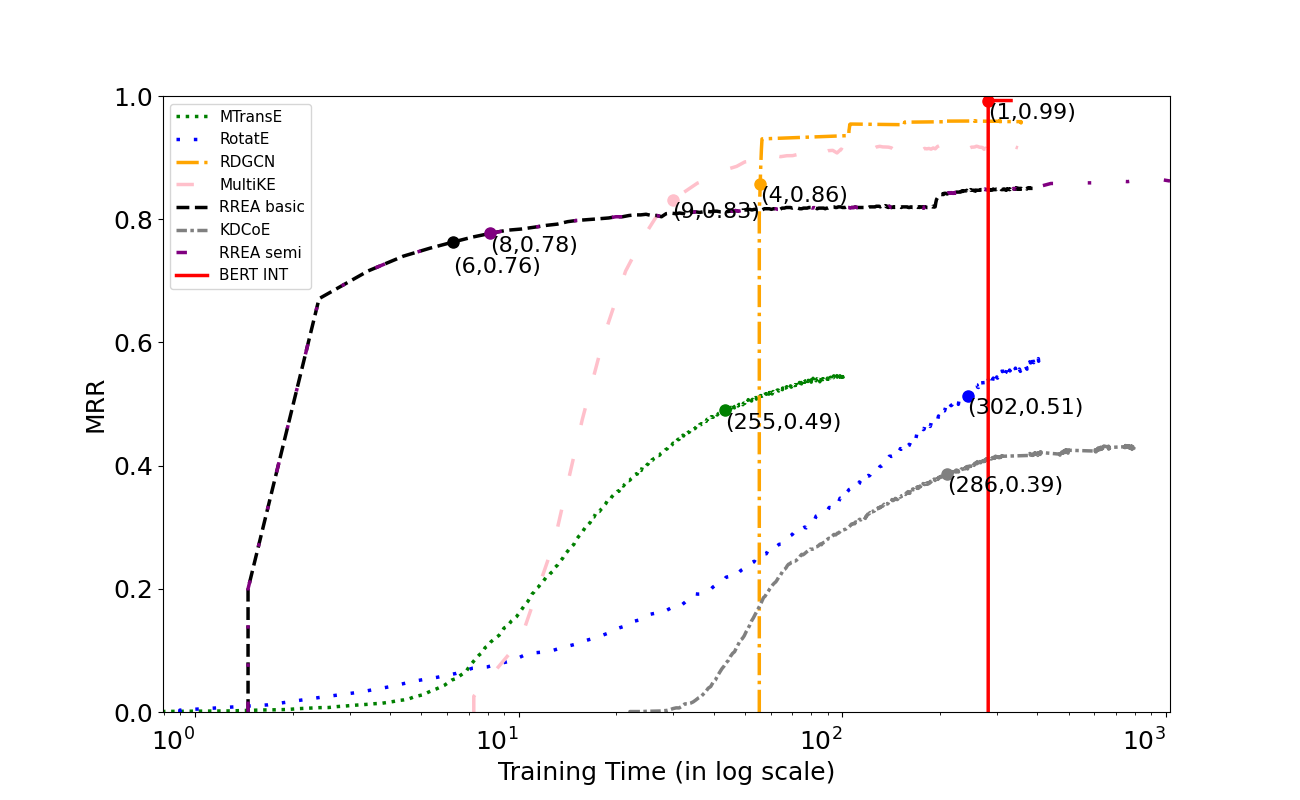}
  \caption{D\_Y\_15K\_V1}  
  \label{fig:TTA_D_Y_15K_V1}
\end{subfigure}
\begin{subfigure}{.5\textwidth}
  \centering
  \includegraphics[width=\textwidth]{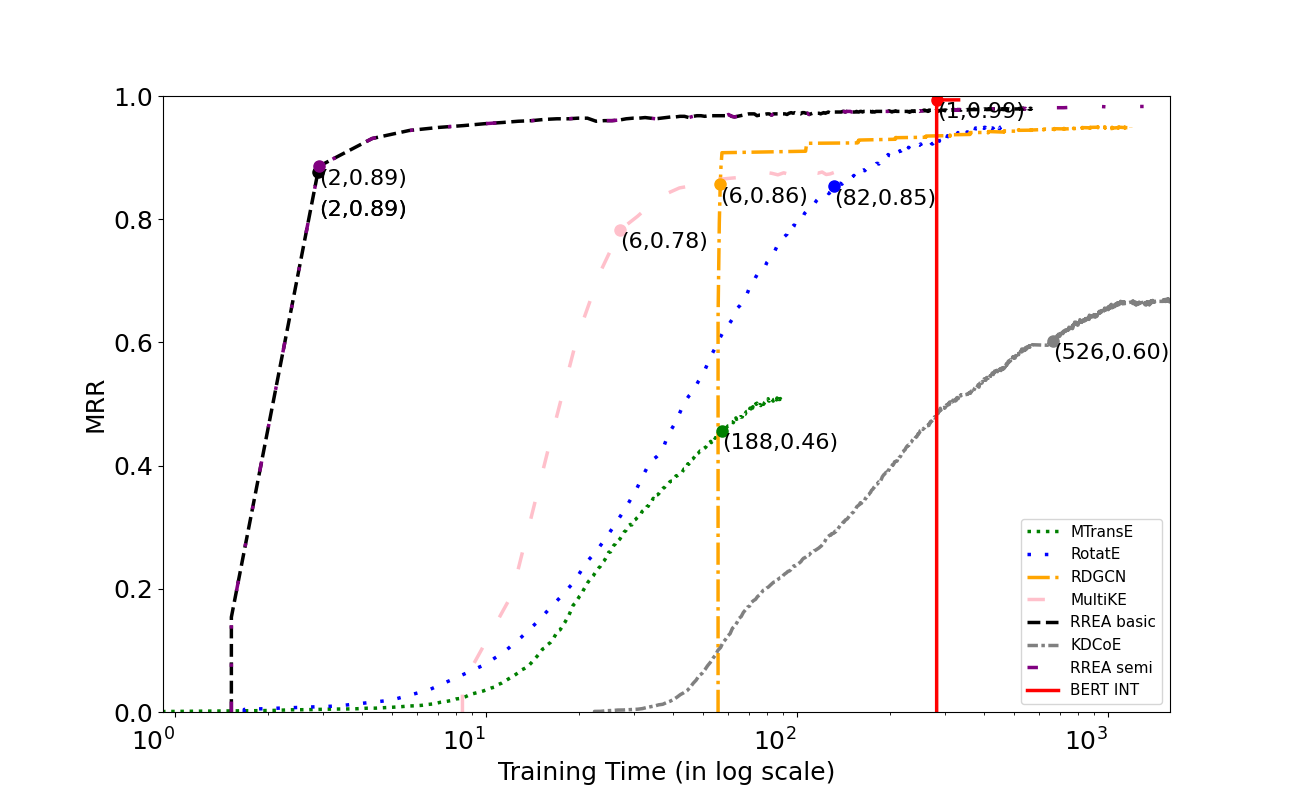}
  \caption{D\_Y\_15K\_V2}  
  \label{fig:TTA_D_Y_15K_V2}
\end{subfigure}
\caption{Methods training time curve for each dataset. Each of the seven points represents the time (measured both in seconds and epochs) that a specific method needs to achieve 90\% of its highest MRR. The x-axis is log-scaled.}
\label{fig:TTA}
\vspace{-0.5cm}
\end{figure}

\begin{table}
\centering
\caption{Coefficient of variation (CV) of methods training epochs.}
\label{tab:coefficient_variation}
\resizebox{\textwidth}{!}{\begin{tabular}{|l|c|c|c|c|c|c|c|c|} 
\hline
\multicolumn{1}{|c|}{\textbf{Dataset}} & \textbf{MTransE} & \textbf{MTransE+RotatE} & \textbf{\textbf{RDGCN}} & \textbf{RREA(basic)} & \textbf{RREA(semi)} & \textbf{KDCoE} & \textbf{\textbf{MultiKE}} & \textbf{\textbf{BERT\_INT}}  \\ 
\hline
D\_W\_15K\_V1                           & 0.031            & 0.009                   & 1.907                   & 0.022                 & 0.018                & 5.283          & 0.019               &   11.94    \\ 
\hline
D\_W\_15K\_V2                           & 0.031            & 0.010                   & 1.789                   & 0.019                 & 0.016                & 4.234          & 0.014              &    11.93    \\ 
\hline
D\_Y\_15K\_V1                           & 0.035            & 0.012                   & 1.955                   & 0.026                 & 0.021                & 6.788          & 0.013            &    11.95      \\ 
\hline
D\_Y\_15K\_V2                           & 0.027            & 0.008                   & 1.849                   & 0.018                 & 0.016                & 7.739          & 0.017  &      11.94              \\ \hline
\end{tabular}}
\end{table}

\subsection{Meta-level Analysis of EA methods}
\label{ssec:meta-learning}

In this subsection, we report the statistically significant correlations between the various meta-features (Table~\ref{metafeatures}) extracted by the KGs of our datasets and the performance achieved by EA methods (Q4). In this experiment, we exclude the Restaurants dataset, since only AttrE was able to run on this dataset.

The null hypothesis $(H_0)$ of our correlation analysis is that ``there is no significant correlation between the performance metric \textbf{X} and the meta-feature \textbf{Y}'', while the alternative hypothesis $(H_a)$ states exactly the opposite. We reject the null hypothesis when the p-values of correlations are less than a significance level $\alpha$. We use the Spearman's correlation, which measures the degree of association between two ranked variables, with a coefficient $r_s=1-\frac{6 \sum d_{i}^{2}}{n\left(n^{2}-1\right)}$, where $d_i$ is the difference between the ranks of the two variables and $n$ is the maximum rank. This choice is motivated by the robustness of the Spearman's correlation to outliers (unlike Pearson's correlation), as it relies on the ranks of the variables and not on their actual values. The closer $r_s$ is to zero, the weaker the association between two ranked variables (negative or positive). A positive Spearman's correlation denotes that when a meta-feature increases the method-specific metric increases too, while negative Spearman's correlation denotes the opposite direction of the association\footnote{Correlation is not causation as the association direction is not known.} between the two ranked variables. 

Table~\ref{tab:correlations} reports the correlations for a significance level $\alpha=0.05$ (otherwise cells are empty) under different categories. To make the results easily comparable across the different performance metrics, we multiply correlations of MR with $-1$, since in contrary to the other metrics, the higher the value of MR the worse. Finally, for meta-features that impact only specific EA methods, e.g., description similarity concerns only KDCoE, their correlations with non relevant methods are omitted (cells are filled with dashes).

\begin{table}
\centering
\caption{Correlation matrix of meta-features with EA methods' performance metrics. Empty cells denote statistically not significant correlation, and dashed cells denote that a meta-feature is not relevant to an EA method.}
\label{tab:correlations}
\resizebox{\textwidth}{!}{\begin{tabular}{cccccccccccc} 
\toprule
\textbf{Methods}                & \textbf{Metrics} & \begin{tabular}[c]{@{}c@{}}\textbf{Seed}\\\textbf{Alignment Size}\end{tabular}                & \multicolumn{4}{c}{\textbf{Density}}                                                                                                                                                                                                                & \multicolumn{5}{c}{\textbf{Heterogeneity}}\\\cmidrule(lr){1-1}\cmidrule(lr){2-2}\cmidrule(lr){3-3}\cmidrule(lr){4-7}\cmidrule(lr){8-12}                                                                                                                                                                                                                                                                                                                                                            
                                &                  & \begin{tabular}[c]{@{}c@{}}\textbf{\#Entity\_ Pairs of}\\\textbf{Seed Alignment}\end{tabular} & \begin{tabular}[c]{@{}c@{}}\textbf{Avg\_ Rels\_}\\\textbf{ per\_}\\\textbf{ Entity }\end{tabular} & \begin{tabular}[c]{@{}c@{}}\textbf{Avg\_ Attrs\_}\\\textbf{ per\_}\\\textbf{ Entity }\end{tabular} & \textbf{Sole\_Rels} & \textbf{Hyper\_Rels} & \begin{tabular}[c]{@{}c@{}}\textbf{\#Ents\_}\\\textbf{ Descr }\end{tabular} & \begin{tabular}[c]{@{}c@{}}\textbf{Descr\_}\\\textbf{ Sim }\end{tabular} & \begin{tabular}[c]{@{}c@{}}\textbf{Ent\_ Name\_}\\\textbf{ Sim }\end{tabular} & \begin{tabular}[c]{@{}c@{}}\textbf{Literal\_}\\\textbf{ Sim}\end{tabular} & \begin{tabular}[c]{@{}c@{}}\textbf{Pred\_ Name}\\\textbf{ Sim}\end{tabular}  \\ 
\hline
\multirow{4}{*}{MTransE}        & H@1              & 0.94                                                                                          & 0.79                                                                                              & -                                                                                                  & -0.78               & 0.78                 & -                                                                           & -                                                                        & -                                                                             & -                                                                         & -                                                                            \\
                                & H@10             & 0.89                                                                                          & 0.74                                                                                              & -                                                                                                  &                     &                      & -                                                                           & -                                                                        & -                                                                             & -                                                                         & -                                                                            \\
                                & MR               &                                                                                               &                                                                                                   & -                                                                                                  &                     &                      & -                                                                           & -                                                                        & -                                                                             & -                                                                         & -                                                                            \\
                                & MRR              & 0.96                                                                                          & 0.79                                                                                              & -                                                                                                  & -0.78               & 0.78                 & -                                                                           & -                                                                        & -                                                                             & -                                                                         & -                                                                            \\ 
\hline
\multirow{4}{*}{MTransE + Rot.} & H@1              & 0.94                                                                                          & 0.90                                                                                              & -                                                                                                  & -0.81               & 0.81                 & -                                                                           & -                                                                        & -                                                                             & -                                                                         & -                                                                            \\
                                & H@10             & 0.94                                                                                          & 0.93                                                                                              & -                                                                                                  & -0.81               & 0.81                 & -                                                                           & -                                                                        & -                                                                             & -                                                                         & -                                                                            \\
                                & MR               &                                                                                               &                                                                                                   & -                                                                                                  &                     &                      & -                                                                           & -                                                                        & -                                                                             & -                                                                         & -                                                                            \\
                                & MRR              & 0.91                                                                                          & 0.93                                                                                              & -                                                                                                  & -0.81               & 0.81                 & -                                                                           & -                                                                        & -                                                                             & -                                                                         & -                                                                            \\ 
\hline
\multirow{4}{*}{RDGCN}          & H@1              & 0.84                                                                                          & 0.81                                                                                              & -                                                                                                  & -0.83               & 0.83                 & -                                                                           & -                                                                        &                                                                               & -                                                                         & -                                                                            \\
                                & H@10             & 0.84                                                                                          & 0.81                                                                                              & -                                                                                                  & -0.83               & 0.83                 & -                                                                           & -                                                                        &                                                                               & -                                                                         & -                                                                            \\
                                & MR               &                                                                                               &                                                                                                   & -                                                                                                  &                     &                      & -                                                                           & -                                                                        & 0.88                                                                          & -                                                                         & -                                                                            \\
                                & MRR              & 0.84                                                                                          & 0.74                                                                                              & -                                                                                                  & -0.78               & 0.78                 & -                                                                           & -                                                                        &                                                                               & -                                                                         & -                                                                            \\ 
\hline
\multirow{4}{*}{RREA(basic)}    & H@1              & 0.94                                                                                          & 0.9                                                                                               & -                                                                                                  & -0.81               & 0.81                 & -                                                                           & -                                                                        & -                                                                             & -                                                                         & -                                                                            \\
                                & H@10             & 0.89                                                                                          & 0.86                                                                                              & -                                                                                                  &                     &                      & -                                                                           & -                                                                        & -                                                                             & -                                                                         & -                                                                            \\
                                & MR               &                                                                                               &                                                                                                   & -                                                                                                  &                     &                      & -                                                                           & -                                                                        & -                                                                             & -                                                                         & -                                                                            \\
                                & MRR              & 0.94                                                                                          & 0.9                                                                                               & -                                                                                                  & -0.81               & 0.81                 & -                                                                           & -                                                                        & -                                                                             & -                                                                         & -                                                                            \\ 
\hline
\multirow{4}{*}{RREA(semi)}     & H@1              & 0.94                                                                                          & 0.9                                                                                               & -                                                                                                  & -0.83               & 0.83                 & -                                                                           & -                                                                        & -                                                                             & -                                                                         & -                                                                            \\
                                & H@10             & 0.89                                                                                          & 0.86                                                                                              & -                                                                                                  &                     &                      & -                                                                           & -                                                                        & -                                                                             & -                                                                         & -                                                                            \\
                                & MR               &                                                                                               &                                                                                                   & -                                                                                                  &                     &                      & -                                                                           & -                                                                        & -                                                                             & -                                                                         & -                                                                            \\
                                & MRR              & 0.94                                                                                          & 0.9                                                                                               & -                                                                                                  & -0.81               & 0.81                 & -                                                                           & -                                                                        & -                                                                             & -                                                                         & -                                                                            \\ 
\hline
\multirow{4}{*}{KDCoE}          & H@1              & 0.85                                                                                          & 0.93                                                                                              &                                                                                                    & -0.90               & 0.90                 & 0.81                                                                        & 0.88                                                                     & -                                                                             & -                                                                         & -                                                                            \\
                                & H@10             & 0.85                                                                                          & 0.93                                                                                              &                                                                                                    & -0.90               & 0.90                 & 0.81                                                                        & 0.88                                                                     & -                                                                             & -                                                                         & -                                                                            \\
                                & MR               &                                                                                               &                                                                                                   &                                                                                                    &                     &                      &                                                                             &                                                                          & -                                                                             & -                                                                         & -                                                                            \\
                                & MRR              & 0.85                                                                                          & 0.93                                                                                              &                                                                                                    & -0.90               & 0.90                 & 0.81                                                                        & 0.88                                                                     & -                                                                             & -                                                                         & -                                                                            \\ 
\hline
\multirow{4}{*}{MultiKE}        & H@1              & 0.89                                                                                          & 0.74                                                                                              &                                                                                                    &                     &                      & -                                                                           & -                                                                        & 0.71                                                                          & -                                                                         & -                                                                            \\
                                & H@10             & 0.81                                                                                          &                                                                                                   &                                                                                                    &                     &                      & -                                                                           & -                                                                        & 0.83                                                                          & -                                                                         & -                                                                            \\
                                & MR               &                                                                                               &                                                                                                   &                                                                                                    &                     &                      & -                                                                           & -                                                                        & 0.9                                                                           & -                                                                         & -                                                                            \\
                                & MRR              & 0.81                                                                                          &                                                                                                   &                                                                                                    &                     &                      & -                                                                           & -                                                                        & 0.83                                                                          & -                                                                         & -                                                                            \\ 
\hline
\multirow{4}{*}{AttrE}          & H@1              & -0.84                                                                                         &                                                                                                   &                                                                                                    &                     &                      & -                                                                           & -                                                                        & -                                                                             & 0.73                                                                      & 0.9                                                                          \\
                                & H@10             & -0.87                                                                                         &                                                                                                   & -0.73                                                                                              &                     &                      & -                                                                           & -                                                                        & -                                                                             &                                                                           & 0.83                                                                         \\
                                & MR               &                                                                                               &                                                                                                   &                                                                                                    &                     &                      & -                                                                           & -                                                                        & -                                                                             &                                                                           &                                                                              \\
                                & MRR              & -0.84                                                                                         &                                                                                                   &                                                                                                    &                     &                      & -                                                                           & -                                                                        & -                                                                             & 0.73                                                                      & 0.9                                                                          \\ 
\hline
\multirow{4}{*}{BERT\_INT}      & H@1              & -0.81                                                                                         &                                                                                                   & -0.9                                                                                               &                     &                      & -                                                                           & -                                                                        & -                                                                             & -                                                                         & -                                                                            \\
                                & H@10             & -0.78                                                                                         &                                                                                                   & -0.9                                                                                               &                     &                      & -                                                                           & -                                                                        & -                                                                             & -                                                                         & -                                                                            \\
                                & MR               &                                                                                               &                                                                                                   &                                                                                                    &                     &                      & -                                                                           & -                                                                        & -                                                                             & -                                                                         & -                                                                            \\
                                & MRR              &                                                                                               &                                                                                                   & -0.8                                                                                               &                     &                      & -                                                                           & -                                                                        & -                                                                             & -                                                                         & -                                                                            \\
\bottomrule
\end{tabular}}
\end{table}

\subsubsection{Seed Alignment Size}
As we can see in Table~\ref{tab:correlations}, the size of the seed alignment (\#Entity\_Pairs) is positively correlated with the supervised and semi-supervised methods MTransE, MTransE+RotatE, RDGCN, RREA(basic), RREA(semi), KDCoE, MultiKE and BERT\_INT and negatively correlated with the unsupervised method AttrE. Increasing the seed alignment and consequently, increasing the space of candidate matches, penalizes unsupervised methods, as it also increases the probability of wrong alignments. On the other hand, the size of the seed alignment is positively correlated with the performance of supervised and semi-supervised methods, as a large number of entity pairs in seed alignment implies larger KGs\footnote{In supervised and semi-supervised methods, all nodes of aligned KGs should be entities of the employed seed alignment.}, and by extension more training data. Although BERT\_INT is a supervised method, in the new datasets (which are smaller, but exhibit higher factual similarity than the OpenEA datasets), it achieves higher performance than in OpenEA datasets, resulting in a negative correlation with the seed alignment size. Note that we only measure correlations, which are not causal relations, as the association direction is not known.

\subsubsection{Density}\label{spa_den}
As we can see in Table~\ref{tab:correlations}, the average number of relations per entity is positively correlated with the performance of MTransE, MTransE+RotatE, RDGCN, RREA(basic), RREA(semi), KDCoE and MultiKE, while all methods are negatively correlated with the sole meta-feature and positively correlated with the hyper meta-feature (see Section \ref{ssec:statistics}). In addition, the average number of attributes per entity is negatively correlated with the performance of AttrE and BERT\_INT.

As we discussed in Sections~\ref{relations} and \ref{negative}, a high number of average relations per entity improves the quality of embeddings. Specifically, entities have the opportunity to minimize their embedding distance with multiple similar entities (neighbors). Also, the more relations we have per entity, the more negative samples we will consider, and as a result, dissimilar entities will be placed farther in the embedding space. Sole and hyper meta-features are essentially alternative measures of KG density. When we have many sole relations, many relation types will never co-occur with others resulting to a low average number of relations per entity (low density). Inversely, when we have many hyper relations, there will be many co-occurring relation types resulting to a high average relations per entity (high density).

The meta-feature measuring the average number of attributes per entity is relevant only to attribute-based methods that exploit similarity of literals for the embeddings, such as MultiKE, AttrE and BERT\_INT. As we can see in Table~\ref{tab:correlations}, the performance of AttrE and BERT\_INT is negatively correlated with this meta-feature. This happens because as the average number of attributes per entity increases, the noise (irrelevant attribute values) per entity is increasing too, especially if the literals similarity is low. The correlation of AttrE with the literals similarity is examined in Section~\ref{hom_het}.

\subsubsection{Heterogeneity}\label{hom_het}
As the only attribute-based semi-supervised method, KDCoE exploits textual descriptions to enrich the seed alignment with new entity pairs that have very similar descriptions. The presence of many entities with highly similar (Descr\_Sim) textual descriptions (\#Ents\_Descr) boosts the performance of KDCoE. For this reason, we observe a positive correlation of these two meta-features with KDCoE. In addition, MultiKE and RDGCN are the only supervised methods that exploit entity names to create and initialize the entity embeddings. Thus, we observe a positive correlation between those methods and entity name similarity. AttrE is the only unsupervised method that is not using pre-trained word embeddings and thus is positively correlated with the meta-feature capturing the similarity of literals (Lit\_Sim). A higher literal similarity is associated with better performance for the method. Finally, as AttrE exploits the Levenshtein distance to align the predicate names of two KGs, its performance is positively correlated with this meta-feature (Pred\_Name\_Sim): the more similar the predicates names of the two KGs, the better for AttrE. Although BERT\_INT, exploits multiple resources of similarity evidence of entities (e.g., textual descriptions, entity names, literal), it uses pre-trained BERT-based word embeddings, so measuring correlations with these meta-features, is irrelevant.

\subsection{Lessons Learned}\label{lessons_learned}

In this section, we summarize the main conclusions drawn from our experiments that shed light on the four open questions of our empirical study.

\begin{enumerate}
\item[\textbf{Q1.}] \emph{What are the critical factors that affect the effectiveness of relation-based and attribute-based methods and how sensitive are the methods to hyperparameters tuning?}

\medskip
\noindent \emph{Negative sampling} and the range of the entity neighborhood are the two critical factors that affect the performance (see Table~\ref{tab:scores}) of the relation-based methods. Specifically, negative sampling helps the methods to exploit the rich semantic information that dense KGs offer, in order to distantiate dissimilar entities in an embedding space with limited dimensions. However, negative sampling harms the effectiveness of methods when the KGs are very sparse (average relations per entity $<$ 3). This is due to the fact that similar entities lie already far in the embedding space and negative sampling further increases their distance: the more negative samples per positive sample (see Table~\ref{tab:categories2}) of KGs, the more the distance increases. 

Moreover, by increasing the \emph{range of the neighborhood} that a method exploits, information from both close and distant neighbors is aggregated to learn entity embeddings. However, an increased number of features for learning embeddings, will also increase the dimensionality of the embedding space under the default neural-network architecture that each method use. Additionally, multi-hop relations entails the risk of modeling noisy information from irrelevant distant neighbors. To alleviate the noise in embeddings, we need either a high number of negative samples or attention mechanisms that weight relevant and important neighbors.

\emph{Interaction-based} methods that compare the pairwise similarity of attributes' values of both close and distant neighbors, are less affected by negative sampling on sparse KGs. They can also cope with the noise introduced by the aggregation of distant neighbors, but they require highly similar attributes, independently of the density of the KGs. As a matter of fact, any additional source of similarity evidence, such as textual descriptions (KDCoE, BERT\_INT), literals (MultiKE, AttrE and BERT\_INT),  entity names (MultiKE, BERT\_INT) or predicate names (AttrE), boosts the performance of attribute-based methods.

\medskip
\item[\textbf{Q2.}] \emph{What is the improvement in the effectiveness of embedding-based entity alignment methods, if we consider not only the structural relations of entities, but also their attribute values?}

\medskip
\noindent The improvement in the effectiveness (see Table~\ref{tab:scores}) of embedding-based EA methods that exploit the factual information of entities depends on the characteristics of the input KGs. In the dense KGs of the OpenEA datasets, where the similarity of factual information (e.g., textual descriptions and entity names) is low, exploiting only the structural neighborhood of entities seems sufficient for a method to achieve a competitive performance. However, when the similarity of factual information is high, the effectiveness of the methods in the dense KGs is significantly improved, approaching the perfect matching. In the sparse KGs of the new datasets containing a high number of textual descriptions and very similar literals and predicates, the attribute-based methods outperform most relation-based methods, especially BERT\_INT that is the overall winning method.

In particular, among all embedding-based methods, both versions of RREA seem to achieve the best performance in dense KGs, when the similarity of factual information (e.g., textual descriptions and entity names) is low, since RREA exploits multi-hop neighborhoods and utilizes attention mechanism that weights relevant and important neighbors. In contrast, when the similarity of factual information is high, BERT\_INT is the best method, since it seems to properly handle it, achieving very high performance. However, in sparse KGs, RREA is outperformed by BERT\_INT that achieves the best performance among all methods in the new datasets, as in these datasets, the literals similarity is higher than in the OpenEA datasets.

The same improvement is also observed in PARIS (see Table~\ref{tab:PARIS}) that outperforms the best embedding-based methods when textual descriptions and entity names similarity is low in dense KGs. However when the textual descriptions and entity names similarity is high, PARIS is outperformed by the best embedding-based method. Nonetheless, even in those cases, PARIS's precision remains high, because it uses only functional attributes. In sparse KGs, when literal similarity is high, PARIS is outperformed by the best embedding-based method.

\medskip
\item[\textbf{Q3.}] \emph{Is the runtime overhead of each method worth paying, with respect to the achieved effectiveness?}

\medskip
\noindent We conclude that the training time of BERT\_INT is a cost worth spending, since the effectiveness improvement is significant, compared to faster methods, like MTransE (Figure~\ref{fig:trade}). Moreover, from Figure~\ref{fig:TTA}, we conclude that BERT\_INT reaches 90\% of its highest MRR faster than the other methods, employing only 1 or 2 epochs. It is worth noticing that in dense KGs with low similarity in factual information (e.g., textual descriptions and entity names), BERT\_INT is not as effective as RREA(basic). In such KGs RREA(basic) should be preferred, although it is slower.

\medskip
\item[\textbf{Q4.}] \emph{To which characteristics of the datasets are supervised, semi-supervised and unsupervised methods sensitive?}

\medskip
\noindent The effectiveness of supervised and semi-supervised methods is positively correlated with the size of the seed alignment and the density of the KGs (see Table~\ref{tab:correlations}). Although RDGCN is a supervised method, it is also affected by the degree of similarity between the entity names, as entity embeddings are initialized using the entity name embeddings. Concerning unsupervised methods, the effectiveness of AttrE is negatively correlated to the size of the seed alignment, since it increases the probability of making wrong alignment decisions without improving its learning. In addition, data characteristics such as the similarity of entity names, predicate names, or textual descriptions and literals, are positively correlated with the corresponding relevant method. More precisely, the effectiveness of KDCoE is positively correlated with the number and similarity of textual descriptions, MultiKE with the entity names similarity and AttrE with the predicate names and literals similarity. The unsupervised nature of AttrE justifies the different correlations found w.r.t. other attribute-based methods, in meta-features that refer to the seed alignment size and average relations per entity. In particular, AttrE relies on  attribute values (literals) rather than seed alignment, to measure the similarity of entities. Its effectiveness is boosted when the aligned entities have not only a high number of attributes, but also their values are highly similar. In the opposite case, the aligning process is adversely penalized by the noise that is created. BERT\_INT exhibits a similar behavior.

\end{enumerate}

 \section{Conclusions and Future Work}\label{section:five}

In this work, we have experimentally evaluated several translational- and Graph Neural Networks-based methods for implementing EA tasks using KG embeddings, as well as, non-learning methods like PARIS. They essentially cover a wide range of supervised, to semi-supervised and unsupervised methods that exploit both relational and factual information of entity descriptions along with different negative sampling and neighborhood range (one- vs multi-hop) strategies. We have measured both the effectiveness and efficiency of all methods over a rich collection of datasets that exhibit different characteristics of entity descriptions provided by KGs.
 
According to our analysis, \emph{negative sampling} proves to be a critical factor that affects the performance of both relation- and attribute-based KG embeddings methods, as it helps distancing dissimilar entities in the embedding space. Another critical factor that affects the performance of relation-based methods is the \emph{attention mechanism} that weights relevant and important neighbors. Moreover, the performance of relation-based methods depends on the \emph{range of the entity neighborhood (one-hop or multi-hop)}. In particular, we observed that multi-hop methods, require an increased number of negative samples or attention mechanisms to compensate noisy information in the produced embeddings from far distant neighbors. As expected, attribute-based methods (independently of the usage of KG embeddings) are more affected by the different types of factual information exploited in embeddings, such as textual descriptions. In addition, \textit{interaction-based} GNNs seem to better cope with the noise of distant neighbors and improve the utilization of attributes for matching, compared to aggregation-based GNNs. We should notice that the unsupervised attribute-based method AttrE, can run only if the predicates of the different KGs have been sufficiently aligned and it does not impose the 1-to-1 mapping assumption. Supervised and semi-supervised methods do not suffer from this limitation. We conclude that BERT\_INT dominates all methods in terms of effectiveness and efficiency overall, especially when the KGs contain high similar factual information. In dense KGs with low factual similarity, RREA and RDGCN achieve the best and the second best performance, respectively, as also reported in~\cite{DBLP:journals/pvldb/SunZHWCAL20,DBLP:conf/coling/ZhangLCCLXZ20,zhao2020experimental}. MTransE is the fastest method (due to the lack of negative sampling and its simple architecture - see Sections~\ref{negative} and~\ref{architecture}), but it suffers from much lower effectiveness compared to the other methods.

As future work, we would like to benchmark embedding-based EA methods exploiting different Neural Network architectures, e.g., Generative Adversarial~\citep{DBLP:conf/sigmod/Vretinaris0EQO21}, Self-Adversarial~\citep{DBLP:conf/iclr/SunDNT19}, Graph Transformer~\citep{DBLP:conf/ijcai/CaiMZJ22}. Moreover, we plan to extend our testbed with datasets proposed in OAEI\footnote{\url{https://oaei.ontologymatching.org/2022/knowledgegraph/index.html}} that exhibit a higher structural heterogeneity between the KGs to be aligned (e.g., in terms of average degrees). Finally, recent works have revealed that KG embedding methods often exhibit \emph{direct} or \emph{indirect} forms of bias, leading to discrimination~\citep{DBLP:conf/cikm/EfthymiouSPC21,fisher-etal-2020-debiasing}. We are particularly interested in assessing the robustness of embedding-based EA methods with respect to increasing structural diversity (i.e., in terms of connected components) of KGs, that reflect indirect forms of bias~\citep{DBLP:conf/eswc/FanourakisECK23}.}

\section*{Declarations}

\subsection*{Funding}
This work has received funding from the Hellenic  Foundation for Research and Innovation (HFRI) and the General Secretariat for Research and Technology (GSRT), under grant agreement No 969.

\subsection*{Conflicts of interest/Competing interests}
The authors have no conflicts of interest to declare that are relevant to the content of this article.

\backmatter


\bibliography{bibliography}

\begin{thebibliography}{59}
\providecommand{\natexlab}[1]{#1}
\providecommand{\url}[1]{{#1}}
\providecommand{\urlprefix}{URL }
\providecommand{\doi}[1]{\url{https://doi.org/#1}}
\providecommand{\eprint}[2][]{\url{#2}}
 \bibcommenthead

\bibitem[{Ahmetaj et~al(2021)Ahmetaj, Efthymiou, Fagin, Kolaitis, Lei,
  {\"{O}}zcan, and Popa}]{DBLP:conf/aaai/AhmetajEFKLO021}
Ahmetaj S, Efthymiou V, Fagin R, et~al (2021) Ontology-enriched query answering
  on relational databases. In: {AAAI}, pp 15247--15254

\bibitem[{Berrendorf et~al(2020)Berrendorf, Faerman, Vermue, and
  Tresp}]{berrendorf2020ambiguity}
Berrendorf M, Faerman E, Vermue L, et~al (2020) On the ambiguity of rank-based
  evaluation of entity alignment or link prediction methods. CoRR
  abs/2002.06914

\bibitem[{Bordes et~al(2013)Bordes, Usunier, Garc{\'{\i}}a{-}Dur{\'{a}}n,
  Weston, and Yakhnenko}]{DBLP:conf/nips/BordesUGWY13}
Bordes A, Usunier N, Garc{\'{\i}}a{-}Dur{\'{a}}n A, et~al (2013) Translating
  embeddings for modeling multi-relational data. In: {NeurIPS}, pp 2787--2795

\bibitem[{Cai et~al(2022)Cai, Ma, Zhan, and Jiang}]{DBLP:conf/ijcai/CaiMZJ22}
Cai W, Ma W, Zhan J, et~al (2022) Entity alignment with reliable path reasoning
  and relation-aware heterogeneous graph transformer. In: {IJCAI}, pp
  1930--1937

\bibitem[{Cao et~al(2019)Cao, Liu, Li, Liu, Li, and
  Chua}]{DBLP:journals/corr/abs-1908-09898}
Cao Y, Liu Z, Li C, et~al (2019) Multi-channel graph neural network for entity
  alignment. In: {ACL}, pp 1452--1461

\bibitem[{Chaurasiya et~al(2022)Chaurasiya, Surisetty, Kumar, Singh, Dey,
  Malhotra, Dhama, and Arora}]{DBLP:journals/corr/abs-2205-08777}
Chaurasiya D, Surisetty A, Kumar N, et~al (2022) Entity alignment for knowledge
  graphs: Progress, challenges, and empirical studies. CoRR abs/2205.08777

\bibitem[{Chen et~al(2017)Chen, Tian, Yang, and
  Zaniolo}]{DBLP:conf/ijcai/ChenTYZ17}
Chen M, Tian Y, Yang M, et~al (2017) Multilingual knowledge graph embeddings
  for cross-lingual knowledge alignment. In: {IJCAI}, pp 1511--1517

\bibitem[{Chen et~al(2018)Chen, Tian, Chang, Skiena, and
  Zaniolo}]{DBLP:conf/ijcai/ChenTCSZ18}
Chen M, Tian Y, Chang K, et~al (2018) Co-training embeddings of knowledge
  graphs and entity descriptions for cross-lingual entity alignment. In:
  {IJCAI}, pp 3998--4004

\bibitem[{Choudhary et~al(2021)Choudhary, Luthra, Mittal, and
  Singh}]{DBLP:journals/corr/abs-2107-07842}
Choudhary S, Luthra T, Mittal A, et~al (2021) A survey of knowledge graph
  embedding and their applications. CoRR abs/2107.07842

\bibitem[{Christophides et~al(2015)Christophides, Efthymiou, and
  Stefanidis}]{DBLP:series/synthesis/2015Christophides}
Christophides V, Efthymiou V, Stefanidis K (2015) Entity Resolution in the Web
  of Data. Synthesis Lectures on the Semantic Web: Theory and Technology,
  Morgan {\&} Claypool Publishers, San Rafael, California

\bibitem[{Christophides et~al(2021)Christophides, Efthymiou, Palpanas,
  Papadakis, and Stefanidis}]{DBLP:journals/csur/ChristophidesEP21}
Christophides V, Efthymiou V, Palpanas T, et~al (2021) An overview of
  end-to-end entity resolution for big data. {ACM} Comput Surv
  53(6):127:1--127:42

\bibitem[{Coleman et~al(2017)Coleman, Narayanan, Kang, Zhao, Zhang, Nardi,
  Bailis, Olukotun, R{\'e}, and Zaharia}]{coleman2017dawnbench}
Coleman C, Narayanan D, Kang D, et~al (2017) Dawnbench: An end-to-end deep
  learning benchmark and competition. Training 100(101):102

\bibitem[{Coleman et~al(2019)Coleman, Kang, Narayanan, Nardi, Zhao, Zhang,
  Bailis, Olukotun, R{\'{e}}, and
  Zaharia}]{DBLP:journals/sigops/ColemanKNNZZBOR19}
Coleman C, Kang D, Narayanan D, et~al (2019) Analysis of dawnbench, a
  time-to-accuracy machine learning performance benchmark. {ACM} {SIGOPS} Oper
  Syst Rev 53(1):14--25

\bibitem[{Demsar(2006)}]{DBLP:journals/jmlr/Demsar06}
Demsar J (2006) Statistical comparisons of classifiers over multiple data sets.
  J Mach Learn Res 7:1--30

\bibitem[{Devlin et~al(2019)Devlin, Chang, Lee, and
  Toutanova}]{DBLP:conf/naacl/DevlinCLT19}
Devlin J, Chang M, Lee K, et~al (2019) {BERT:} pre-training of deep
  bidirectional transformers for language understanding. In: {NAACL-HLT}, pp
  4171--4186

\bibitem[{Dong et~al(2014)Dong, Gabrilovich, Heitz, Horn, Lao, Murphy,
  Strohmann, Sun, and Zhang}]{DBLP:conf/kdd/0001GHHLMSSZ14}
Dong X, Gabrilovich E, Heitz G, et~al (2014) Knowledge vault: a web-scale
  approach to probabilistic knowledge fusion. In: {SIGKDD}, pp 601--610

\bibitem[{Efthymiou et~al(2015)Efthymiou, Stefanidis, and
  Christophides}]{DBLP:conf/bigdataconf/EfthymiouSC15}
Efthymiou V, Stefanidis K, Christophides V (2015) Big data entity resolution:
  From highly to somehow similar entity descriptions in the web. In: {IEEE} Big
  Data, pp 401--410

\bibitem[{Efthymiou et~al(2022)Efthymiou, Stefanidis, Pitoura, and
  Christophides}]{DBLP:conf/cikm/EfthymiouSPC21}
Efthymiou V, Stefanidis K, Pitoura E, et~al (2022) {FairER}: Entity resolution
  with fairness constraints. In: {CIKM}, pp 3004--3008

\bibitem[{Fanourakis et~al(2023)Fanourakis, Efthymiou, Christophides, Kotzinos,
  Pitoura, and Stefanidis}]{DBLP:conf/eswc/FanourakisECK23}
Fanourakis N, Efthymiou V, Christophides V, et~al (2023) Structural bias in
  knowledge graphs for the entity alignment task. In: {ESWC}

\bibitem[{Fisher et~al(2020)Fisher, Mittal, Palfrey, and
  Christodoulopoulos}]{fisher-etal-2020-debiasing}
Fisher J, Mittal A, Palfrey D, et~al (2020) Debiasing knowledge graph
  embeddings. In: EMNLP, pp 7332--7345

\bibitem[{Jiang et~al(2021)Jiang, Li, and Gu}]{DBLP:conf/dsc/JiangLG21}
Jiang J, Li M, Gu Z (2021) A survey on translating embedding based entity
  alignment in knowledge graphs. In: {DSC}, pp 187--194

\bibitem[{Kamigaito and Hayashi(2022)}]{DBLP:conf/icml/KamigaitoH22}
Kamigaito H, Hayashi K (2022) Comprehensive analysis of negative sampling in
  knowledge graph representation learning. In: {ICML}, pp 10661--10675

\bibitem[{Kipf and Welling(2017)}]{DBLP:conf/iclr/KipfW17}
Kipf TN, Welling M (2017) Semi-supervised classification with graph
  convolutional networks. In: {ICLR}

\bibitem[{Kocmi and Bojar(2017)}]{DBLP:conf/icon-nlp/KocmiB17}
Kocmi T, Bojar O (2017) An exploration of word embedding initialization in
  deep-learning tasks. In: {ICON}, pp 56--64

\bibitem[{Lehmann et~al(2015)Lehmann, Isele, Jakob, Jentzsch, Kontokostas,
  Mendes, Hellmann, Morsey, van Kleef, Auer, and
  Bizer}]{DBLP:journals/semweb/LehmannIJJKMHMK15}
Lehmann J, Isele R, Jakob M, et~al (2015) Dbpedia - {A} large-scale,
  multilingual knowledge base extracted from wikipedia. Semantic Web
  6(2):167--195

\bibitem[{Leone et~al(2022)Leone, Huber, Arora, Garc{\'{\i}}a{-}Dur{\'{a}}n,
  and West}]{DBLP:journals/pvldb/LeoneHAGW22}
Leone M, Huber S, Arora A, et~al (2022) A critical re-evaluation of neural
  methods for entity alignment. {PVLDB} 15(8):1712--1725

\bibitem[{Lin(1991)}]{DBLP:journals/tit/Lin91}
Lin J (1991) Divergence measures based on the shannon entropy. {IEEE} Trans Inf
  Theory 37(1):145--151

\bibitem[{Mao et~al(2020{\natexlab{a}})Mao, Wang, Xu, Lan, and
  Wu}]{mao2020mraea}
Mao X, Wang W, Xu H, et~al (2020{\natexlab{a}}) {MRAEA}: an efficient and
  robust entity alignment approach for cross-lingual knowledge graph. In:
  {WSDM}, pp 420--428

\bibitem[{Mao et~al(2020{\natexlab{b}})Mao, Wang, Xu, Wu, and
  Lan}]{DBLP:conf/cikm/MaoWXWL20}
Mao X, Wang W, Xu H, et~al (2020{\natexlab{b}}) Relational reflection entity
  alignment. In: {CIKM}, pp 1095--1104

\bibitem[{Million(2007)}]{million2007hadamard}
Million E (2007) The {Hadamard} product. Course Notes 3(6)

\bibitem[{Nemenyi(1963)}]{nemenyi1963distribution}
Nemenyi P (1963) Distribution-free Multiple Comparisons. Princeton University,
  Princeton

\bibitem[{Nickel and Kiela(2017)}]{DBLP:conf/nips/NickelK17}
Nickel M, Kiela D (2017) Poincar{\'{e}} embeddings for learning hierarchical
  representations. In: {NeurIPS}, pp 6338--6347

\bibitem[{Obraczka et~al(2021)Obraczka, Schuchart, and
  Rahm}]{DBLP:journals/corr/abs-2101-06126}
Obraczka D, Schuchart J, Rahm E (2021) {EAGER:} embedding-assisted entity
  resolution for knowledge graphs. CoRR abs/2101.06126

\bibitem[{Parisi et~al(2022)Parisi, Neagu, Ma, and
  Campean}]{DBLP:journals/eswa/ParisiNMC22}
Parisi L, Neagu D, Ma R, et~al (2022) Quantum relu activation for convolutional
  neural networks to improve diagnosis of parkinson's disease and {COVID-19}.
  Expert Syst Appl 187:115892

\bibitem[{Rebele et~al(2016)Rebele, Suchanek, Hoffart, Biega, Kuzey, and
  Weikum}]{DBLP:conf/semweb/RebeleSHBKW16}
Rebele T, Suchanek FM, Hoffart J, et~al (2016) {YAGO:} {A} multilingual
  knowledge base from wikipedia, wordnet, and geonames. In: {ISWC}, pp 177--185

\bibitem[{Suchanek et~al(2011)Suchanek, Abiteboul, and
  Senellart}]{DBLP:journals/pvldb/SuchanekAS11}
Suchanek FM, Abiteboul S, Senellart P (2011) {PARIS:} probabilistic alignment
  of relations, instances, and schema. {PVLDB} 5(3):157--168

\bibitem[{Sun et~al(2018)Sun, Hu, Zhang, and Qu}]{DBLP:conf/ijcai/SunHZQ18}
Sun Z, Hu W, Zhang Q, et~al (2018) Bootstrapping entity alignment with
  knowledge graph embedding. In: {IJCAI}, pp 4396--4402

\bibitem[{Sun et~al(2019)Sun, Deng, Nie, and Tang}]{DBLP:conf/iclr/SunDNT19}
Sun Z, Deng Z, Nie J, et~al (2019) Rotate: Knowledge graph embedding by
  relational rotation in complex space. In: {ICLR}

\bibitem[{Sun et~al(2020)Sun, Zhang, Hu, Wang, Chen, Akrami, and
  Li}]{DBLP:journals/pvldb/SunZHWCAL20}
Sun Z, Zhang Q, Hu W, et~al (2020) A benchmarking study of embedding-based
  entity alignment for knowledge graphs. {PVLDB} 13(11):2326--2340

\bibitem[{Tang et~al(2020)Tang, Zhang, Chen, Yang, Chen, and
  Li}]{DBLP:conf/ijcai/Tang0C00L20}
Tang X, Zhang J, Chen B, et~al (2020) {BERT-INT:} {A} bert-based interaction
  model for knowledge graph alignment. In: {IJCAI}, pp 3174--3180

\bibitem[{Tarus et~al(2018)Tarus, Niu, and
  Mustafa}]{DBLP:journals/air/TarusNM18}
Tarus JK, Niu Z, Mustafa G (2018) Knowledge-based recommendation: a review of
  ontology-based recommender systems for e-learning. Artif Intell Rev
  50(1):21--48

\bibitem[{Trisedya et~al(2019)Trisedya, Qi, and
  Zhang}]{DBLP:conf/aaai/TrisedyaQZ19}
Trisedya BD, Qi J, Zhang R (2019) Entity alignment between knowledge graphs
  using attribute embeddings. In: {AAAI}, pp 297--304

\bibitem[{Trouillon et~al(2016)Trouillon, Welbl, Riedel, Gaussier, and
  Bouchard}]{DBLP:conf/icml/TrouillonWRGB16}
Trouillon T, Welbl J, Riedel S, et~al (2016) Complex embeddings for simple link
  prediction. In: {ICML}, pp 2071--2080

\bibitem[{Velickovic et~al(2018)Velickovic, Cucurull, Casanova, Romero,
  Li{\`{o}}, and Bengio}]{DBLP:journals/corr/abs-1710-10903}
Velickovic P, Cucurull G, Casanova A, et~al (2018) Graph attention networks.
  In: {ICLR}

\bibitem[{Vrandecic and Kr{\"{o}}tzsch(2014)}]{DBLP:journals/cacm/VrandecicK14}
Vrandecic D, Kr{\"{o}}tzsch M (2014) Wikidata: a free collaborative
  knowledgebase. Commun {ACM} 57:78--85.
  \urlprefix\url{https://doi.org/10.1145/2629489}

\bibitem[{Vretinaris et~al(2021)Vretinaris, Lei, Efthymiou, Qin, and
  {\"{O}}zcan}]{DBLP:conf/sigmod/Vretinaris0EQO21}
Vretinaris A, Lei C, Efthymiou V, et~al (2021) Medical entity disambiguation
  using graph neural networks. In: {SIGMOD}, pp 2310--2318

\bibitem[{Wang et~al(2017)Wang, Mao, Wang, and
  Guo}]{DBLP:journals/tkde/WangMWG17}
Wang Q, Mao Z, Wang B, et~al (2017) Knowledge graph embedding: {A} survey of
  approaches and applications. {IEEE} Trans Knowl Data Eng 29(12):2724--2743

\bibitem[{Wang et~al(2020)Wang, Yang, and Ye}]{DBLP:conf/emnlp/WangYY20}
Wang Z, Yang J, Ye X (2020) Knowledge graph alignment with entity-pair
  embedding. In: {EMNLP}, pp 1672--1680

\bibitem[{Wang et~al(2021)Wang, Li, and Gu}]{DBLP:conf/dsc/WangLG21}
Wang Z, Li M, Gu Z (2021) A review of entity alignment based on graph
  convolutional neural network. In: {DSC}, pp 144--151

\bibitem[{Wu et~al(2019)Wu, Liu, Feng, Wang, Yan, and
  Zhao}]{DBLP:conf/ijcai/WuLF0Y019}
Wu Y, Liu X, Feng Y, et~al (2019) Relation-aware entity alignment for
  heterogeneous knowledge graphs. In: {IJCAI}, pp 5278--5284

\bibitem[{Xiong et~al(2017)Xiong, Dai, Callan, Liu, and
  Power}]{DBLP:conf/sigir/XiongDCLP17}
Xiong C, Dai Z, Callan J, et~al (2017) End-to-end neural ad-hoc ranking with
  kernel pooling. In: {SIGIR}, pp 55--64

\bibitem[{Yang et~al(2015)Yang, Yih, He, Gao, and
  Deng}]{DBLP:journals/corr/YangYHGD14a}
Yang B, Yih W, He X, et~al (2015) Embedding entities and relations for learning
  and inference in knowledge bases. In: {ICLR}

\bibitem[{Zeng et~al(2021)Zeng, Li, Hou, Li, and Feng}]{zeng2021comprehensive}
Zeng K, Li C, Hou L, et~al (2021) A comprehensive survey of entity alignment
  for knowledge graphs. AI Open 2:1--13

\bibitem[{Zhang et~al(2017)Zhang, Zhou, Han, Hu, and Ji}]{zhang2017knowledge}
Zhang C, Zhou M, Han X, et~al (2017) Knowledge graph embedding for
  hyper-relational data. Tsinghua Science and Technology 22(2):185--197

\bibitem[{Zhang et~al(2019)Zhang, Sun, Hu, Chen, Guo, and
  Qu}]{DBLP:conf/ijcai/ZhangSHCGQ19}
Zhang Q, Sun Z, Hu W, et~al (2019) Multi-view knowledge graph embedding for
  entity alignment. In: {IJCAI}, pp 5429--5435

\bibitem[{Zhang et~al(2022)Zhang, Trisedya, Li, Jiang, and
  Qi}]{DBLP:journals/vldb/ZhangTLJQ22}
Zhang R, Trisedya BD, Li M, et~al (2022) A benchmark and comprehensive survey
  on knowledge graph entity alignment via representation learning. {VLDB}J
  31(5):1143--1168

\bibitem[{Zhang et~al(2020)Zhang, Liu, Chen, Chen, Liu, Xiang, and
  Zheng}]{DBLP:conf/coling/ZhangLCCLXZ20}
Zhang Z, Liu H, Chen J, et~al (2020) An industry evaluation of embedding-based
  entity alignment. In: {COLING}, pp 179--189

\bibitem[{Zhao et~al(2022)Zhao, Zeng, Tang, Wang, and
  Suchanek}]{zhao2020experimental}
Zhao X, Zeng W, Tang J, et~al (2022) An experimental study of state-of-the-art
  entity alignment approaches. {IEEE} Trans Knowl Data Eng 34(6):2610--2625

\bibitem[{Zhu et~al(2019)Zhu, Zhou, Wu, Tan, and
  Guo}]{DBLP:conf/ijcai/ZhuZ0TG19}
Zhu Q, Zhou X, Wu J, et~al (2019) Neighborhood-aware attentional representation
  for multilingual knowledge graphs. In: {IJCAI}, pp 1943--1949

\end{thebibliography}

\end{document}